\documentclass[submit]{smj}

\usepackage[utf8]{inputenc} 
\usepackage[T1]{fontenc}    
\usepackage{url}            
\usepackage{booktabs}       
\usepackage{amsfonts}       
\usepackage{nicefrac}       
\usepackage{microtype}      
\usepackage{lipsum}
\usepackage{natbib}

\usepackage[english]{babel}
\usepackage{amsmath}
\usepackage{graphicx}
\usepackage{multirow}
\usepackage{amsthm}
\usepackage{amssymb}
\usepackage{bbm}
\usepackage{caption}
\usepackage{float}
\usepackage{subcaption}
\usepackage{multirow}
\usepackage{makecell}
\usepackage{mathtools}
\usepackage{algorithmic}
\usepackage{bm}
\usepackage{physics}
\usepackage{appendix}
\usepackage{enumitem}
\allowdisplaybreaks

\numberwithin{thm}{section}
\numberwithin{defn}{section}
\numberwithin{property}{section}
\numberwithin{lemma}{section}
\numberwithin{cor}{section}

\newcommand{\indFct}[1]{\ensuremath{\mathbbm{I}\left(#1\right)}}

\renewcommand{\v}[1]{\ensuremath{\bm{#1}}}

\newcommand{\vM}[3]{\ensuremath{f_{\mathcal{V}}\left(#1 \mid #2, #3\right)}}
\newcommand{\vMF}[3]{\ensuremath{f_{v\mathcal{MF}}\left(#1 \mid #2, #3\right)}}

\renewcommand{\v}[1]{\ensuremath{\bm{#1}}}

\Author{Rayleigh Lei\Affil{1}\ORCID{0000-0002-0444-9708}, 
        \and Long Nguyen\Affil{2}
}
\AuthorRunning{Rayleigh Lei \textrm{et al.}}

\Affiliations{

\item Department of Statistics, 
      University of Washington,
      Seattle, WA, USA

\item Department of Statistics,
      University of Michigan, 
      Ann Arbor, MI, USA

}   

\CorrAddress{Rayleigh Lei, 
             Department of Statistics, 
             University of Washington, 
             Padelford Hall C-324, 
             4110 E Stevens Way NE, 
             Seattle, WA 98195}
\CorrEmail{rlei13@uw.edu}
\CorrPhone{(+1)\;206\;543\;7237}

\Title{Modeling random directions of changes in simplex-valued data}
\TitleRunning{Modeling random directions}

\Abstract{
We propose models and algorithms for learning about random directions in simplex-valued data. The models are applied to the study of income level proportions and their changes over time in a geostatistical area. There are several notable challenges in the analysis of simplex-valued data: the measurements must respect the simplex constraint and the changes exhibit spatiotemporal smoothness and may be heterogeneous. To that end, we propose Bayesian models that draw from and expand upon building blocks in circular and spatial statistics by exploiting a suitable transformation for the simplex-valued data. Our models also account for spatial correlation across locations in the simplex and the heterogeneous patterns via mixture modeling. We describe some properties of the models and model fitting via MCMC techniques. Our models and methods are applied to an analysis of movements and trends of income categories using the Home Mortgage Disclosure Act data. 
}


\begin{document}

\maketitle







\section{Introduction}
Modeling the variation and change of measurements that lie on the simplex is useful, but difficult. These measurements are non-negative proportions and have shown up in studies from a variety of fields, such as microbiology \citep{HolmesEtAlDirichletMultinomialMixtures2012, MaoMaDirichlettreeMultinomialMixtures2020}, geology \citep{AitchisonNewApproachNull1981, IyengarDeyBoxCoxTransformations1998}, demography \citep{MartinezEtAlBayesianAnalysisPseudocompositional2020, TsagrisStewartFoldedModelCompositional2020}, and economics \citep{FryEtAlCompositionalDataAnalysis2000,  FilzmoserEtAlAppliedCompositionalData2018}. Indeed, the motivating data set for this paper is a collection of income proportions for each census tract of Los Angeles County from 1990 to 2010 observed in the Home Mortgage Disclosure Act (HMDA) data. Understanding how these income proportions change will help sociologists and policymakers better comprehend the effects of policy changes and socioeconomic events. However, because the proportions must sum up to one, i.e. the simplicial constraint, a change in a few proportions might affect the other proportions. Thus, a model might detect correlation even if there is none in the data. Pearson called this "spurious correlation" \citep{PearsonMathematicalContributionsTheory1896}. There are also challenges with traditional techniques used to analyze the simplex-valued data. Customary compositional data approaches rely on "log-ratio" transformations to convert the data from the simplex, which is denoted by $\Delta^D$, to the unbounded $D$-dimensional space $\mathbbm{R}^{D}$ \citep{AitchisonStatisticalAnalysisCompositional1982, je2003}. Ordinary statistical techniques can then analyze the changes in transformed data, such as with a time series analysis \citep{RavishankerEtAlCompositionalTimeSeries2001}. However, if any of the proportions are zero, then the transformation becomes undefined.

To resolve these difficulties and to find a more interpretable representation of these changes, we propose the following modeling framework. The year to year changes observed in the data can be modeled as movements of points lying in the simplex. Such a model is applicable to data in the interior and at the boundary of the simplex, while avoiding spurious correlation imposed by the simplex constraints. Then, because of the isomorphism between a simplex and a positive orthant, this movement can be represented by a geodesic in the orthant that connects the (higher dimensional) spherical coordinates of the movement's starting and ending points. This geodesic can also be parametrized with the (higher dimensional) spherical coordinates if they are defined with respect to vectors other than the usual Cartesian coordinate system. Let $\v{x}_{\ell}$ be one year's point in a simplex and $\v{x'}_{\ell}$ to be the next year's point. Under the choice of coordinates such that $\v{x}_{\ell}$ is a pole, we can assign (higher dimensional) spherical coordinates, $(\theta'_2, \v{y}_{\ell})$, to $\sqrt{\v{x'}_{\ell}}$. Here, $\theta'_2 \in [0, \frac{\pi}{2}]$ and $y_{\ell} \in [0, 2\pi)$ in the 2D simplex case. In higher dimensions, $y_{\ell, d} \in [0, \pi]$ for $d = 1, 2, \ldots, D - 2$ and $y_{\ell, d} \in [0, 2\pi)$ for $d = D - 1$. Because of how $y_{\ell}$ is defined, $y_{\ell}$ is a "direction" according to which $\v{x}_{\ell}$ moves toward $\v{x'}_{\ell}$. The geodesic's length is a function of $\theta'_2$ so $\theta'_2$ designates how "far" $\v{x}_{\ell}$ goes in that direction to reach $\v{x'}_{\ell}$. Any geodesic from one point to another within the simplex can be parametrized in this way. As a result, we now have two latent variables that are interpretable and intrinsic to changes within a simplex to model. 

\begin{figure}[!tb]
\centering
\begin{subfigure}{.24\textwidth}
\centering
\includegraphics[width = 0.9\textwidth]{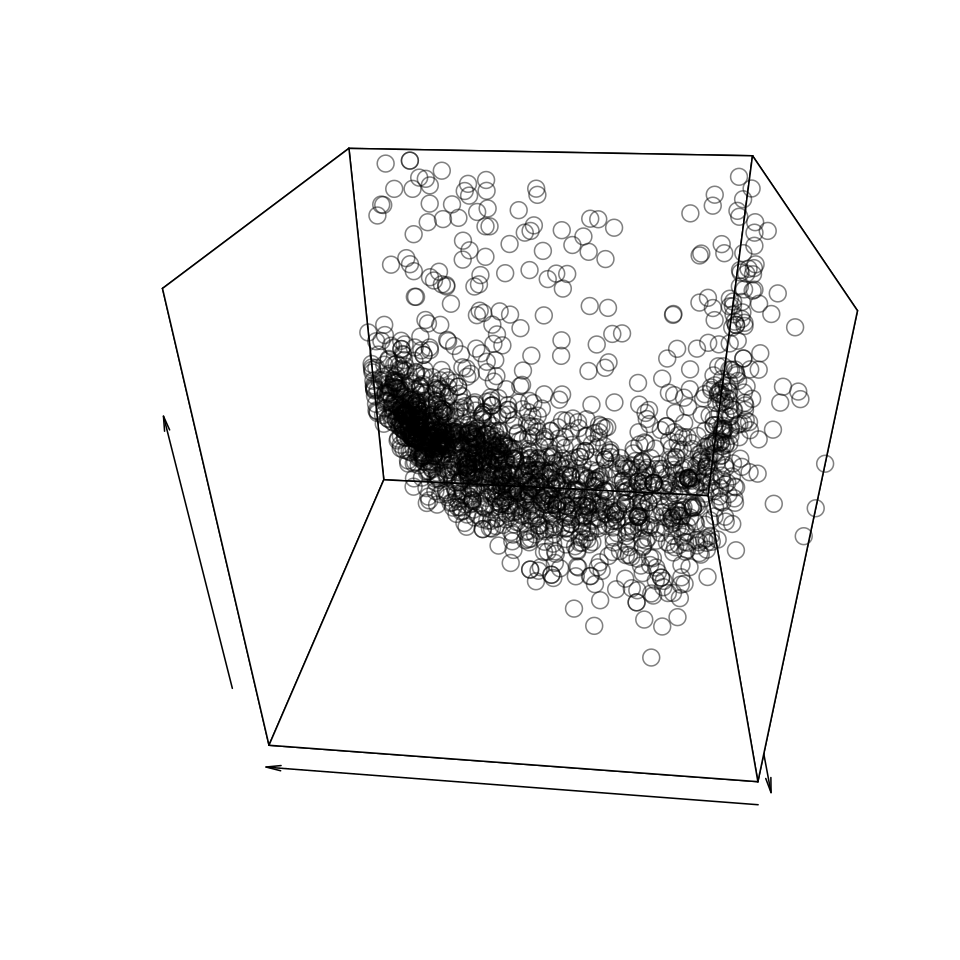}
\caption{2003-2004}
\label{fig:real_data_example_one_comp}
\end{subfigure}
\begin{subfigure}{.24\textwidth}
\centering
\includegraphics[width = 0.9\textwidth]{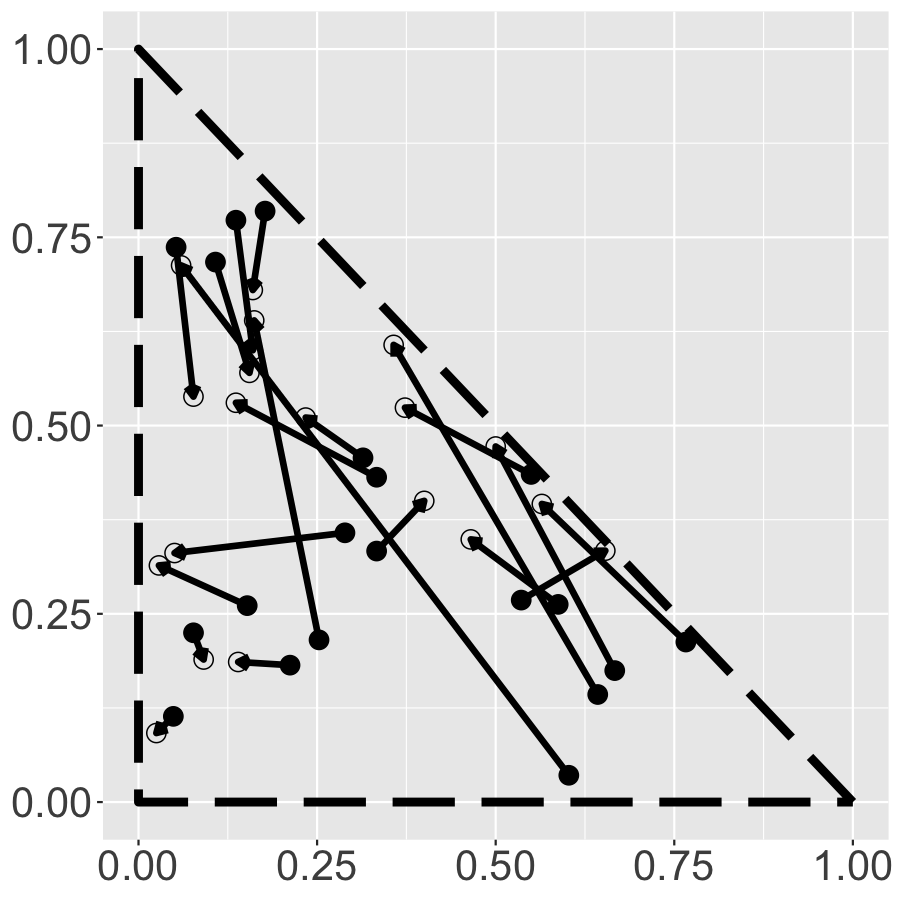}
\caption{2003-2004}
\label{fig:real_data_example_one_comp_start_end}
\end{subfigure}
\begin{subfigure}{.24\textwidth}
\centering
\includegraphics[width = 0.9\textwidth]{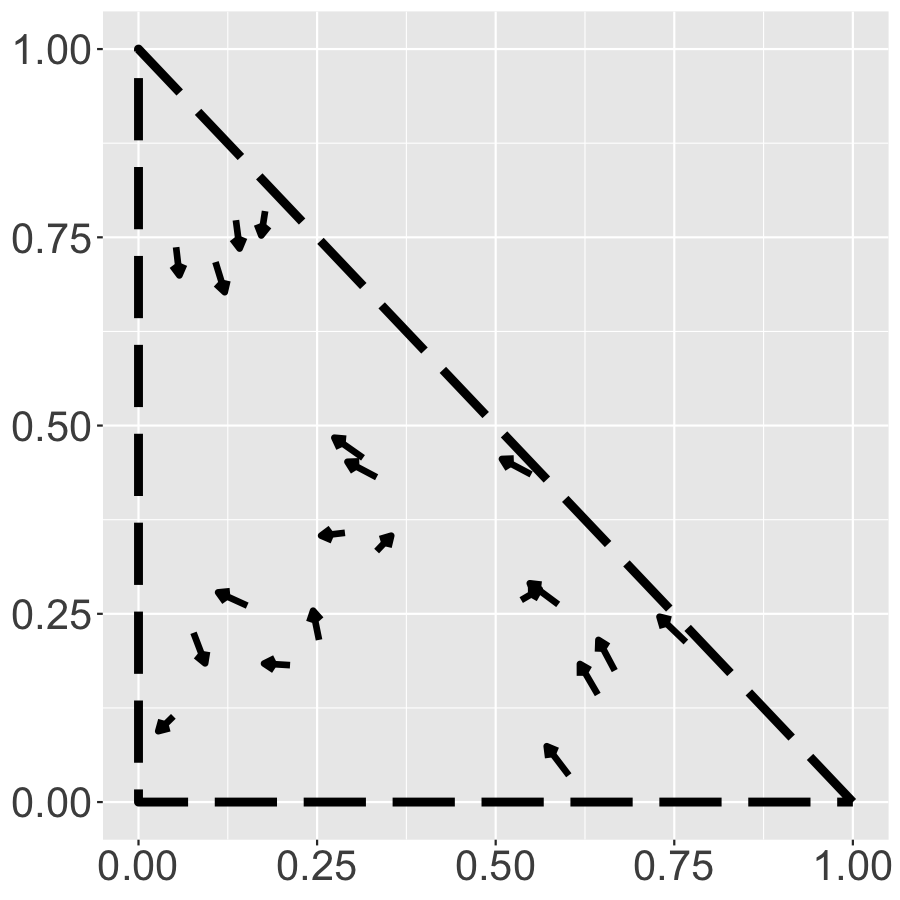}
\caption{2003-2004}
\label{fig:real_data_example_one_comp_subset}
\end{subfigure}
\begin{subfigure}{.24\textwidth}
\centering
\includegraphics[width = 0.9\textwidth]{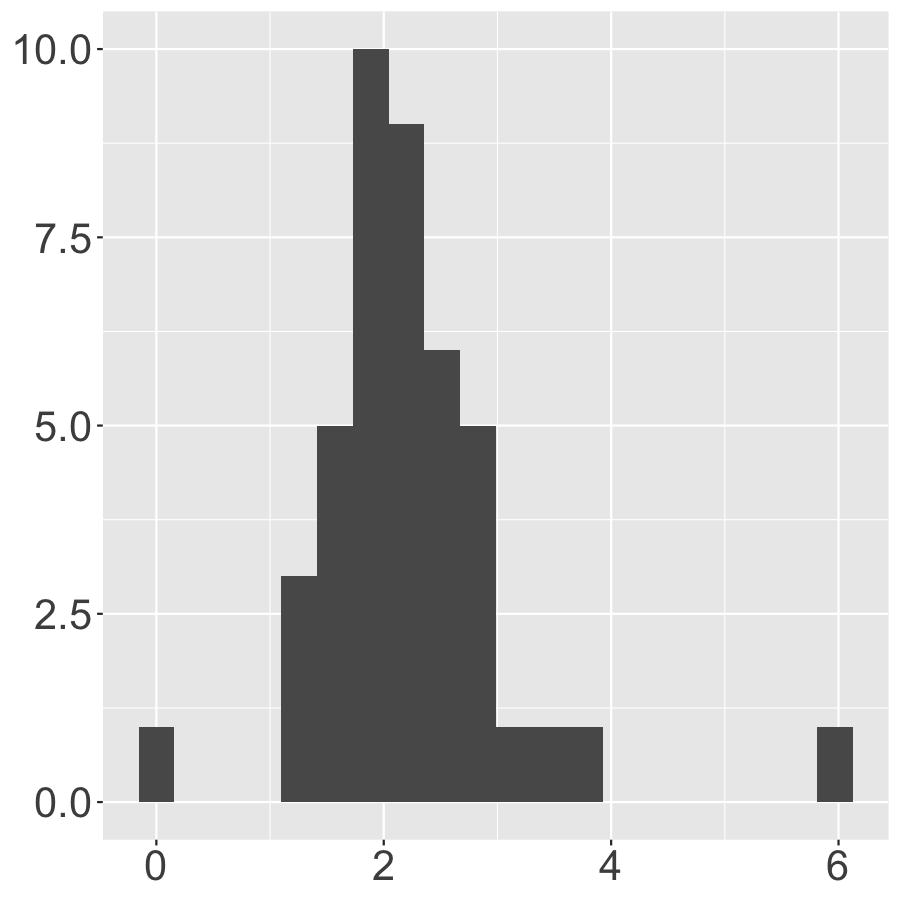}
\caption{2003-2004}
\label{fig:real_data_example_one_comp_hist}
\end{subfigure}\\
\begin{subfigure}{.24\textwidth}
\centering
\includegraphics[width = 0.9\textwidth]{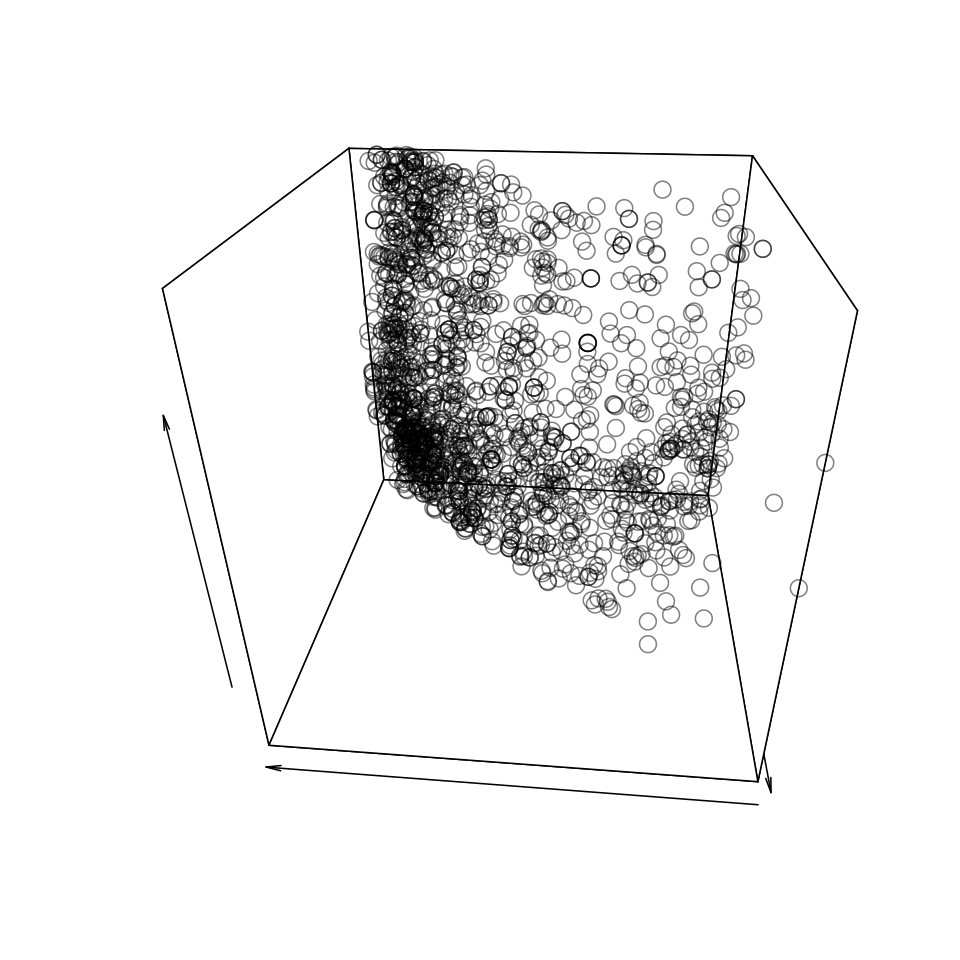}
\caption{1998-1999}
\label{fig:real_data_example_mixture}
\end{subfigure}
\begin{subfigure}{.24\textwidth}
\centering
\includegraphics[width = 0.9\textwidth]{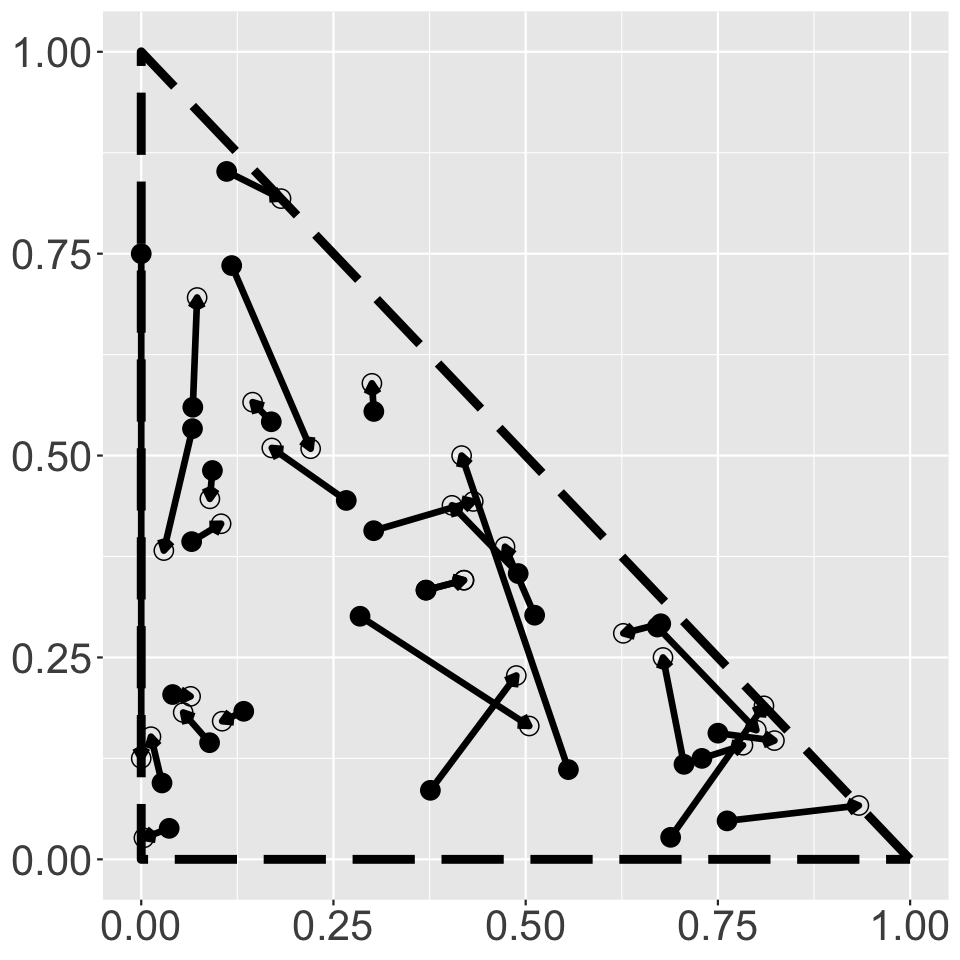}
\caption{1998-1999}
\label{fig:real_data_example_mixture_start_end}
\end{subfigure}
\begin{subfigure}{.24\textwidth}
\centering
\includegraphics[width = 0.9\textwidth]{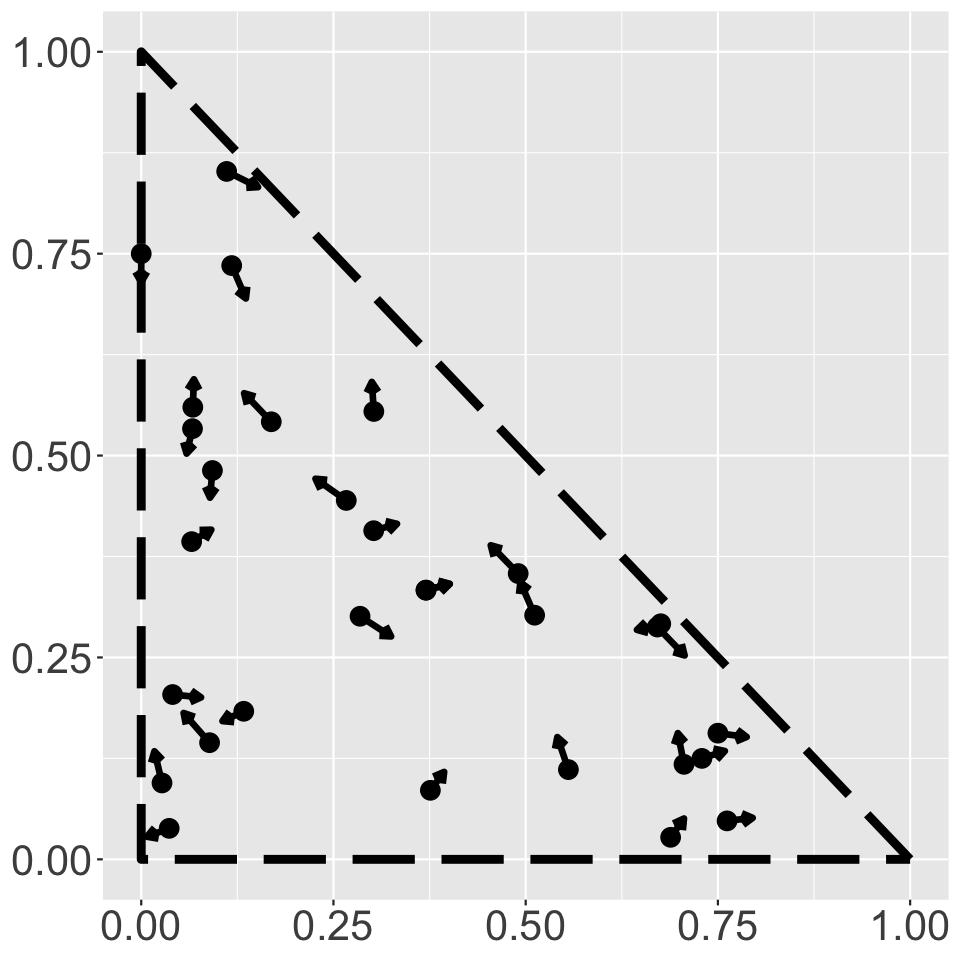}
\caption{1998-1999}
\label{fig:real_data_example_mixture_subset}
\end{subfigure}
\begin{subfigure}{.24\textwidth}
\centering
\includegraphics[width = 0.9\textwidth]{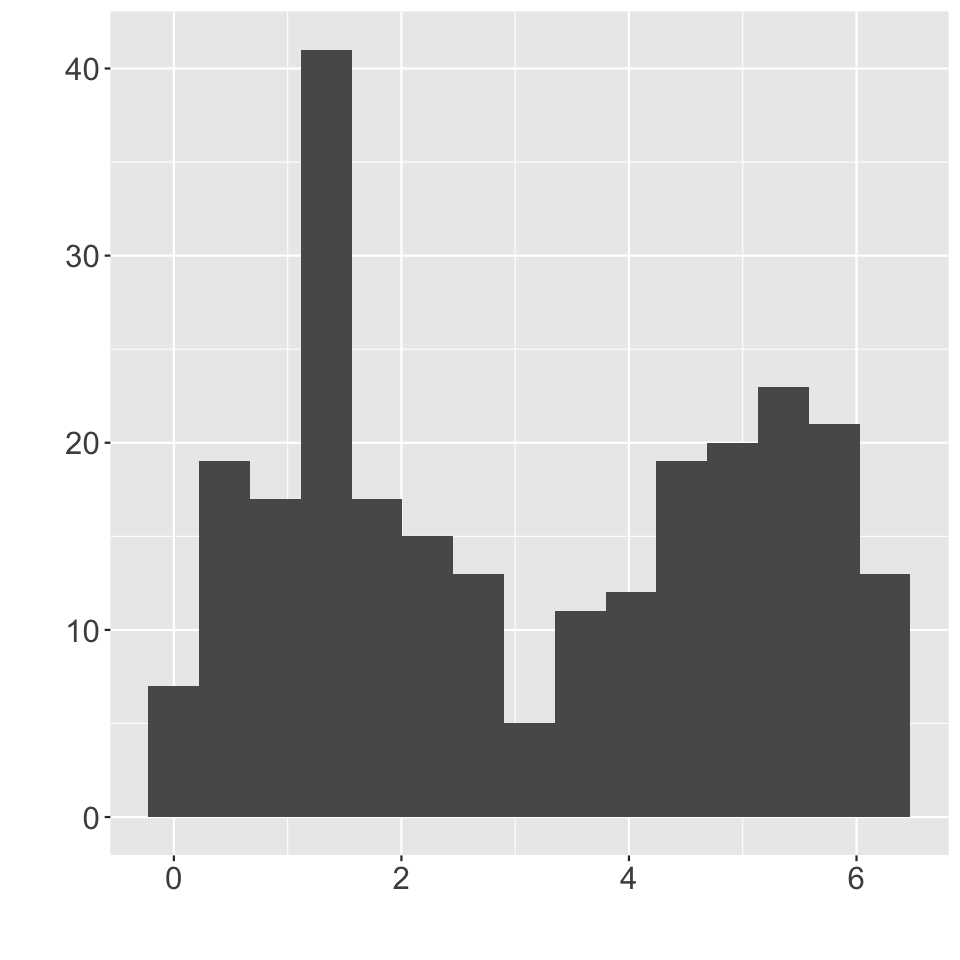}
\caption{1998-1999}
\label{fig:real_data_example_mixture_hist}
\end{subfigure}
\caption{\small Random direction plots for different years. The $x-y$ coordinates represent the income proportion in the first two categories. If the plot is three dimensional, the z coordinate is the random direction divided by $2\pi$ to lie in $[0, 1)$. The two leftmost plots show all random directions for the years listed below. The middle plots show the movement for a subset of 20-30 "locations" and vectors of equal length corresponding to the extracted directions from these movements. The two rightmost plots display a histogram of the random directions associated with locations that are within $0.05$ $L_2$-distance of $(0.39, 0.56, 0.05)$ for 2003-2004 and of $(0.85, 0.10, 0.05)$ for 1998-1999.}
\label{real_data_example}
\end{figure}

A sensible two-step approach to model these quantities is to first model $\v{y}_{\ell}$ and then $\theta'$ conditioned on $\v{y}_{\ell}$. The magnitude of a point's movement is bounded by the maximum distance between the point and the boundary point in a given direction. However, modeling $y_{\ell}$ presents non-trivial difficulties. To understand them, we reduce the number of income categories in our motivating data set to three: \$0-100 000, \$100 000-\$200 000, and \$200 000+. Here, a point on the 2D simplex corresponds to the proportions observed in each income category. For instance, $(\frac{1}{3}, \frac{1}{3}, \frac{1}{3})$ on the simplex represents a tract such that a third of all individuals fall into each category. Figure \ref{real_data_example} shows the extracted $y_{\ell}$'s for the years 1998-1999 and 2003-2004. Not only can we not assume that directions are uniformly distributed, but also there may be evidently preferred directions depending on the tracts' current income proportions. These preferred directions are correlated such that the change in their pattern is smooth as we pass across the simplex in this figure. There also appears to be an unimodal or a mixture of unimodal empirical distribution for the random directions associated with the nearby points on the simplex. For example, Figure \ref{fig:real_data_example_one_comp_hist} shows such a distribution for the points in the simplex, $(0.39, 0.56, 0.05)$. Hence, we must assign probability to these directions. In doing so, we denote these directions as "random directions". One naive approach is to assign probability to some random variable $z \in \mathbbm{R}$ and using the inverse logit function to transform $z$ to that interval. Such an approach is problematic because the endpoints of the interval and values near the end points are mapped near their respective $\pm\infty$ and are far apart. Meanwhile, the end points for the random angle's interval, $0$ and $2\pi$, denote the same direction and should not be so far apart. The inverse logit function that maps $\mathbb{R}$ to the angles $[0,2\pi)$ does not possess a continuous inverse even though the function is continuous and one-to-one. In addition, this approach does not generalize to directions from higher dimensional simplices. Thus, these challenges motivate us to model the random direction of movements from one year to the next in this paper. Doing so already allows us to discover meaningful patterns in the changes of Los Angeles County income proportions.

To model the random direction of movements for simplex-valued data, we will leverage and expand upon the building blocks advocated by~\citep{MardiaJuppDirectionalStatistics2010,RasmussenWilliamsGaussianProcessesMachine2006} and the techniques of~\citep{WangGelfandDirectionalDataAnalysis2013,WangGelfandModelingSpaceSpaceTime2014}. In particular, we assume that the observed random directions are distributed according to $K$ von Mises distribution for one dimensional directions and $K$ von Mises-Fisher distribution for multidimensional directions for some $K \geq 1$. We then correlate each of the von Mises or von Mises-Fisher's distributions' means with a projected Gaussian process of appropriate dimensions. Such a choice makes sense because the mean of the von Mises or von Mises-Fisher distribution can be thought of as a vector on the unit circle, which is what a projected Gaussian process outputs. In addition, it allows both the prior and likelihood to recognize the geometry of angles. Not only can this model harness the power of Gaussian processes to "spatially" correlate random directions of similar income proportions and handle noisy directions, but also it estimates each component's average random directions. We can interpret this average to understand the year to year changes in income proportions. Indeed, when we apply the model to directions extracted from a three income proportion version of our motivating data set, understanding the results enables us to discover trends consistent with larger macroeconomic ones and provides further information on the year to year changes. We can also expand upon our comprehension of a year to year change by interpreting results from our models when applied to directions from a six income category version. Thus, our new parameterization of changes to data that lie on a simplex allows us to use ideas from directional statistics to analyze these changes in an interpretable way.




Notice that our approach is a circular or spherical version of a mixture of Gaussian process with Gaussian white noise. Although Gaussian processes are a standard modeling tool in spatial statistics, we wish to emphasize that our use for modeling changes in the data that lie on a simplex is somewhat non-standard. Specifically, while our motivating data sets and similar such data sets have information on physical locations, the index space for the Gaussian process model in our method is not that space of locations. Instead, the index space corresponds to the starting measurements in the simplex, and the Gaussian process is in effect a model for the \emph{velocity vector field} of the year to year measurement vectors. While it would be interesting to additionally consider incorporating the physical locations of the tracts into the Gaussian process modeling, there is too much inhomogeneity in this type of spatial dependence in our motivating data set, and so we do not pursue this spatial modeling approach here. 


The rest of the paper is organized as following. First, we give details about how to extract the random directions from data that lie on a simplex in Section \ref{section:rand_dir_extract} in both the 2D and higher dimensional simplex cases. We then go over the distributions that we will use in Section \ref{background} because these distributions, particularly the higher dimensional versions, are not well known. Next, our models will be introduced in the Section \ref{Model}. We then discuss how to fit these models in Section \ref{sec:model_fitting}. After briefly discussing our simulation study in Section \ref{section:sim}, we introduce the motivating data set and interpret the results from fitting our models to two versions of the data set in Section \ref{sec:real_data_results}. Finally, Section \ref{sec:conclusion} highlights the contribution of our work and discusses possible extensions.

\section{Extracting random direction}
\label{section:rand_dir_extract}
We will provide further details about extracting random directions from data that lie on a simplex in this section. Because of our motivating data set, we will also discuss what these directions mean in the context of income proportions. To simplify the presentation, we leave some of the technical details to the supplementary material.

We extract the random direction in the following way. Let $\v{x}_{\ell}$ be the income proportion for one year and $\v{x'}_{\ell}$ be the proportion for the next. First, we need the spherical coordinates for $\sqrt{\v{x}_{\ell}}$. Next, we use these spherical coordinates to construct an orthogonal matrix, $\v{\mathcal{O}_p}$. The last column of $\v{\mathcal{O}_p}$ is $\sqrt{\v{x}_{\ell}}$ in the 2D simplex case whereas the first column of $\v{\mathcal{O}_p}$ is $\sqrt{\v{x}_{\ell}}$ in the higher dimensional case because of how we define angles. We then number the remaining columns. For each column, we add $\frac{\pi}{2}$ to the spherical coordinate of $\sqrt{\v{x}_{\ell}}$ corresponding to the column number and set the previous coordinates to $\frac{\pi}{2}$. For instance, the second column of $\v{\mathcal{O}_p}$ in the 2D simplex case is a vector with spherical coordinates $\frac{\pi}{2}$ and the second spherical coordinate of $\sqrt{\v{x}_{\ell}}$ plus $\frac{\pi}{2}$ because the second column is the second remaining other column. Finally, we extract the spherical coordinates of $\v{\mathcal{O}_p}^{-1} \sqrt{\v{x'}_{\ell}}$ to derive the random directions. 

\begin{figure}[!tp]
    \centering
    \includegraphics[width = 0.45\textwidth]{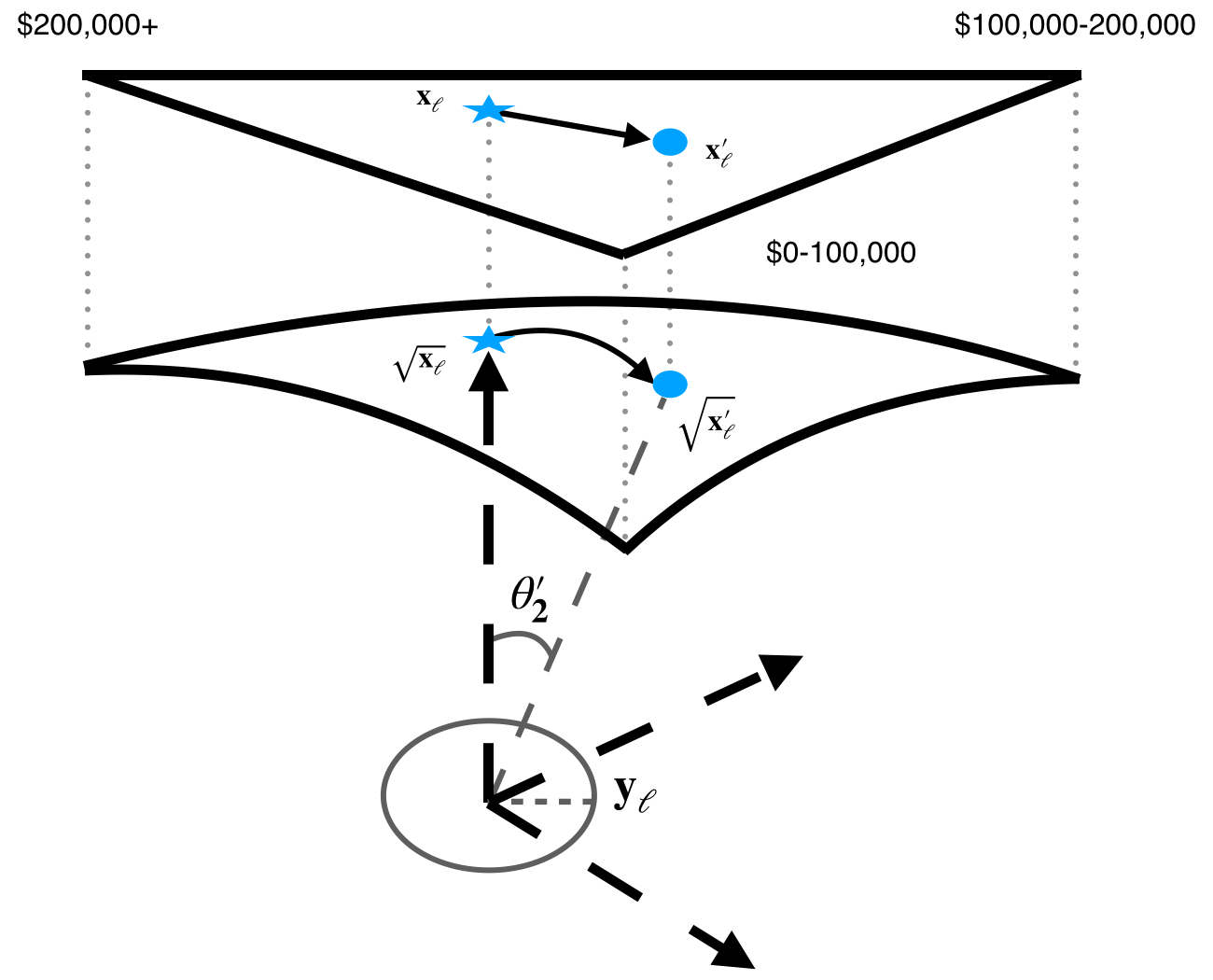}
    \caption{\small Figure displaying the connection between movement of points on a 2D simplex and random directions. The income proportions observed for a census tract in one year are shown as a star whereas the income proportions observed for a census tract in the next year are shown as a circle. The coordinate system we define in Section \ref{section:rand_dir_extract} is also displayed with dashed lines ending in arrows. Note that $\theta'_2$ is defined in \eqref{eq:rand_dir_info_2D}.}
    \label{fig:rand_dir_overview}
\end{figure}

We further break down this last step based on the dimension of the simplex. In the two dimensional case, set $\theta'_2 \in [0, \frac{\pi}{2})$ and $y_{\ell} \in [0, 2\pi)$ to be the angles such that
\begin{align}
    \theta'_2 = \arccos(\left(O_p \sqrt{\v{\widetilde{x}'_\ell}}\right)_3), & & y_{\ell} = \arctan^*\left(\left(O_p \sqrt{\v{\widetilde{x}'_\ell}}\right)_1, \left(O_p \sqrt{\v{\widetilde{x}'_\ell}}\right)_2\right).
    \label{eq:rand_dir_info_2D}
\end{align}
Here, $\arctan^*(\cdot, \cdot)$ is the modified $\arctan$ function for $z_1, z_2 \in \mathbbm{R}$ such that for $z_1^2 + z_2^2 = 1$,
\begin{align}
\textrm{arctan}^*(z_1, z_2) &=
    \begin{cases}
        \textrm{arctan}(\frac{z_2}{z_1}) & z_1 \geq 0, z_2 \geq 0\\
        \textrm{arctan}(\frac{z_2}{z_1}) + 2\pi & z_1 \geq 0, z_2 < 0 \\
        \textrm{arctan}(\frac{z_2}{z_1}) + \pi & z_1 < 0.
    \end{cases}
    \label{fct:arctan_star}
\end{align}
Meanwhile, in the higher dimensional case, define $\theta_2' \in [0, 2\pi)$ and $\v{y}_\ell \in [0, \pi]^{D - 2} \times [0, 2\pi)$ to be the following quantities:
\begin{align}
    \theta_2' = \mathcal{ST}^{-1}(\mathcal{O}^{-1}_p \sqrt{\v{x}'_{\ell}})_1, & & \v{y}_\ell = \{\mathcal{ST}^{-1}(\mathcal{O}^{-1}_p \sqrt{\v{x}'_{\ell}})_d\}_{d = 2}^{D}.
\end{align}
We define $\mathcal{ST}^{-1}(\cdot)$ to be the following transform for points on the sphere, $\v{s} \in \mathbbm{S}^{D}$, to an angle, $\v{a} \in [0, \pi]^{D - 1} \times [0, 2\pi)$, with a few exceptions:
\begin{align}
    a_d = \acos\left(\frac{s_d}{\sqrt{\sum_{d' = d}^{D + 1} s^2_{d'}}}\right), d = 1, 2, \ldots, D - 1, & & a_{d} = \begin{cases}
        \acos\left(\frac{s_{D}}{\sqrt{s_D^2 + s_{D + 1}^2}}\right) & s_d \geq 0,\\
        2\pi - \acos\left(\frac{s_{D}}{\sqrt{s_D^2 + s_{D + 1}^2}}\right) & s_d < 0.\\
    \end{cases} 
\end{align}
The exceptions occur when $\sum_{d' = d}^D s^2_{d'} = 0$ for some $d \in 1, 2, \ldots, D$. In those cases, set $a_{d'} = 0$ for $d \leq d' \leq D$. Finally, as an abuse of notation, we also let $\mathcal{ST}^{-1}(\v{s})$ represent the spherical coordinates themselves.
Then, $\v{y_{\ell}}$ is the random direction and $\theta'_2$ represents how "far" $\v{x}_\ell$ goes in that random direction. An illustration of these variables is shown in Figure \ref{fig:rand_dir_overview}.

These random directions are interpretable based on their interactions with the columns of $\v{\mathcal{O}_p}$. In the two dimensional case, there is only one random direction. Because the first and second coordinate include $\cos(y_{\ell})$ and $\sin(y_{\ell})$ respectively, this direction interacts with the first two columns of $\v{\mathcal{O}_p}$. By construction, the first column represents a push toward the third income category. Meanwhile, the second column is a pull toward the second income category. Then, to understand the random direction, we examine $0$, $\frac{\pi}{2}$, $\pi$, and $\frac{3\pi}{2}$. At $0$ and $\pi$, the first coordinate will be $1$ and $-1$ and the second will be zero by definition. In other words, $0$ "confirms" and $\pi$ "negates" the first column. As a result, a random direction of $0$ is a push away from the third income category whereas a random direction of $\pi$ is a pull toward that category. On the other hand, the first coordinate is zero and the second coordinate is $1$ and $-1$ at $\frac{\pi}{2}$ and $\frac{3\pi}{2}$. This means that a random direction of $\frac{\pi}{2}$ and $\frac{3\pi}{2}$ represent a pull toward and push away from the second income category.

We can also understand the higher dimensional random directions in a similar way. Here, the random directions interact with every column of $\v{\mathcal{O}_p}$ except the first. Again by construction, all but the last of the interacted column represent a push away from the income category one less than the column number toward higher income categories. The last column is a pull toward the highest income category. Then, for $d = 1, 2, \ldots, D - 2$, $\v{y_{\ell, d}}$ affects all columns after $d$. If we alter $\v{y_{\ell, d}}$, we have that for some constants $c, c_{d'} \in \mathbbm{R}$ for $d' \in {d + 1, d + 2, \ldots, D - 2}$, 
\begin{align*}
  \mathcal{O}_{p, d + 1} \cos(y_{\ell, d}) c & & \mathcal{O}_{p, d'} \sin(y_{\ell, d}) c_{d'}, \quad d' \in {d + 1, d + 2, \ldots, D - 2}.
\end{align*}
The natural values to examine are $0, \frac{\pi}{2}, \pi$. Because $\cos(0) = 1$ and $\cos(\pi) = -1$ and $\sin(0) = \sin(\pi) = 0$, $\v{y_{\ell, d}}$ switches between a "confirmation" and a push away from income category $d$ to a "negation" and a pull towards income category $d$. Meanwhile, the random direction can indicate "support" for the changes occurring to higher income categories as $\cos(\frac{\pi}{2}) = 0$ and $\sin(\frac{\pi}{2}) = 1$. 

The interpretation of the last random direction, $\v{y}_{\ell, D - 1}$, is different. Note that $\v{y}_{\ell, D - 1} \in [0, 2\pi)$. It is associated with the last two columns because we multiply the penultimate column by $\cos(\v{y}_{\ell, D - 1})$ and the last column by $\sin(\v{y}_{\ell, D - 1})$. Thus, $\v{y}_{\ell, D - 1} = 0$ represents a push away from the third highest income category and $\v{y}_{\ell, D - 1} = \pi$ represents a pull toward that category. However, because the last column of $O_{p}$ is a pull toward the highest income categories, $\v{y}_{\ell, D - 1} = \frac{\pi}{2}$ is a pull toward the highest income category and $\v{y}_{\ell, D - 1} = \frac{3\pi}{2}$ is a push away from that category. 

\begin{table}[!t]
\centering
\resizebox{.9\textwidth}{!}{%
 \begin{tabular}{c c c c c c} 
 \hline
 & $\v{\mathcal{O}_{p, 2}}$ & $\v{\mathcal{O}_{p, 3}}$ & $\v{\mathcal{O}_{p, 4}}$ & $\v{\mathcal{O}_{p, 5}}$ & $\v{\mathcal{O}_{p, 6}}$ \\ [0.5ex] 
 \hline
  & & &\\
 \textbf{Income category} & \$0 - \$25 000 & \$25 - \$50 000 & \$50 000 - \$100 000 & \$100 000 - \$150 000 & \$200 000+ \\ 
 & & &\\
 \textbf{Column Meaning} & Push & Push & Push & Push & Pull \\ 
 & & &\\
 \textbf{Random direction} & $y_{\ell, 1}$ & $y_{\ell, 2}$ & $y_{\ell, 3}$ & $y_{\ell, 4}$ & $y_{\ell, 4}$ \\ 
 & & &\\
 \hline
  & & &\\
 \end{tabular}
 }
 \caption{\small Table displaying rotation matrix's columns and their associated income category, random direction, and meaning for the six income category data set.}
 \label{table:real_data_random_dir_trans}
\end{table}

To help make the interpretation of the higher dimensional angles more concrete, Table \ref{table:real_data_random_dir_trans} shows which random directions are directly related to which columns in a six income category data set.

\section{Background on distributions}
\label{background}
This section introduces the distributions that we will use to model these random directions. For multidimensional random directions, it is easier to first work with the points on the appropriate higher dimensional sphere that corresponds to the random direction. After we do so, it is possible to write the distribution in terms of the random directions if needed. Because the distribution for one dimensional random directions are written in terms of the random directions, we discuss the multidimensional versions first. In doing so, we demonstrate how the one dimensional forms of the distribution can be derived from the multidimensional types.


\subsection{von Mises and von Mises-Fisher distributions}
A natural choice to model $\v{z} \in \mathbbm{R}^{D}$ is the multivariate Gaussian distribution of dimension $D$. In order to make it applicable to data on the n-sphere, we assume that $\v{z}$ and the mean of the Gaussian distribution, $\v{\widetilde{m}}$, lie on the higher dimensional sphere, $\mathbbm{S}^{D - 1}$. In other words, $\norm{\v{z}}_2 = 1$ and $\norm{\v{\widetilde{m}}}_2 = 1$. Further, assume that the covariance matrix is $\rho^{-1} \v{\mathbbm{I}_{D \times D}}$ for $\rho \in \mathbbm{R}^{+}$. The purpose of these assumptions will become clear shortly. Then, we obtain the following:
\begin{align*}
    p(\v{z} \mid \v{\widetilde{m}},\rho^{-1} \v{\mathbbm{I}_{D \times D}}) &\propto \exp\left(-\frac{\rho}{2}(z - \v{\widetilde{m}})^2\right)\\
    &= \exp\left(-\frac{\rho}{2}(\v{z}^T \v{z} + \v{\widetilde{m}}^T\v{\widetilde{m}} - 2 \v{\widetilde{m}}^T \v{z})\right)\\
    &= \exp\left(\rho\v{\widetilde{m}}^T \v{z} - \rho\right)
\end{align*}
If we re-normalize this by integrating over all $\v{z} \in \mathbbm{S}^{D - 1}$, we get the density for the von Mises-Fisher distribution with mean $\v{\widetilde{m}}$ and concentration parameter, $\rho$, \citep{MardiaJuppDirectionalStatistics2010}:
\begin{align}
    \vMF{\v{z}}{\v{\widetilde{m}}}{\rho} &= \frac{\rho^{D / 2 - 1}}{(2\pi)^{D / 2} I_{D / 2 - 1}(\rho)} \exp\left(\rho\v{\widetilde{m}}^T \v{z}\right).
\end{align}
Here, $I_{D / 2 - 1}(\rho)$ is the modified Bessel function of the first kind and of order $\frac{D}{2} - 1$. The modified Bessel function of the first kind with order $n$ is defined to be the following:
\begin{equation}
    I_n(\rho) := \frac{1}{2\pi} \int_0^{2\pi} \cos(n a) \textrm{e}^{\rho \cos(y)} d a.
    \label{eq:bessel}    
\end{equation}

We can derive the von Mises distribution when $D = 2$ \citep{MardiaJuppDirectionalStatistics2010}. We can use the polar coordinate transform. In other words, there exists angle, $y_{\ell} \in [0, 2\pi)$, and $r \in \mathbbm{R}^+$ such that $\v{z} = (r\cos(y_{\ell}), r\sin(y_{\ell}))$. This angle can be found using the $\arctan^*$ function defined in \eqref{fct:arctan_star}. We can also use this function to find $m \in [0, 2\pi)$ such that $\v{\widetilde{m}} = (\cos(m), \sin(m))$. Meanwhile, by assumption, $r = 1$ with probability 1. Then, if we re-write the distribution in terms of $r$, $y_{\ell}$, and $m$, the Jacobian of the transform is $r$. This gives us the density for the von Mises distribution with mean angle, $m$, and concentration parameter, $\rho$:
\begin{equation}
    \begin{aligned}
        \vM{y_{\ell}}{m}{\rho} &= \frac{r \rho^{D / 2 - 1}}{(2\pi)^{D / 2} I_{D / 2 - 1}(\rho)} \exp\left(\rho\v{\widetilde{m}}^T \v{z}\right)\\ 
        &= \frac{r}{2\pi I_0(\rho)}\exp\left(r \rho(\cos(m)\cos(y_\ell) + \sin(m)\sin(y_\ell))\right)\\
        &= \frac{r}{2\pi I_0(\rho)}\exp\left(r\rho\cos(m - y_\ell)\right)\\
        &= \frac{1}{2\pi I_0(\rho)}\exp\left(\rho\cos(m - y_\ell)\right).
    \end{aligned}
\end{equation}

We make a few remarks. Because the density function is proportional and restricted to a multivariate Gaussian centered at a point on the sphere, if $\rho > 0$, the von Mises-Fisher distribution is unimodal and symmetric around its mean. The distribution becomes uniform if $\rho = 0$.  

\subsection{Projected Gaussian process}
\label{ssection:pgp}
For our problem, it is of interest to define a stochastic process of random directions indexed in a general domain, $\Omega$. One way we can accomplish this is to transform the Gaussian process, a popular stochastic process. However, we cannot transform the processs with the assumptions needed to derive a von-Mises Fisher distribution because the assumptions are very strong. Instead, inspired by the projected normal \citep{WangGelfandDirectionalDataAnalysis2013, MardiaJuppDirectionalStatistics2010}, we will "project" the draws from a Gaussian process onto the appropriate spheres. This idea was studied by \citep{WangGelfandModelingSpaceSpaceTime2014}. A simpler version of this idea will be presented here.



We begin by discussing a powerful modeling tool for spatio-temporal data, Gaussian processes \citep{CressieWikleStatisticsSpatiotemporalData2011,BanerjeeEtAlHierarchicalModelingAnalysis2015,RasmussenWilliamsGaussianProcessesMachine2006}.
Given observations $z_1, z_2, \dots z_N$ indexed by the corresponding locations $x_1, x_2, \dots, x_N \in \Omega$, we assume that these observations are realizations of a stochastic process $\{Z(x) |x \in \Omega\}$, i.e., $z_\ell = Z(x_{\ell})$ for $\ell=1, 2, \ldots, N$. 
To account for the spatial dependence of these observations, one may assume that $\v{z}$ is a Gaussian process, which is parameterized by a mean function $\mu(\cdot)$ and a covariance function $K(\cdot, \cdot)$ on $\Omega$. Abusing notation, let $\v{\mu}$ denote the mean function applied to every location such that $\mu_\ell = \mu(x_\ell)$. If $\v{\Sigma}$ is a $\mathbbm{R}^{N \cross N}$ matrix such that $\Sigma_{\ell, \ell'} = K(x_{\ell}, x_{\ell'})$ for $\ell, \ell' \in 1, 2, \ldots, N$, we will denote this as $\v{z} \sim \textrm{GP}(\v{\mu}, \v{\Sigma})$ because $p(\v{z} \mid \v{\mu}, \v{\Sigma}) = \textrm{N}(\v{z} \mid \v{\mu}, \v{\Sigma})$.

We describe how to convert the Gaussian process to the projected Gaussian process. Suppose that $\v{z_1} \sim \textrm{GP}(\v{\mu_1}, \v{\Sigma_1}), \v{z_2} \sim \textrm{GP}(\v{\mu_2}, \v{\Sigma_2}), \ldots, \v{z_D} \sim \textrm{GP}(\v{\mu_D}, \v{\Sigma_D})$. For identifiability of the stochastic process, one can assume that for $\v{\Sigma_1}, \v{\Sigma_2}, \ldots, \v{\Sigma_D}$,
\begin{align*}
    \begin{pmatrix}
        \v{\Sigma_1} & 0  & \ldots & 0\\
         0 & \v{\Sigma_2} & \ldots & 0\\
         \vdots  & \vdots & \ddots & 0\\
         0 & 0 & 0 & \v{\Sigma_D}\\
\end{pmatrix} =
    \v{\mathbbm{I}_{D \times D}} \otimes \Sigma.
\end{align*}
Here, $\v{\Sigma}$ is the covariance matrix discussed previously in this subsection. Instead of using $\v{\mathbbm{I}_{D \times D}}$, another matrix defined according to what Wang and Gelfand or Hernandez-Stumpfhauser et al. propose can be used \citep{WangGelfandModelingSpaceSpaceTime2014,Hernandez-StumpfhauserEtAlGeneralProjectedNormal2017}. In our case, due to the lack of additional information and the ordering imposed by the income proportions, we use an identity matrix. Then, we perform the following operation to get a stochastic process for points on the appropriate sphere. For location $\ell = 1, 2, \ldots, N$ and $d = 1, 2, \ldots, D$, let 
\begin{align}
   \widetilde{m}_{\ell, d} = \frac{z_{d, \ell}}{\sqrt{\sum_{d = 1}^D z_{d, \ell}^2}}.
\end{align}

This is a generative description of the process and the output for a location perfectly matches the mean parameter for the von Mises-Fisher distribution. However, to see how we can sample this distribution and write down its probability, we need to write this as an angle-valued stochastic process, $\v{\mathcal{A}}$. To do so, we first transform a draw into its spherical coordinates in higher dimensions and radii. In other words, rewrite $z_{d, \ell}$ as $r_\ell \mathcal{ST}(\v{y}_\ell)$ for some $r_\ell \in \mathbbm{R}$ and some higher dimensional angle, $\v{y}_\ell \in [0, \pi]^{D - 2} \times [0, 2\pi)$. Here, $\mathcal{ST}(\cdot)$ is the following transformation for a higher dimensional angle, $\v{a} \in [0, \pi]^{D - 1} \times [0, 2\pi)$, and a point on a higher dimensional sphere, $\v{s} \in \mathbbm{S}^{D}$:
\begin{equation}
\begin{aligned}
   s_1 &= \cos(a_1),\\
   s_2 &= \sin(a_1)\cos(a_2),\\
   s_3 &= \sin(a_1)\sin(a_2)\cos(a_3),\\
   &\vdots\\
   s_{D} &= \left(\prod_{d = 1}^{D - 1} \sin(a_d)\right)\cos(a_{D}),
   s_{D + 1} &= \left(\prod_{d = 1}^{D - 1} \sin(a_d)\right)\sin(a_{D}).   
\end{aligned}
\label{eq:spherical_transform}    
\end{equation}
We also need the Jacobian of this transform. For $D > 3$, the Jacobian is the following \citep{Hernandez-StumpfhauserEtAlGeneralProjectedNormal2017}:
\[
\prod_{\ell = 1}^N r_{\ell}^{D - 1} \prod_{d = 1}^{D - 2} \sin^{D - 1 - d}(y_{\ell, d}).
\]

We now have all the pieces. However, before writing the distribution, we introduce some more notation to make it easier to write the distribution. Set $\v{r} \in \mathbbm{R}^N$ to be the vector comprised of $r_\ell$ at each location. Then, let $\v{r} \{\mathcal{ST}(\v{y}_\ell)_d\}_{\ell = 1}^N$ denote the vector such that $(\v{r} \{\mathcal{ST}(\v{y}_\ell)_d\}_{\ell = 1}^N)_\ell = (r_\ell \mathcal{ST}(\v{y}_\ell))_{d}$ for $\ell = 1, 2, \ldots, N$ and $d = 1, 2, \ldots, D$. 
We obtain the following probability:
\begin{align}
    p(\v{\mathcal{A}}, \v{r} \mid \v{\mu_1}, \v{\mu_2}, \ldots, \v{\mu_D}, \v{\Sigma}) = \left(\prod_{d = 1}^D \textrm{N}(\v{r} \{\mathcal{ST}(\v{y}_\ell)_d\}_{\ell = 1}^N \mid \v{\mu_d}, \v{\Sigma})\right) \prod_{\ell = 1}^N r_{\ell}^{D - 1} \prod_{d = 1}^{D - 2} \sin^{D - 1 - d}(y_{\ell, d}).
\end{align}
The density for the projected Gaussian process is:
\begin{align}
    p(\v{\mathcal{A}} \mid \v{\mu_1}, \v{\mu_2}, \ldots, \v{\mu_D}, \v{\Sigma}) &= \int_{\v{r} > 0} p(\v{\mathcal{A}}, \v{r} \mid \v{\mu_1}, \v{\mu_2}, \ldots, \v{\mu_D}, \v{\Sigma}) d\v{r}.
\end{align}

Meanwhile, the distribution can be written slightly more explicitly in the circular case. We apply the polar coordinate transformation element-wise defined in \eqref{fct:arctan_star} such that $z_{1, \ell} = r_\ell\cos(\mathcal{A}_{\ell})$ and $z_{2, \ell} = r_\ell\sin(\mathcal{A}_{\ell})$ for $\ell = 1, 2, \ldots, N$. As a shorthand, define $\cos(\mathcal{A})$ and $\sin(\mathcal{A})$ to be the vectors such that $\cos(\mathcal{A})_\ell = \cos(\mathcal{A}_\ell)$ and $\sin(\mathcal{A})_\ell = \sin(\mathcal{A}_\ell)$ for $\ell = 1, 2, \ldots, N$. Since the Jacobian of this transformation is $\prod_\ell r_\ell$, we have that
\begin{align}
    p(\v{\mathcal{A}}&, \v{r} \mid \v{\mu_1}, \v{\mu_2}, \v{\Sigma_1}, \v{\Sigma_2}) = \frac{1}{\sqrt{(2\pi)^N\abs{\v{\Sigma_1}}}}\exp\left(-\frac{1}{2}(\v{r}\cos(\v{\mathcal{A}}) - \v{\mu_1})^T\Sigma^{-1}(\v{r}\cos(\v{\mathcal{A}}) - \v{\mu_1})\right)\nonumber\\
    &\quad \times \frac{1}{\sqrt{(2\pi)^N\abs{\v{\Sigma_2}}}}\exp\left(-\frac{1}{2}(\v{r}\sin(\v{\mathcal{A}}) - \v{\mu_2})^T\Sigma^{-1}(\v{r}\sin(\v{\mathcal{A}}) - \v{\mu_2})\right) \left(\prod_{\ell} r_{\ell}\right)
\end{align}
The projected Gaussian process in two dimensions is the marginal distribution of $\v{\mathcal{A}}$:
\begin{align}
    p(\v{\mathcal{A}} \mid \v{\mu_1}, \v{\mu_2}, \v{\Sigma}) &= \int_{\v{r} > 0} p(\v{\mathcal{A}}, \v{r} \mid \v{\mu_1}, \v{\mu_2}, \v{\Sigma}) d\v{r}.
\end{align}

In all dimensions, the integral for the marginal distribution of $\v{\mathcal{A}}$ is intractable. Instead, we have to sample for $\v{r}$ implicitly while sampling for $\v{\mathcal{A}}$ or its element-wise transform to points on a higher dimensional sphere. Because of how the distribution is defined, this is still a valid way to sample from the projected Gaussian process.

\section{Modeling Random Directions} \label{Model}
Our approach aims to integrate spatial information into the means of the von Mises or von Mises-Fisher distributions. We call this model the \textit{Spatially varying von Mises component mixture} model or \textit{SvM-c}. To begin, we assume that there $N$ locations and observed angles. Let $\v{x}_\ell \in \Delta^D$ denote a location and $\v{y}_\ell$ an observation for that location for $\ell = 1, 2, \dots, N$. Here, $y_\ell \in [0, 2\pi)$ if $D = 2$ whereas $y_{\ell, d} \in [0, \pi]$ for $d = 1, 2, \ldots, D - 2$ and $y_{\ell, D - 1} \in [0, 2\pi)$ if $D > 2$.  According to this model specification, each observation may be distributed by one of $K$ von Mises-Fisher distributions with probability $\lambda_k$ regardless of its location. We will use $\zeta_\ell \in \{1, 2, \ldots, K\}$ to label which von Mises distribution the observation is associated with and $\v{\zeta}$ to denote the entire vector of labels. Each distribution's mean parameters, $\v{\widetilde{m_{k, \ell}}} \in \mathbbm{S}^{D - 1}$ at location $\v{x}_\ell$, are transformed from $D$ draws, $\v{z_{k, 1}}, \v{z_{k, 2}}, \ldots, \v{z_{k, D}} \in \mathbbm{R}^N$, from $D$ Gaussian process with its own mean, $\v{\mu}_{k, d} \in \mathbbm{R}^N$, and covariance matrix, $\v{\Sigma} \in \mathbbm{R}^{N \times N}$. This transformation is accomplished by element-wise projecting each location's draws from the various Gaussian process onto the $D$-sphere. A von Mises or von Mises-Fisher distribution's concentration parameters, $\v{\rho_k} \in \mathbbm{R}^{N+}$, are random variables, $\v{\varphi_k} \in \mathbbm{R}^N$, that have been elementwise exponentiated. These random variables are distributed according to a hierarchical normal distribution. At a lower level, they are conditionally distributed according to a normal distribution with the same standard deviation, $\varsigma \in \mathbbm{R}^+$, but with different hierarchical means, $\nu_k \in \mathbbm{R}^+$. These hierarchical means, $\nu_k$, are given the same hyperprior, $\textrm{N}(0, \tau)$. To summarize, we have that
\begin{align*}
    \v{z_{k, d}} &\sim \textrm{GP}(\cdot \mid \v{\mu_{k, d}}, \v{\Sigma_k}), & k = 1, 2, \ldots, K\\
    & & d = 1, 2, \ldots, D\\
    \v{\widetilde{m_{k, \ell}}} &= \frac{\{z_{k, d, \ell}\}_{d = 1}^D}{\norm{\{z_{k, d, \ell}\}_{d = 1}^D}}, & k = 1, 2, \ldots, K\\
    & & \ell = 1, 2, \ldots, N\\
    \v{\varphi_k} &\stackrel{iid}{\sim} \textrm{N}(\cdot \mid \nu_k, \varsigma^2), & k = 1, 2, \ldots, K \stepcounter{equation}\tag{\theequation}\label{model:SvM-c-higher}\\
    \v{\rho_k} &= \exp{\v{\varphi_k}}, & k = 1, 2, \ldots, K\\
    \zeta_\ell \mid \lambda_1, \lambda_2, \ldots, \lambda_K &\stackrel{iid}{\sim} \textrm{Cat}(\cdot \mid \lambda_1, \lambda_2, \ldots, \lambda_K), & \ell = 1, 2, \ldots, N\\
    y_\ell \mid \zeta_\ell = k, \v{\widetilde{m_{k, \ell}}}, \rho_{k, \ell} & \sim \vMF{\cdot}{\v{\widetilde{m_{k, \ell}}}}{\rho_{k, \ell}}, & \ell = 1, 2, \ldots, N.
\end{align*}

In the one dimensional random direction case, we work with the random angles:
\begin{align*}
    \v{z_{k, 1}} &\sim \textrm{GP}(\cdot \mid \v{\mu_{k, 1}}, \v{\Sigma_k}), & k = 1, 2, \ldots, K\\
    \v{z_{k, 2}} &\sim \textrm{GP}(\cdot \mid \v{\mu_{k, 2}}, \v{\Sigma_k}), & k = 1, 2, \ldots, K\\
    \v{m_k} &= \textrm{arctan}^*(\v{z_{k, 1}}, \v{z_{k, 2}}), & k = 1, 2, \ldots, K \stepcounter{equation}\tag{\theequation}\label{model:SvM-c}\\
    \v{\varphi_k} &\stackrel{iid}{\sim} \textrm{N}(\cdot \mid \nu_k, \varsigma^2), & k = 1, 2, \ldots, K\\
    \v{\rho_k} &= \exp{\v{\varphi_k}}, & k = 1, 2, \ldots, K\\
    \zeta_\ell \mid \lambda_1, \lambda_2, \ldots, \lambda_K &\stackrel{iid}{\sim} \textrm{Cat}(\cdot \mid \lambda_1, \lambda_2, \ldots, \lambda_K), & \ell = 1, 2, \ldots, N\\
    y_\ell \mid \zeta_\ell = k, m_{k, \ell}, \rho_{k, \ell} & \sim \vM{\cdot}{m_{k, \ell}}{\rho_{k, \ell}}, & \ell = 1, 2, \ldots, N.
\end{align*}
Here, we use a von Mises distribution instead of a von Mises-Fisher distribution. Because the von Mises distribution requires a mean angle, we element-wise transform draws from the Gaussian process using the $\arctan^*$ function. The parametrization for the concentration parameter and the mixing probability remain the same otherwise.

We use the following hierarchical prior for \textit{SvM-c}'s concentration parameters because it is a compromise between assigning an individual and a global concentration parameter:
\begin{align*}
    \nu_k &\sim \textrm{N}(\cdot \mid 0, \tau^2),\\
    \v{\varphi_k} &\stackrel{iid}{\sim} \textrm{N}(\cdot \mid \nu_k, \varsigma^2).
\end{align*}
Using a global parameter $\rho$ will affect the estimates of the mean if the variances differ significantly because the model cannot adjust the concentration parameter. Conversely, assigning an individual parameter makes the model too flexible. This will negatively affect the model's ability to spatially correlate the observations. This concern also leads us to set the standard deviation for the lower term, $\varsigma$, to a small value instead of sampling for it. Even with a tight prior on $\varsigma$, the variance of the lower terms will be greater if we sample for the standard deviation. We also do not use another Gaussian process to model the variance parameter $\varphi_\ell$ for computation reasons and to avoid making the model too rich. Still, a normal distribution is useful because it will allow us to separately sample the hierarchical mean, $\nu_k$, from the lower term, $\varphi_{k, \ell}$.

We discuss some notation related to these models. We will denote the number of von Mises distributions after the model if we need to specify $K$. For instance, \textit{SvM-c-3} indicates the model \textit{SvM-c} with $K = 3$. The one exception to this guideline is \textit{SvM}, which is \textit{SvM-c} with $K = 1$ and introduced in the supplementary material. 

\section{Posterior Inference}
\label{sec:model_fitting}
We now describe how to fit our models through a three part blocked Gibbs sampler. First, we sample for the labels, $\zeta_\ell$ given the other parameters and $y_{\ell}$ for $\ell = 1, 2, \ldots, N$. Next, we creatively use the elliptical slice sampler to sample for $\v{z_{k, 1}}, \v{z_{k, 2}}, \ldots, \v{z_{k, D}}$ given the other parameters and $\v{y}_{\ell}$ for $\ell = 1, 2, \ldots, N$. This also gives us samples of $\v{\widetilde{m_{\ell}}}$ given the other parameters and $\v{y}_{\ell}$ for $\ell = 1, 2, \ldots, N$ or in the one dimensional random direction case, $m_{\ell}$ given the other parameters and $y_{\ell}$ for $\ell = 1, 2, \ldots, N$. Finally, we use Hamiltonian Monte Carlo (HMC) to sample for the concentration parameters and its hierarchical means given the other parameters and $y_{\ell}$ for $\ell = 1, 2, \ldots, N$.

We discuss the elliptical slice sampler in further detail while leaving details of the other two steps to the appendix because the other steps are more standard. In particular, we begin with the one dimensional random direction case. The sampler is designed to sample parameters that have a normal prior with mean $0$ and an arbitrary likelihood. For $d = 1, 2, \ldots, D$, we can sample $\v{z_{k, 1}} - \v{\mu_{k,1}}$ and $\v{z_{k, 2}} - \v{\mu_{k,2}}$ using the elliptical slice sampler because 
\begin{align*}
\begin{pmatrix} 
\v{z_{k, 1}} - \v{\mu_{k,1}}\\ 
\v{z_{k, 2}} - \v{\mu_{k,2}}\\
\end{pmatrix} 
\sim
\textrm{N}(\cdot \mid \v{0}, \v{\mathbbm{I}_{D \times D}} \otimes \v{\Sigma}).
\end{align*}
Here, $\begin{pmatrix} 
        \v{z_{k, 1}} - \v{\mu_{k,1}}\\ 
        \v{z_{k, 2}} - \v{\mu_{k,2}}\\ 
    \end{pmatrix} \in \mathbbm{R}^{2N}$ such that
\begin{align*}   
    \left\{\begin{pmatrix} 
        \v{z_{k, 1}} - \v{\mu_{k,1}}\\ 
        \v{z_{k, 2}} - \v{\mu_{k,2}}\\ 
    \end{pmatrix}\right\}_{\ell = 1}^N =  \v{z_{k, 1}} - \v{\mu_{k,1}}, & &
    \left\{\begin{pmatrix} 
        \v{z_{k, 1}} - \v{\mu_{k,1}}\\ 
        \v{z_{k, 2}} - \v{\mu_{k,2}}\\ 
    \end{pmatrix}\right\}_{\ell = N + 1}^{2N} = \v{z_{k, 2}} - \v{\mu_{k,2}}.
\end{align*}
Then, the likelihood is the following:
\[
L(y_\ell \mid \v{z_{k, 1}}, \v{z_{k, 2}}, \v{\rho}, \v{\zeta}) = \prod_\ell (\vM{y_\ell}{\arctan^*(z_{k, 1, \ell}, z_{k, 2, \ell})}{\rho_{k, \ell}})^{\indFct{\zeta_\ell = k}}.
\]

The elliptical slice sampler then works as following \citep{MurrayEtAlEllipticalSliceSampling2010}. Draw a vector, $\v{u}$, from a distribution, $\textrm{N}(\cdot \mid \v{0}, \mathbbm{I}_{2 \times 2} \otimes \v{\Sigma_k})$; a random cutoff, $c$, from a distribution, Unif(0, 1); and a random angle, $a$, from a distribution, Unif(0, $2\pi$). Propose a new $\v{z'_{k, 1}}$ and $\v{z'_{k, 2}}$ such that
\[
\begin{pmatrix} 
        \v{z'_{k, 1}} - \v{\mu_{k,1}}\\ 
        \v{z'_{k, 2}} - \v{\mu_{k,2}}\\ 
\end{pmatrix} =
\begin{pmatrix} 
        \v{z_{k, 1}} - \v{\mu_{k,1}}\\ 
        \v{z_{k, 2}} - \v{\mu_{k,2}}\\ 
\end{pmatrix}\cos(a) + \v{u}\sin(a).
\]
We accept this proposal if $\frac{L(y_\ell \mid \v{z'_{k, 1}}, \v{z'_{k, 2}}, \v{\rho}, \v{\zeta})}{L(y_\ell \mid \v{z_{k, 1}}, \v{z_{k, 2}}, \v{\rho}, \v{\zeta})} \geq c$. If the proposal is not accepted, we propose a new $a' \in (a - 2\pi, a)$ while keeping $\v{u}$ and $c$. This leads to another proposed $\v{z'_{k, 1}}, \v{z'_{k, 2}}$, which is either accepted or rejected. If the proposal is again rejected, a new $a'$ is sampled from a shrunken support based on $(a - 2\pi, a)$. This process is repeated until either a proposal is accepted or the support for $a$ becomes empty.

We can extend this idea to the multidimensional case. However, because the geometry of angles is different and we assume independence between the Gaussian processes for each dimension, we use the elliptical slice sampler to sample $\v{z_{k,d}}$ for $d = 1, 2, 3, \ldots, D$ conditioned on draws from the other Gaussian processes. This is still valid because $\v{z_{k,d}} - \v{\mu_{k, d}} \sim \textrm{N}(\cdot \mid \v{0}, \v{\Sigma})$. Another difference in this case is the likelihood. Set $\widetilde{m}_{k, \ell, d} = \frac{z_{k, d, \ell}}{\sqrt{\sum z_{k, d, \ell}^2}}$ for $\ell = 1, 2, \ldots, N$ and $d = 1, 2, \ldots, D$. The likelihood becomes:
\[
L(y_\ell \mid \v{z_{k, 1}}, \v{z_{k, 2}}, \ldots, \v{z_{k, D}}, \v{\rho}, \v{\zeta}) = \prod_\ell (\vMF{y_\ell}{\v{\widetilde{m}_{k, \ell}}}{\rho_{k, \ell}})^{\indFct{\zeta_\ell = k}}.
\]
We can then use the elliptical slice sampler with this new likelihood and normal prior.

We make a few more comments before continuing. First, to help the samplers, we used initial values obtained via a regularized version of Expectation Maximization algorithm derived from \textit{SvM-c}. We leave the details for these algorithms to the supplementary material. Second, for this paper, we do not sample the hyperparameters of the covariance matrices for the Gaussian processes. While this places restrictions on the model, we do so due to computational reasons. In our motivating data set, there are 1884 to 2310 observed angles in the three income version and around 2000 observed angles in the six income version. This makes it challenging to invert the covariance matrix. Instead, we fit our models using various hyperparameters and kernels and then model select to pick the best options. 

\section{Simulations}
\label{section:sim}
We conduct a thorough simulation study for the introduced models that we leave to the supplementary materials. In addition to examining the homogeneous and heterogeneous versions of our models in all dimensions, we also look at a von Mises distribution and mixture of von Mises distribution in all dimensions to study the effect of including spatial information. This study shows that when used for inference, our models correctly recover the model parameters for data generated according to their respective models. However, homogeneous models struggled to represent heterogeneous random directions. Further, the von Mises distribution and mixture of von Mises distribution had difficulty capturing the random directions generated from spatially correlated means. Our study also demonstrated that we can use the posterior predictive probability computed on 10\% of the data to model select. The posterior predictive probability is:
\begin{align}
p(\v{y^*} \mid x, x^*, \v{y}) := \int \int \textrm{p}(y^* \mid \vartheta^*) \textrm{p}(\vartheta^* \mid \vartheta, x, x^*)  \textrm{p}(\vartheta \mid x, y) d\vartheta^* d\vartheta,
\label{eq:post_pred_prob}    
\end{align} 
where $x^*$ represent the withheld locations, $y^*$ the withheld data, $\vartheta$ a posterior draw for the parameters based on $x$ and $y$, and $\vartheta^*$ a draw for the parameters for $x^*$ and $y^*$. The posterior predictive probability enabled us to select the model that generated the data. All in all, the simulation studies gave us confidence in our model's ability to capture the random direction patterns and the posterior predictive probability's guidance in selecting the best fitting model.

\section{Results}
\label{sec:real_data_results}
In this section, we briefly go over the motivating data set and how we recombine the income proportions in two ways. We then discuss fitting our models to the two versions and the interpretation of our models' results.

\subsection{Data overview}
We now introduce the motivating data set, i.e. the income proportions in Los Angeles County, and explain why we analyze the proportions. While the HMDA data is publicly available, the dataset we worked with is not because it is fused with data purchased from a private company. We choose to examine these proportions because the number of mortgages recorded in a year differ per census tracts. Analyzing proportions allows us to potentially ignore the biases that might arise from these differences. We also assume that these proportions observed are the true income proportions for a census tract. This assumption is reasonable because people are likely to move into tracts or neighborhoods with demographic characteristics similar to their own. 

The dataset itself has sixteen income categories: \$0 -- \$10 000; \$10 000 -- 15 000; \$15 000 -- \$20 000; \$20 000 -- \$25 000; \$25 000 -- \$30 000; \$30 000 -- \$35 000; \$35 000 -- \$40 000; \$40 000 -- \$45 000; \$45 000 -- \$50 000; \$50 000 -- 60 000; \$60 000 -- 75 000; \$75 000 -- 100 000; \$100 000 -- \$125 000; \$125 000 -- \$150 000; \$150 000 -- \$200 000; and \$200 000+. We recombine these income categories in the data set in two ways. The first is to merge the data set into three income categories: \$0 to \$100 000, \$100 000 to \$200 000, and greater than \$200 000. These values are chosen because one natural income category is the \$200 000+ category and we want to split the remaining amount evenly among the other two categories. Further, this allows us to visualize how the random directions change across the entire simplex, enabling us to categorize the different phases of changes. Meanwhile, because the goal is to understand one year's change in greater detail and it can be hard to understand the changes to all sixteen income categories, we again reduce the number of income categories. Ideally, we would split the \$0 to \$100 000 and \$100 000 to \$200 000 income categories in half. However, nine of the original income categories are between \$0-\$50 000. To avoid one category representing so many of the original income categories, we further divide the \$0-\$50 000 in half. As a result, in the higher dimensional case, there are six income categories to analyze: \$0-25 000, \$25 000-\$50 000, \$50-100 000, \$100 000-\$150 000, \$150 000-\$200 000, and \$200 000+. 

\begin{figure}[!tb]
\centering
\begin{subfigure}[t]{.4\textwidth}
\centering
\includegraphics[width = 0.53\textwidth]{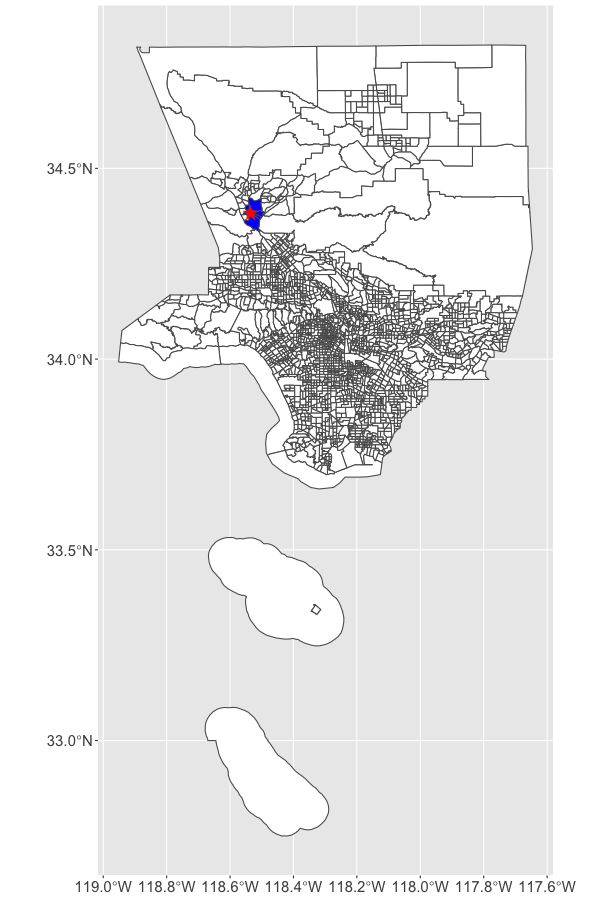}
\label{fig:real_data_Census_tract}
\end{subfigure}
\begin{subfigure}[t]{.4\textwidth}
\centering
\includegraphics[width = 0.8\textwidth]{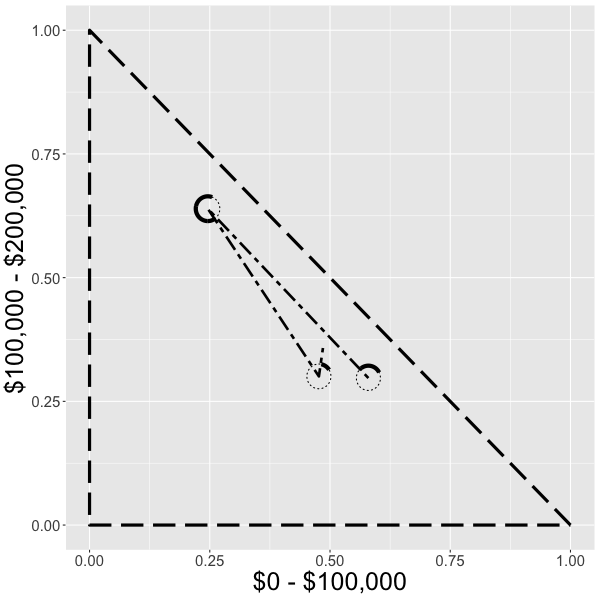}
\label{fig:real_data_subset_evo}
\end{subfigure}
\caption{\small Plots showing information about census tract 6037920336. Left plot shows the tract's location in red and the neighboring tracts in blue. The right plot connects the changes in observed income proportions for census tract 6037920336 from 1995 to 1998. The right plot also illustrates the idea of direction because the directions are represented by the solid lines on the circle at the observed income proportions for 1995 to 1997.}
\label{fig:real_data_Census_tract_example}
\end{figure}

For both versions, we apply the appropriate procedure outlined in Section \ref{section:rand_dir_extract} to extract the random directions. Figure \ref{fig:real_data_Census_tract_example} displays all Census tracts and the random direction in the context of changes to the income proportion of a Census tract 6037920336.

\subsection{Data analysis (Two dimensions)}
We now fit the proposed models to the random directions observed in the year to year three income proportion changes in the Home Mortgage Disclosure Act (HMDA) data for each census tract of Los Angeles County. We do so with one additional pre-processing step. We removed duplicated directions so that each location has at most one observation. Duplicated directions at the same location should happen with probability zero according to our model and they can be easily identified. The number of observations are reduced from between 2295 to 2347 per year to between 2120 to 2310 for the first 18 years. The number of observations during the last two years change from 2281 to 1884 and 2297 to 1879. 

For our models, we also had to select an appropriate kernel. Two popular choices are the squared exponential kernel,  
\begin{align}
k(\v{x}_{\ell}, \v{x_{\ell'}}) &= \sigma^2 \exp{-\frac{(\v{x}_{\ell} - \v{x_{\ell'}})^2}{2\omega^2}},
\label{eq:sq_exp_kernel}
\end{align}
and the Matern kernel,
\begin{align}
    k_{\mathcal{M}}(\v{x}_{\ell}, \v{x_{\ell'}}) &= \frac{2^{1 - v}}{\Gamma(v)}\left(\sqrt{2v}\frac{\sqrt{(\v{x}_{\ell} - \v{x_{\ell'}})^2}}{\omega}\right)^v K_v\left(\sqrt{2v}\frac{\sqrt{(\v{x}_{\ell} - \v{x_{\ell'}})^2}}{\omega}\right)
    \label{eq:matern_kernel}
\end{align}
for some $\sigma, \omega, v \in \mathbbm{R}^+$. To pick, we ran an extensive sensitivity analysis on the hyperparameters and compared our choice of kernel. While we defer the full sensitivity analysis to the supplementary material, we selected the squared exponential kernel with $\sigma = 0.5$ and $\omega = 0.5$ based on the posterior preditive probability.

We then used these hyperparameter choices and the log posterior predictive probability given in \eqref{eq:post_pred_prob} to select the number of components and model. \textit{SvM-c-3} perform the best for all years except 1990-1991, 1997-1999, and 2004-2007. For those years years, \textit{SvM-c-2} performs the best. Further, the posterior predictive probability for \textit{SvM-c-2} is close to \textit{SvM-c-3} in 1990-1991 and 2006-2007. 


\begin{figure}[!tb]
\centering
\begin{subfigure}{.3\textwidth}
\centering
\includegraphics[width = 0.8\textwidth]{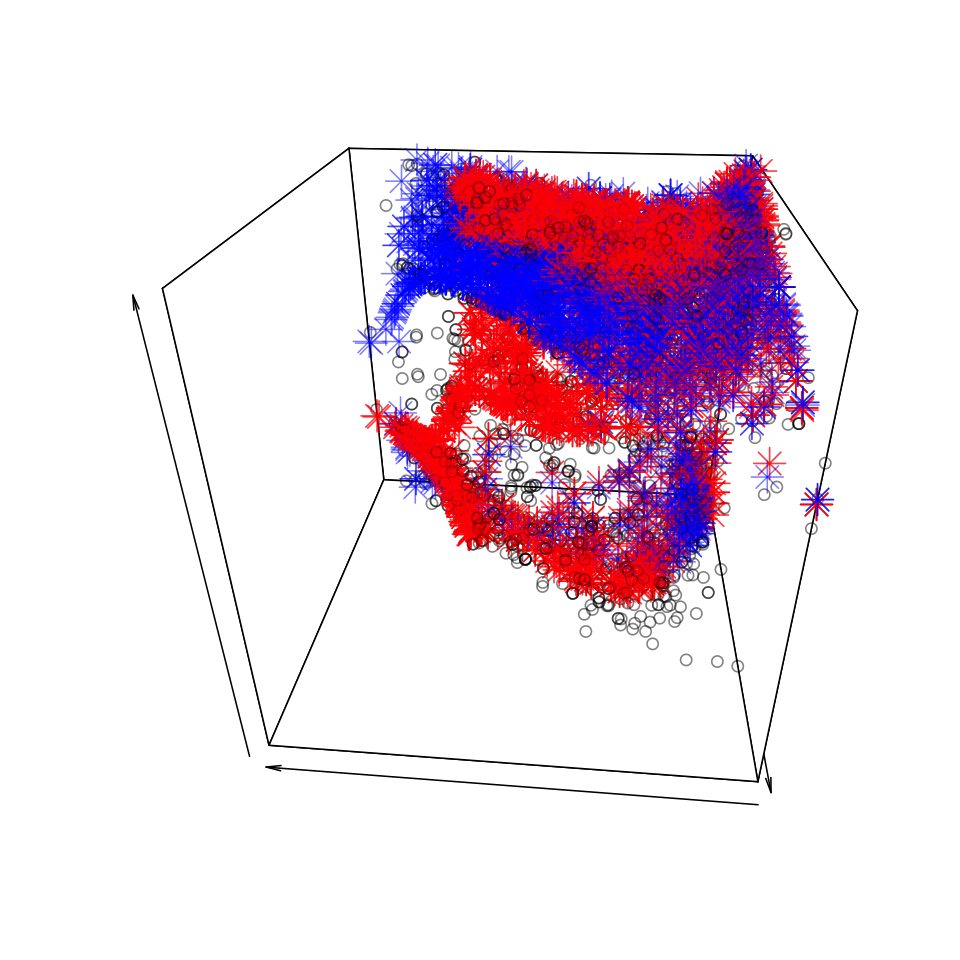}
\caption{1990-1991\\ (\textit{SvM-c-2})}
\label{fig:real_data_fitted_results_1}
\end{subfigure}
\centering
\begin{subfigure}{.3\textwidth}
\centering
\includegraphics[width = 0.8\textwidth]{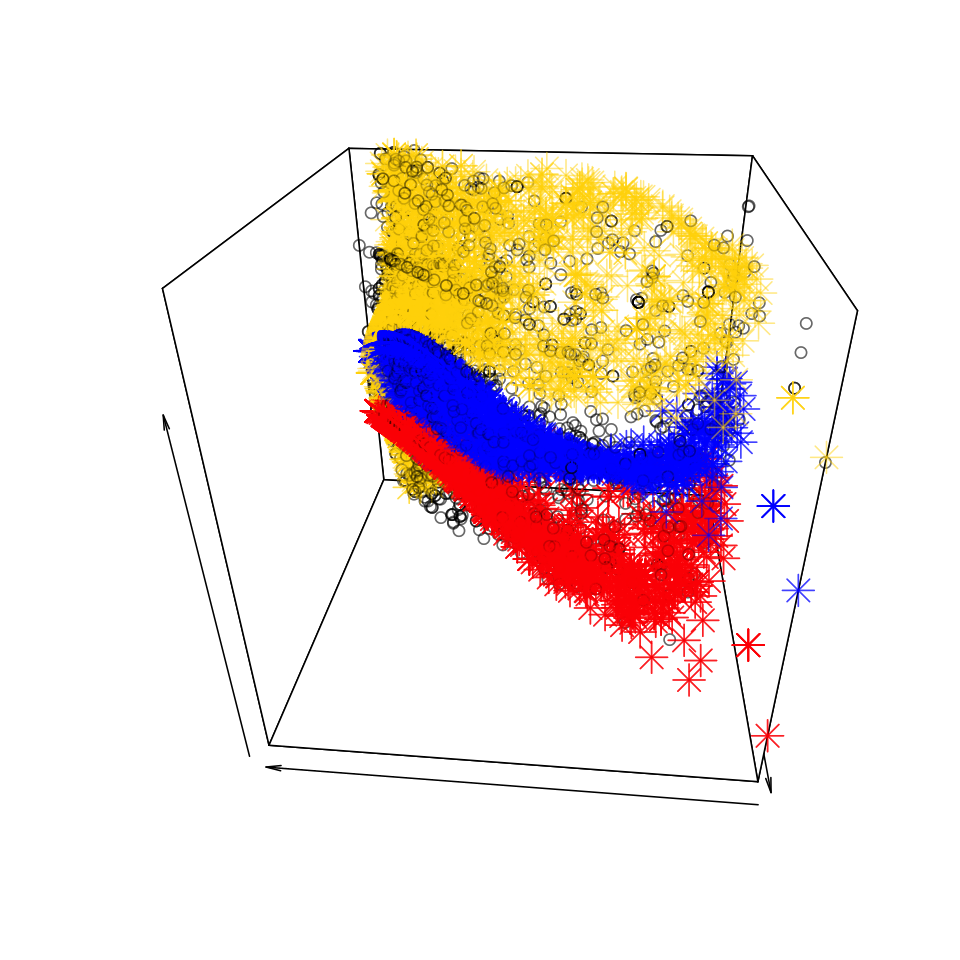}
\caption{1996-1997\\(\textit{SvM-c-3})}
\label{fig:real_data_fitted_results_3}
\end{subfigure}
\centering
\begin{subfigure}{.3\textwidth}
\centering
\includegraphics[width = 0.8\textwidth]{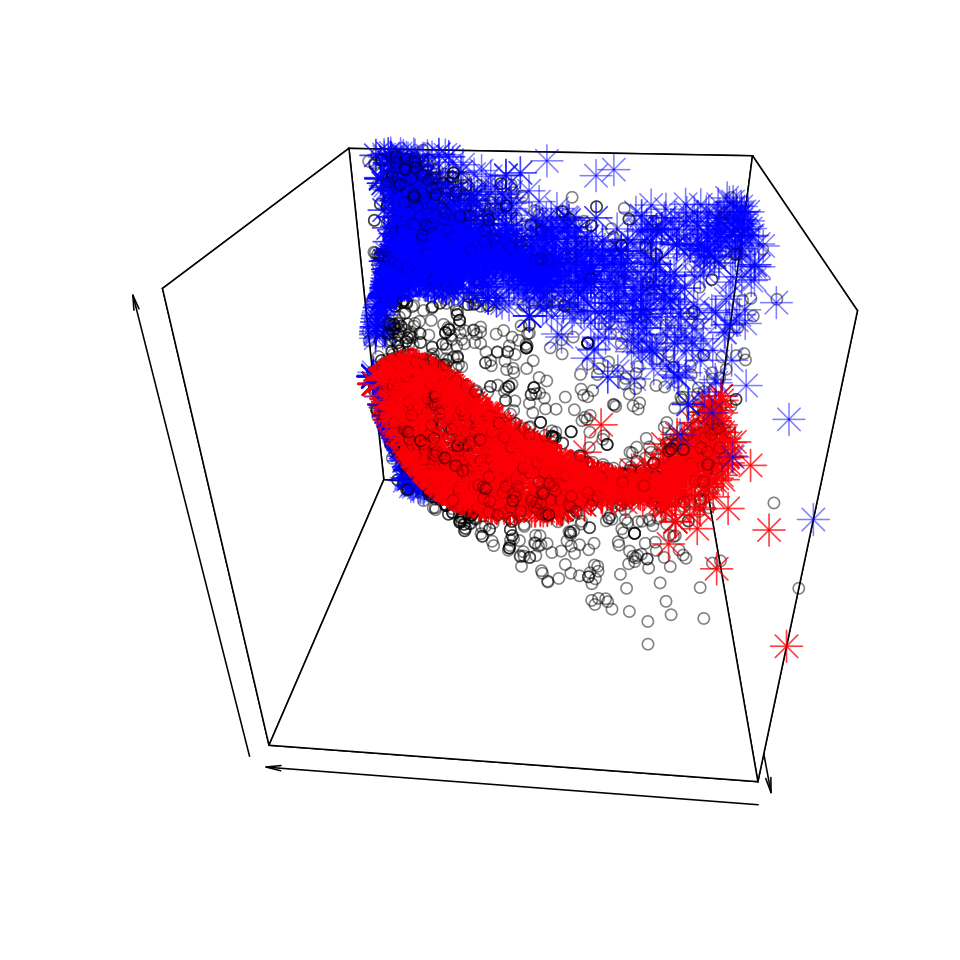}
\caption{1998-1999\\(\textit{SvM-c-2})}
\label{fig:real_data_fitted_results_9}
\end{subfigure}\\
\centering
\begin{subfigure}{.3\textwidth}
\centering
\includegraphics[width = .8\textwidth]{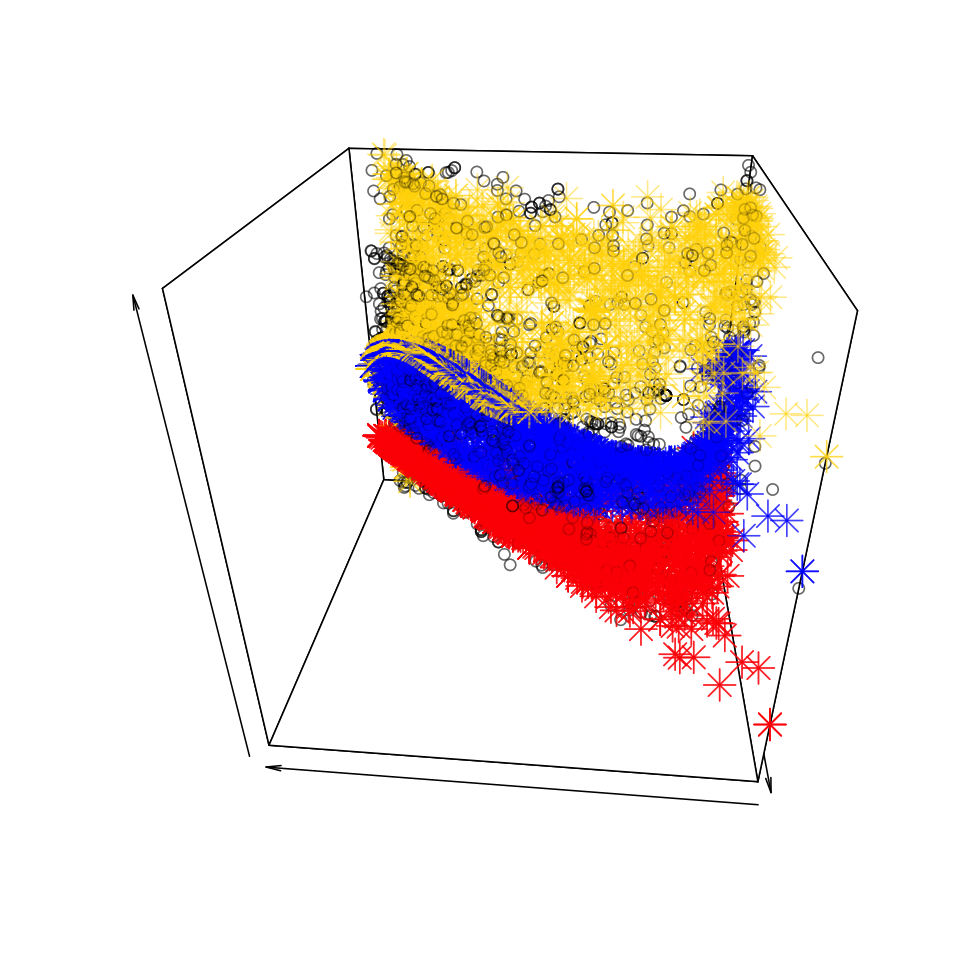}
\caption{2000-2001\\(\textit{SvM-c-3})}
\label{fig:real_data_fitted_results_16}
\end{subfigure}
\centering
\begin{subfigure}{.3\textwidth}
\centering
\includegraphics[width = .8\textwidth]{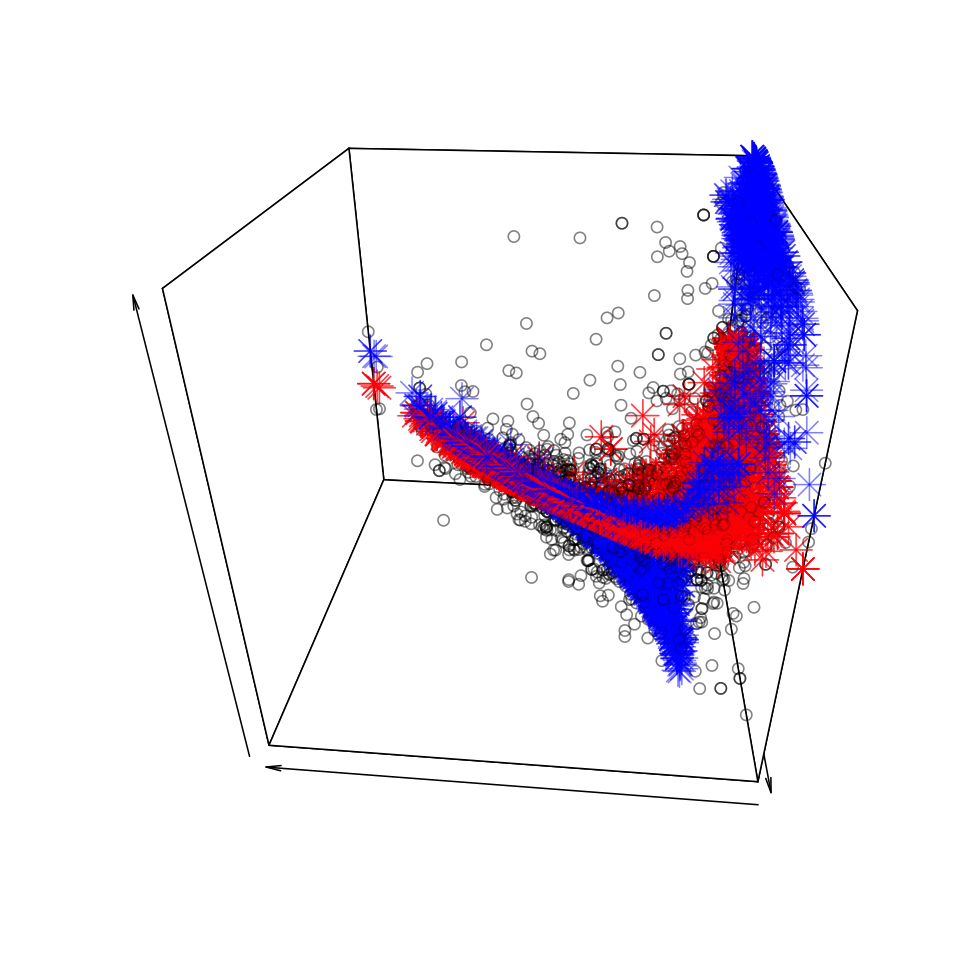}
\caption{2005-2006\\(\textit{SvM-c-2})}
\label{fig:real_data_fitted_results_16_svm_c}
\end{subfigure}
\centering
\begin{subfigure}{.3\textwidth}
\centering
\includegraphics[width = .8\textwidth]{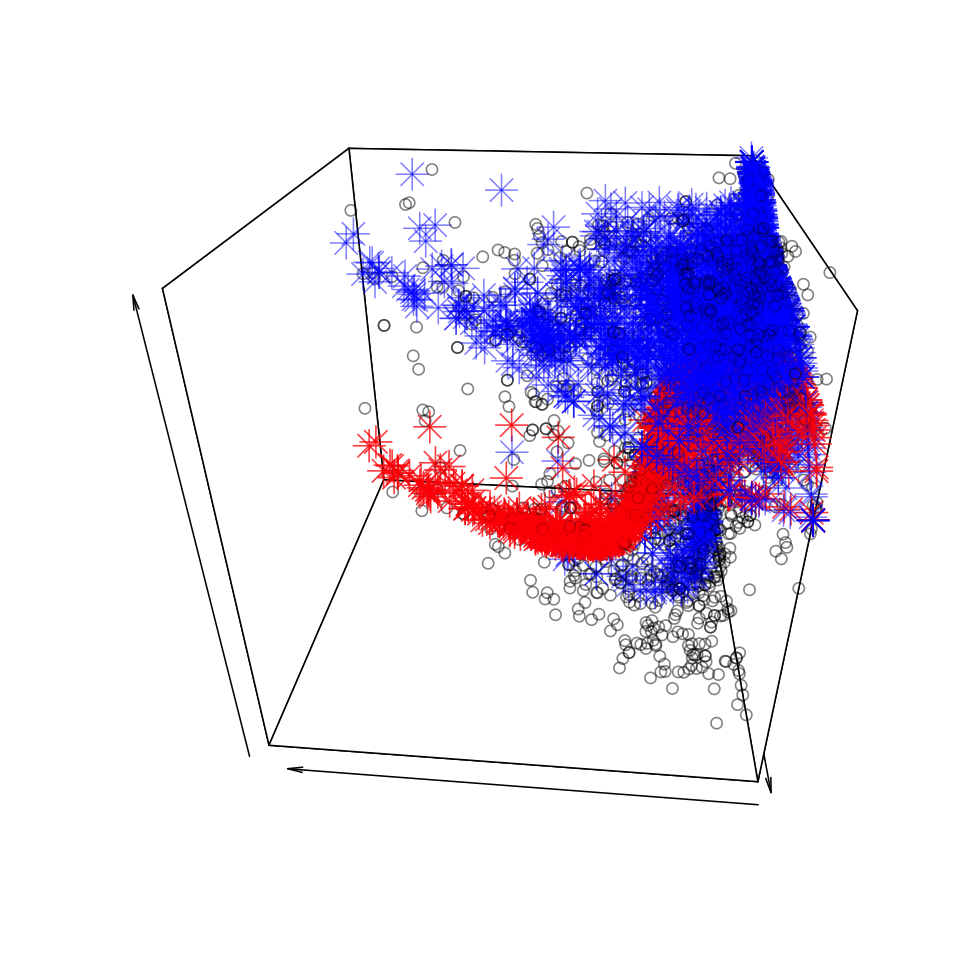}
\caption{2006-2007\\(\textit{SvM-c-2})}
\label{fig:real_data_fitted_results_17_svm_c}
\end{subfigure}
\caption{\small Plots showing the observed random direction not withheld and the fitted mean surface of the model selected by the posterior predictive log probability in \eqref{eq:post_pred_prob}. The front axis represents the proportion in the first income category and the side axis represents the proportion in the second income category. The up-down axis represents the direction. The start of the arrow indicates a value of zero whereas the end indicates a value of $2\pi$ for angles and 1 for proportions.}
\label{fig:real_data_fitted_results_example}
\end{figure}

Then, based on the best model and components and the fitted mean surfaces, the change in income proportions can be divided into four phases. Representative examples of each phase can be found in Figure \ref{fig:real_data_fitted_results_example}. All fitted mean surfaces are displayed in the appendix due to space constraints. The first phase is 1990-1992. While SvM-c is the favored model, the number of components is different. Still, the mean surfaces appear similar because there is a lower surface and an upper mean surface bisected by an upper mean surface from another component. The next phase takes place from 1992 to 2000. All fitted mean surface have a common upper and lower mean surface. The common lower surface is a tight tube-like blob that spirals upwards from the lowest to the highest income category. Meanwhile, the common upper surface is much more diffuse and dome-like. Because the fitted mean surfaces from 1997-1999 has these features, we include it in this phase despite the log posterior predictive probability favoring only two components. We then lump 2000-2002 into its own phase because the mean surfaces are unlike any before or after that time period. Indeed, the two mean surfaces during that period are different from each other because SvM-p-3 is favored during 2000-2001 and SvM-c-3 is favored during 2001-2002. Finally, while we will discuss the year to year changes in greater detail in the next paragraph, 2002-2010 are linked because there is a temporal evolution in the mean surfaces. In particular, the middle component from 2002-2003 grows for the second income category before disappearing after 2006-2007. Meanwhile, though as not as large as the increase in the middle component, the lower surface from 2002-2003 also increases until 2006-2007. It then flattens out from 2007-20010. Finally, the top surface from 2002-2003, flattens in 2003-2004, vanishes from 2004 to 2007, and reappears as a similar surface after 2007. Interestingly enough, these phases correspond to the early 1990 recession, the economic boom in the 1990s, the dot com bubble bursting, and the recovery from the dot com bubble with the subsequent housing market crash respectively. 


\begin{figure}[!t]
\centering
\begin{subfigure}{.45\textwidth}
\centering
\includegraphics[width = 0.6\textwidth]{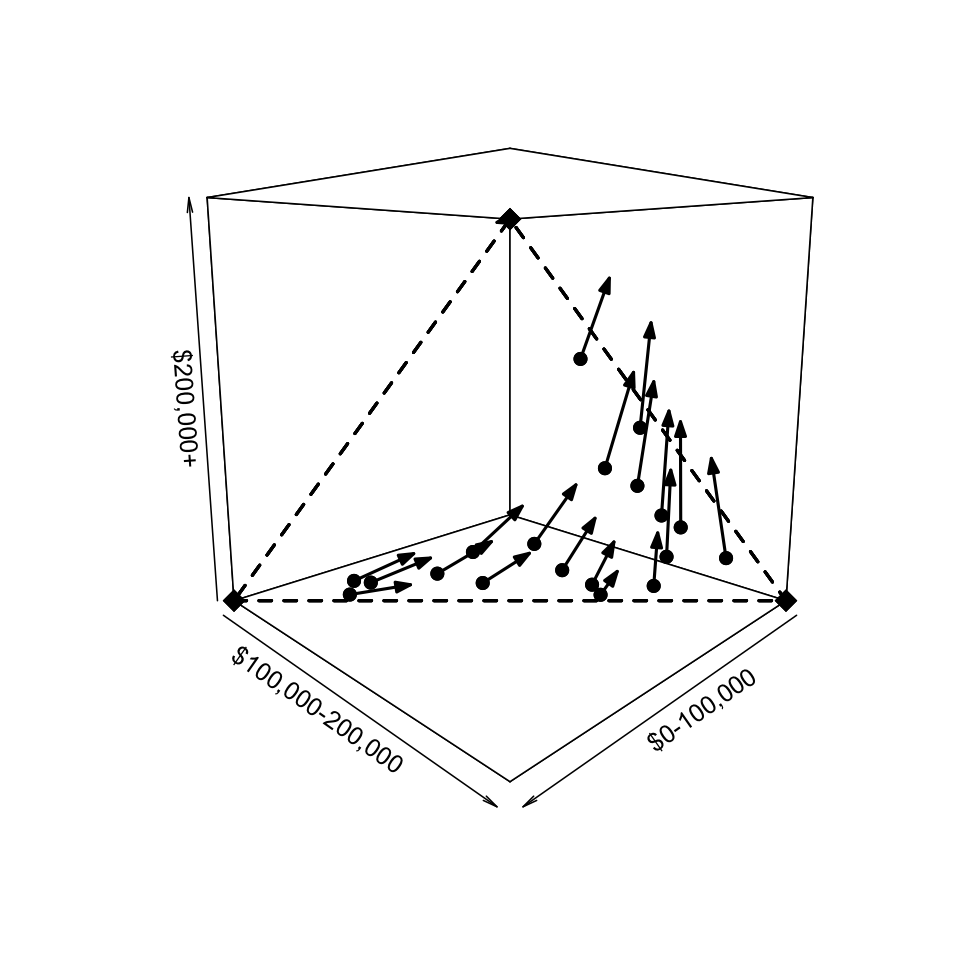}
\vspace{-5mm}
\caption{2005-2006 Red Component}
\label{fig:2005-2006_red}
\end{subfigure}
\begin{subfigure}{.45\textwidth}
\centering
\includegraphics[width = 0.6\textwidth]{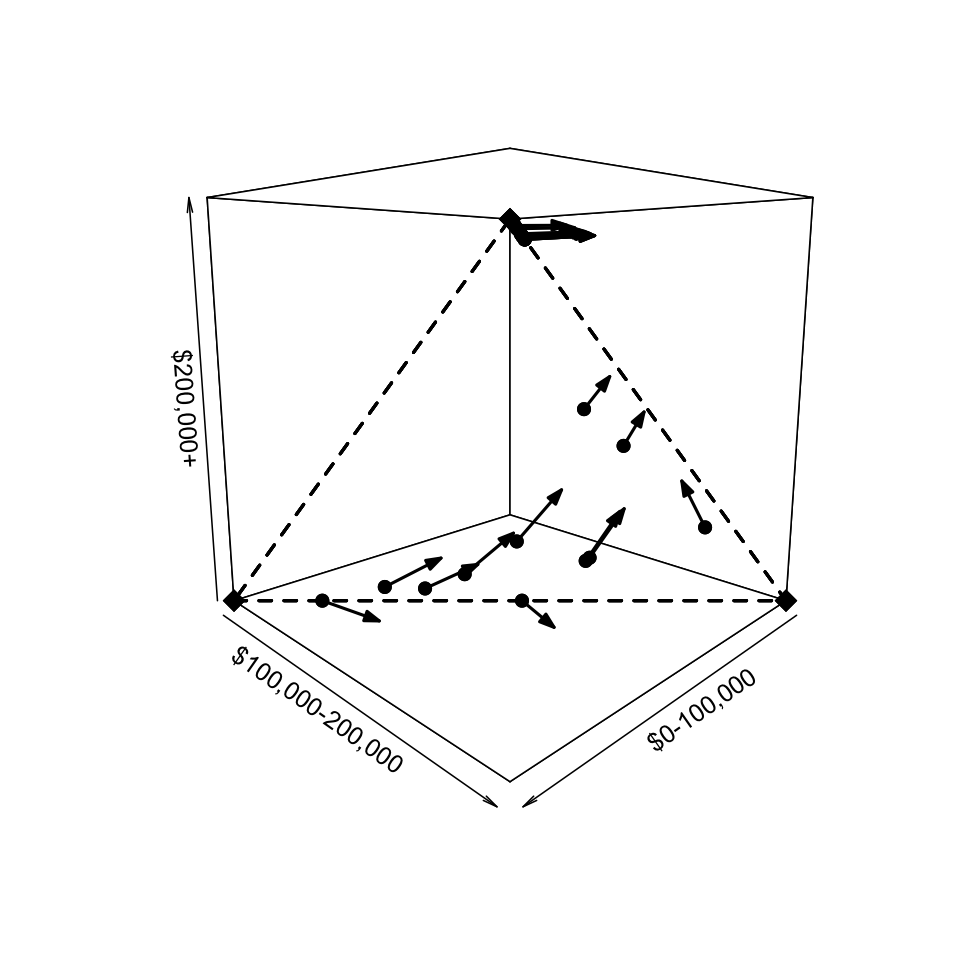}
\vspace{-5mm}
\caption{2005-2006 Blue Component}
\end{subfigure}\\
\begin{subfigure}{.45\textwidth}
\centering
\includegraphics[width = 0.6\textwidth]{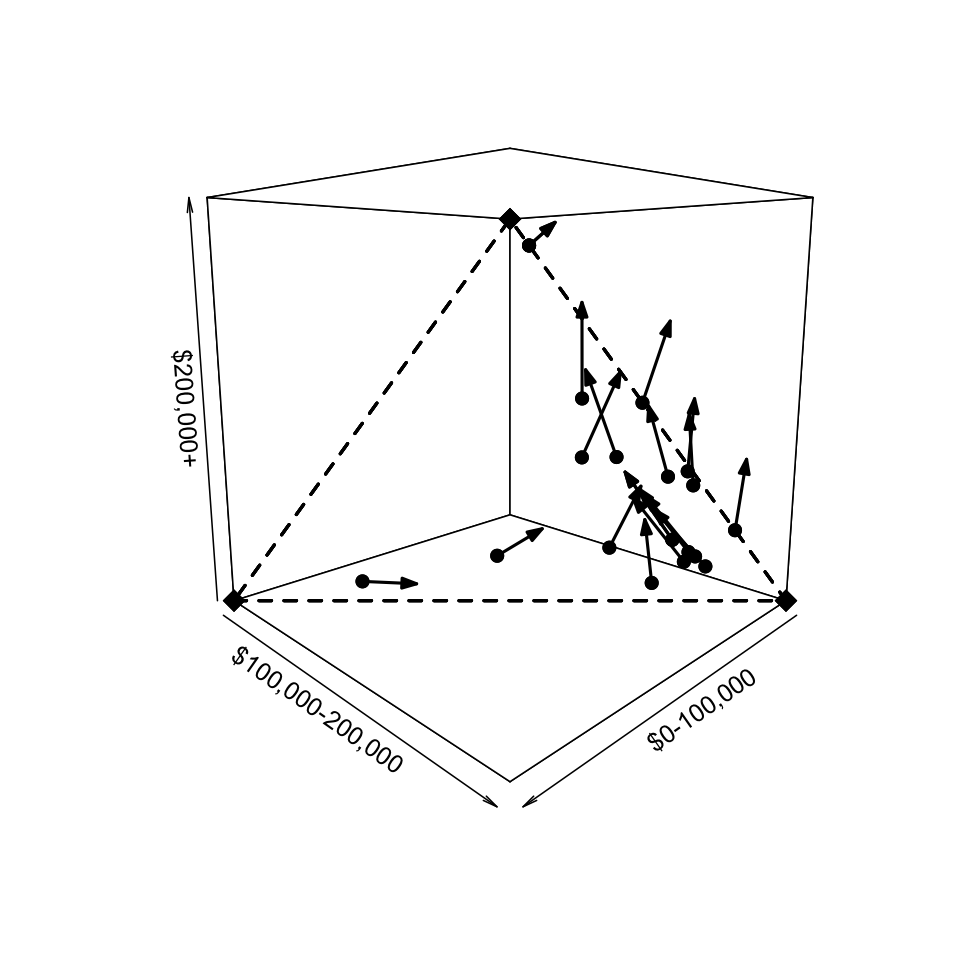}
\vspace{-5mm}
\caption{2006-2007 Red Component}
\end{subfigure}
\begin{subfigure}{.45\textwidth}
\centering
\includegraphics[width = 0.6\textwidth]{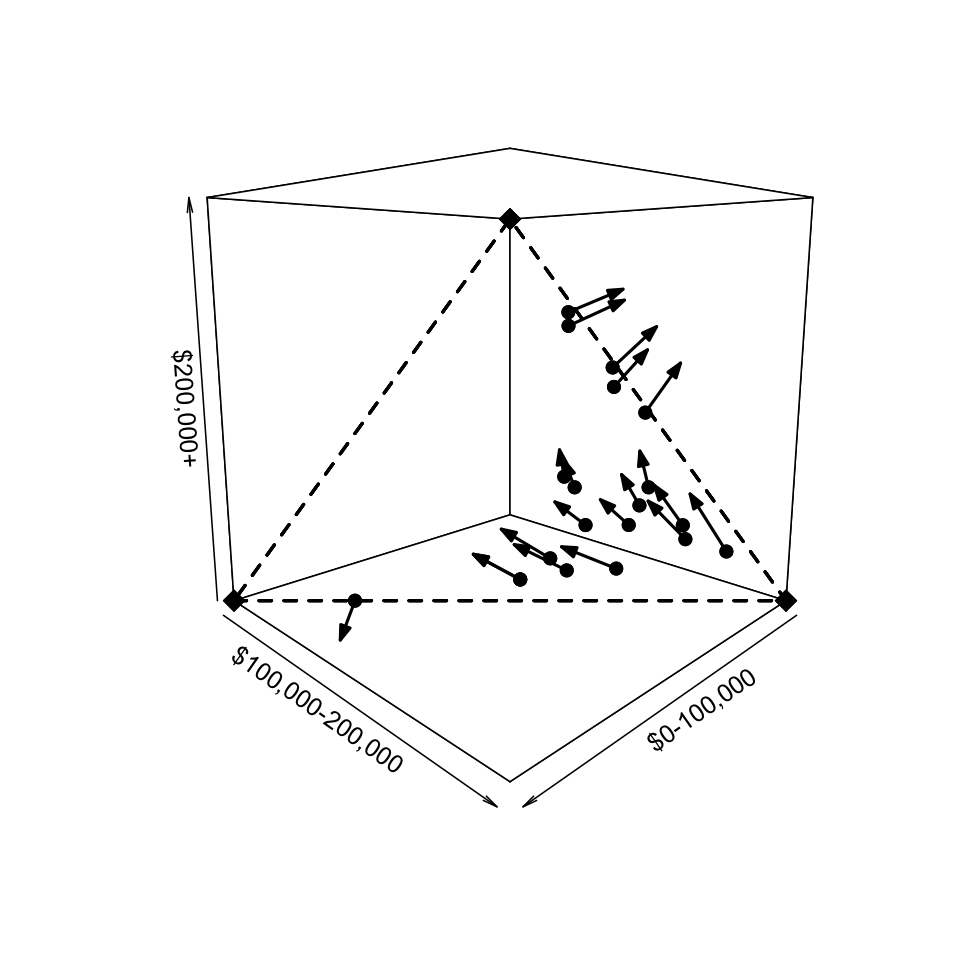}
\vspace{-5mm}
\caption{2006-2007 Blue Component}
\end{subfigure}
\caption{\small Plots showing the posterior mean random angles as their corresponding "movement vectors" for the top twenty locations associated with each component based on the mixing probability computed with the posterior means of the parameters and the random directions at the locations. These "movement vectors" were calculated with $\v{\mathcal{O}_p}$ and the spherical coordinates using the posterior mean random angles as $\phi_2$ and $\theta_2$ = 0.1. The color refers to the components' colors in Figures \ref{fig:real_data_fitted_results_16_svm_c} and \ref{fig:real_data_fitted_results_17_svm_c}.}
\label{fig:random_dir_vector_plot}
\end{figure}

As an example of the stories these surfaces can tell us, consider 2005 to 2006 and 2006 to 2007. We illustrate the entire surface in Figures \ref{fig:real_data_fitted_results_16_svm_c} and \ref{fig:real_data_fitted_results_17_svm_c}. Further, the top twenty posterior average angles based on the posterior mixing probability for each component are displayed in Figure \ref{fig:random_dir_vector_plot}. Like the year before 2005, there is one surface that is a curved, half spiral increasing from $\frac{\pi}{2}$ to around $\pi$ if we follow it from neighborhoods with income proportions largely below \$100 000 to neighborhoods with income proportions largely between \$100 000 and \$200 000 and then to neighborhoods with income proportions largely greater than \$200 000. This suggests that for these years, the income distributions for all neighborhoods are being pulled up a category. Indeed, this phenomenon is illustrated in Figure \ref{fig:2005-2006_red}. However, there is another surface that while similar for neighborhoods of lower income, is centered around zero for neighborhoods with income proportions in the second and third categories. This suggests that there already is a push away from the third income category for these neighborhoods, two years before the housing market crash. Because the posterior predictive probabilities are similar for \textit{SvM-c-2} and \textit{SvM-c-3} and there appears to be a surface that is zeroed out for \textit{SvM-c-3}, we examine the results from \textit{SvM-c-2}. Interestingly enough, the half spiral from the year increase from around $\frac{\pi}{3}$ to $\frac{3\pi}{2}$ if we follow it from neighborhoods with income proportions largely below \$100 000 to neighborhoods with income proportions largely between \$100 000 and \$200 000. It then decreases to $\pi$ if we follow it onward to neighborhoods with income proportions greater than \$200 000. This suggests that for these years, the income distributions for all neighborhoods are still being pulled up a category. However, the growth for lower income neighborhoods may not be as strong because there now may be a push away from the third or highest income category. Meanwhile, the other surface is now concentrated at around 0 and there is no longer a link through the "middle" or $\pi$. Indeed, it ranges between $\frac{3\pi}{2}$ and $\frac{\pi}{3}$ with a particular concentration around 0 for neighborhoods primarily in the second and third income categories. Again, this suggests that there already is a push away from the third income category for these neighborhoods. Unlike the previous year, this component's probability is larger than the "growth" component, which indicates the upcoming housing bubble burst. 

\subsection{Data analysis (Higher dimensions)}
Because of the spiral observed in the fitted mean surface for the changes in income proportions from 2005 to 2006, we examined the alterations in greater detail. Now, the number of categories are now reduced to six: \$0-\$25 000, \$25 000-\$50 000, \$50 000-\$100 000, \$100 000-\$150 000, \$150 000-\$200 000, and \$200 000+. This results in a data set with 2335 observations. After removing the 16 duplicates, we withheld 10\% of the data to compute the posterior predictive probability. This led to a training set of 2088 locations and observed angles and a test set of 231 locations and observed angles. We have fewer duplicates because there are more income categories. 

As we did in simulation, we ran the sampler for 25 000 iterations. We also had to model select the hyperparameters for the Gaussian process using the posterior predictive probability. Based on the previous subsection, we used $\omega = 0.5$ and $\sigma = 0.5$ as the reference set of parameters. For all but one set of hyperparameters, we used $\epsilon = 0.1$ for the Hamiltonian Monte Carlo step in order to avoid degenerate values for the concentration parameters. To avoid overfitting to one year, we looked at a time period to help us pick hyperparameters. In particular, the model was fitted to income proportions from 2002 to 2007, i.e. the housing bubble years. We found that the posterior predictive probability is generally highest for $\omega = 0.5$ and $\sigma = 1$. Note that this set required $\epsilon = 0.01$. Once we decided our choice of hyperparameters, we additionally had to select the number of mixing components. When the posterior predictive probability for SvM, SvM-c-2, SvM-c-3, and SvM-c-4 are compared, SvM-c-3 has the highest probability. Interestingly enough, this suggests that a mean direction surface is lost when we reduce the dimension of the data set. We did not check the kernel choice because of its poor performance in the three income category case.

\begin{figure}[!tb]
\centering
\begin{subfigure}{.24\textwidth}
\centering
\includegraphics[width = 1\textwidth]{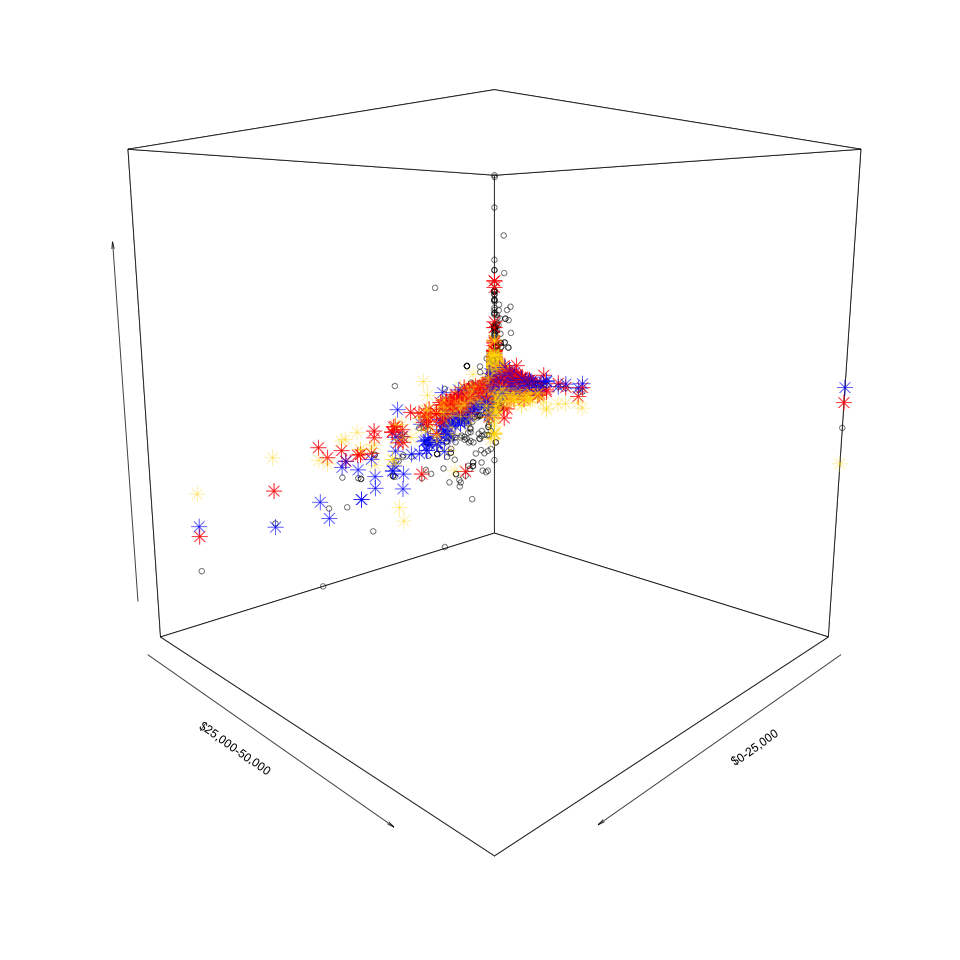}
\caption{First random direction}
\label{fig:real_data_higher_dim_first_dir}
\end{subfigure}
\centering
\begin{subfigure}{.24\textwidth}
\centering
\includegraphics[width = 1\textwidth]{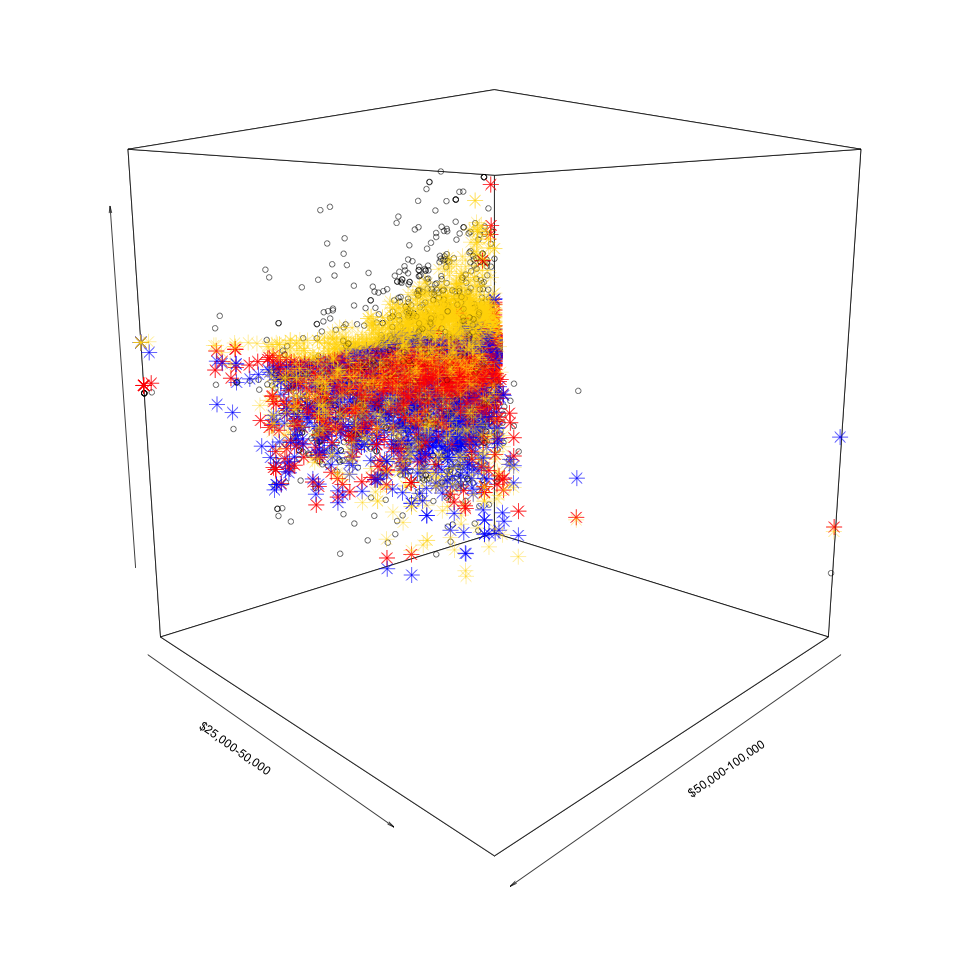}
\caption{Second random direction}
\label{fig:real_data_higher_dim_second_dir}
\end{subfigure}
\centering
\begin{subfigure}{.24\textwidth}
\centering
\includegraphics[width = 1\textwidth]{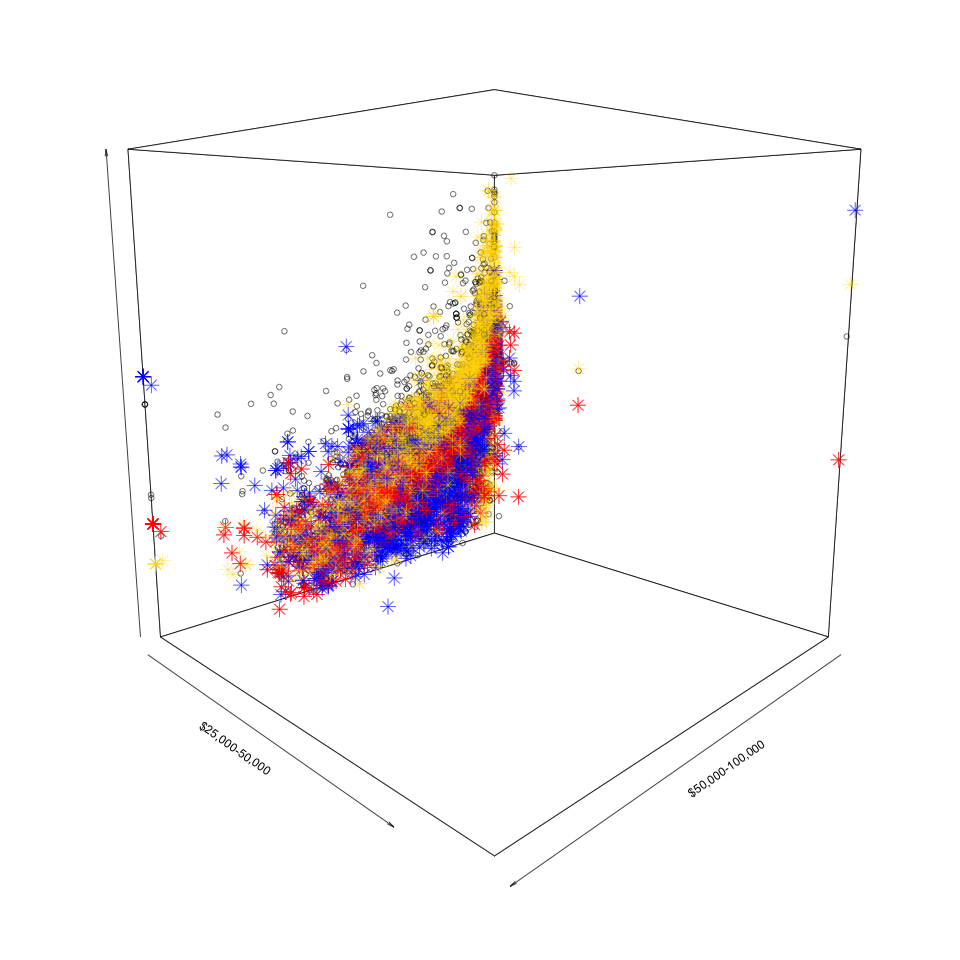}
\caption{Third random direction}
\label{fig:real_data_higher_dim_third_dir}
\end{subfigure}\\
\centering
\begin{subfigure}{.24\textwidth}
\centering
\includegraphics[width = 1\textwidth]{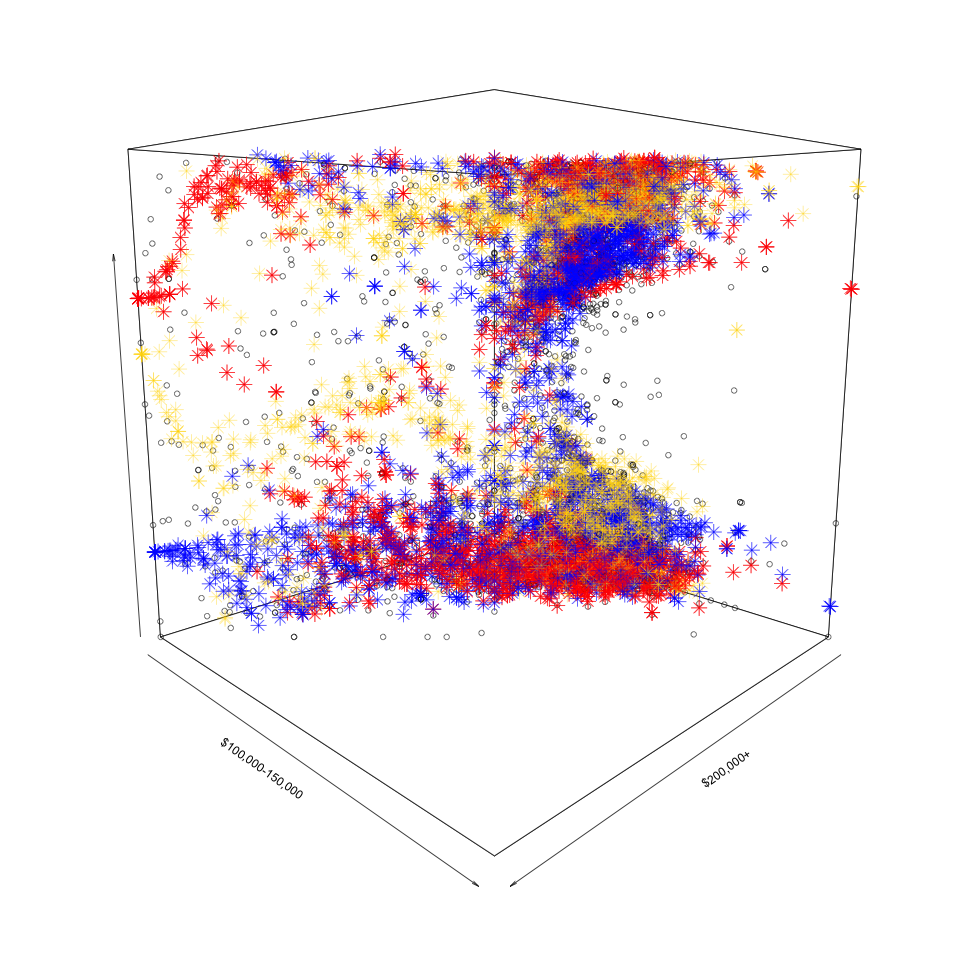}
\caption{Fourth random direction}
\label{fig:real_data_higher_dim_fourth_dir}
\end{subfigure}
\centering
\begin{subfigure}{.24\textwidth}
\centering
\includegraphics[width = 1\textwidth]{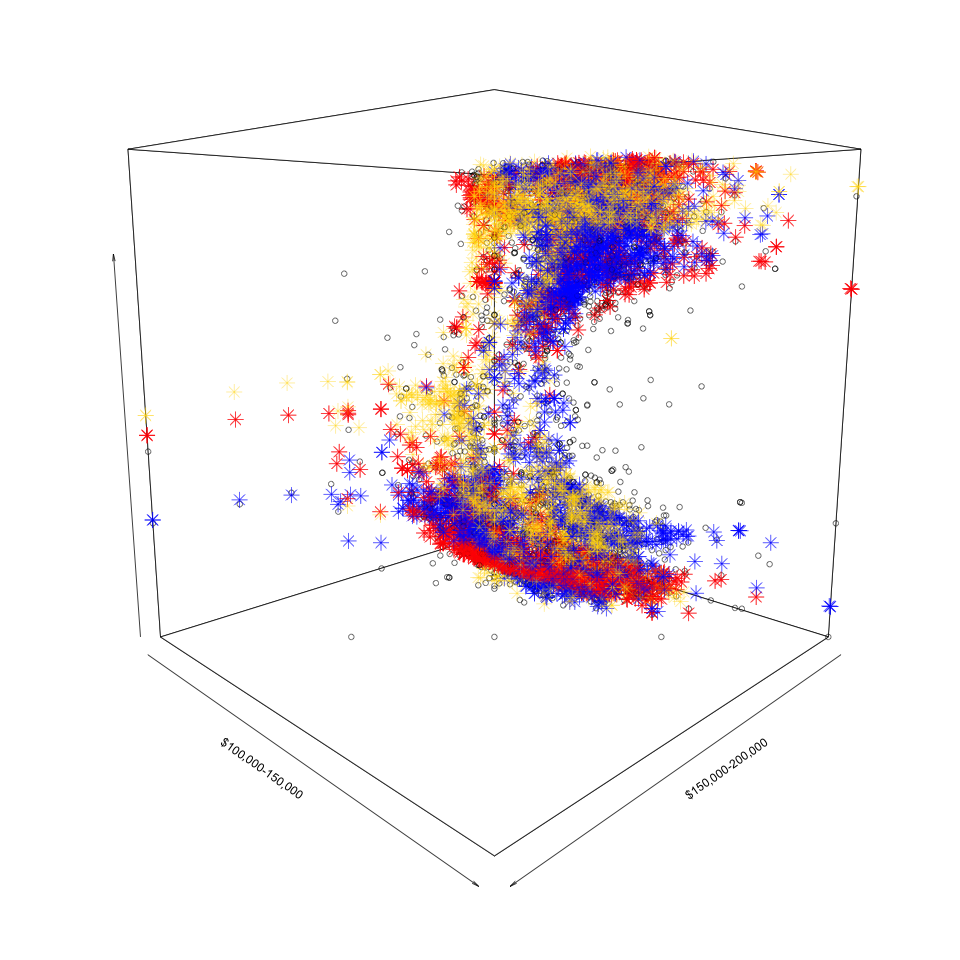}
\caption{Fourth random direction}
\label{fig:real_data_higher_dim_fourth_dir_alt}
\end{subfigure}
\caption{\small Plots showing the observed random direction not withheld and the fitted mean surface of the model selected by the posterior predictive log probability in \eqref{eq:post_pred_prob}. The up-down axis represents the direction. The start of the arrow indicates a value of zero whereas the end indicates a value of $\pi$ or $2\pi$ depending on the angle.}
\label{fig:real_data_fitted_results_example_higher_dim}
\end{figure}

We visualize the fitted mean surface to understand why and to better understand the fitted mean surface in the previous subsection. In particular, Figure \ref{fig:real_data_fitted_results_example_higher_dim} displays how the fitted surface for a random direction changes across the two income categories most relevant to that direction. First, we examine the mean surfaces associated with the first three random directions. These directions are associated with the three lowest income categories. These categories are related because we split the lowest income category from the three dimensional data set into the three lowest income categories for the higher dimensional data set. Previously, it had appeared that census tracts largely comprised of the lowest income category were increasing in wealth. The surfaces for the higher dimensional reveal a more nuanced picture. The first random direction's surface are mostly straight lines at $\frac{\pi}{2}$, suggesting no information about the changes. This makes sense because the income proportions for \$0-\$25 000 is mostly zero. The second random direction's surface has one surface largely above $\frac{\pi}{2}$, one concentrated at values slightly below $\frac{\pi}{2}$, and one largely below $\frac{\pi}{2}$. In other words, any change is possible for \$25 000-\$50 000. It is only the third random direction that might supports this characterization. Figure \ref{fig:real_data_higher_dim_third_dir} shows three banana or parabolic surfaces. Because it is the most visible, we discuss the gold surface. The gold one starts at $\pi$ for tracts in which the \$50 000-\$100 000 income proportions is zero and drops to $0$ for tracts in which the \$50 000-\$100 000 income proportions is one. In other words, this suggests a pull toward the third income category for tracts in which the \$50 000-\$100 000 income proportions is zero that becomes a push away from the third income category for tracts in which the \$50 000-\$100 000 income proportions is one. The other surfaces follow a similar trend, but may start and/or end at different places. This suggests a less dramatic change in the push and pull away from the third income proportions at either extremes of the proportion. One final note is that the \$50 000-\$100 000 income proportion is the largest constituent of the \$0-\$100 000 income proportion. As a result, this income's proportion might drive most of the changes observed in the lower dimensional changes.

Next, we examine the mean surfaces associated with the fourth random direction. Here, the results are more consistent with what we discovered in the previous subsection. Figure \ref{fig:real_data_higher_dim_fourth_dir_alt} displays how the fourth random direction changes with respect to the \$100 000-\$150 000 and the \$150 000-\$200 000, i.e. the categories that make the second income category. We see multiple surfaces that are centered at zero or $2\pi$ and connected through the middle with different surfaces for different combinations of the two income categories. This implies a push away from the \$100 000-\$150 000. It is consistent with what we observed in the lower dimensional data set because we observed a push away from the \$100 000-\$200 000 income category. If we examine this change with respect to the \$100 000-\$150 000 and the \$200 000+, i.e. the two income categories associated with directions, we still observe this trend for tracts whose income proportions are largely between \$100 000 and \$150 000. Meanwhile, for tracts whose income proportions are mostly above \$200 000, there are surfaces between $\pi$ and $\frac{3\pi}{2}$. This suggests a push away from the largest income category. 

We conclude this subsection by speculating why there are only two surfaces in the lower dimensional case. Figure \ref{fig:real_data_fitted_results_example_higher_dim} presents a "marginal" view of the surfaces because we only observe how the surface changes for one direction across two income categories. In other words, we reduce the dimensions of the space and the surface observed. This might make the surfaces appear "closer" to each other than they actually are in the 5D simplex. Still, even with this caveat, there appears to be significant overlap between the surfaces. In particular, the red and blue surfaces have the most overlap whereas the gold surface is somewhat distinct. As a result, we postulate that when the dimension of the data set is reduced, the red and blue surfaces are combined into one mean surface and the gold surface becomes another surface.

\section{Conclusion and Future Directions}
\label{sec:conclusion}
In this paper, as part of a new modeling framework for changes in data that lie on a simplex, we introduced new hierarchical models for the random direction associated with these changes and efficient samplers that recognize the geometry of angles to fit these models. Not only is there an unity to our approach across dimensions, but also these directions are of interest because they are fundamentally related to the movement of points within simplices. Indeed, we also explain in this paper the three steps needed to extract directions from any year to year changes of data that lie on a simplex. The other benefit of this approach is that this makes interpretation easier. For instance, when we analyzed our motivating data set of income proportions and extracted random directions for a set of census tracts in LA County from 1990 to 2010, the patterns our models discovered matches and clarifies real world economic trends during the same time period. We were then able to expand upon the trends observed in 2005-2006 by looking at a six income version of the data set.

There are several directions worth exploring moving forward. To better sample \textit{SvM-c} or the multidimensional equivalents, we might combine Multiple Try MCMC with elliptical slice sampling. This could provide a principled approach to consider the entire range of angles during each proposal step for the next mean angles and thus result in better mixing. In addition, we might merge our random direction models with a model for the magnitude in order to model the random movement. As discussed earlier, the two components of the movement were separated to achieve greater modeling flexibility and to better deal with the challenges posed by them. For instance, modeling the magnitudes and directions on the simplex's boundary require special care because the range of valid movements and directions are limited. In addition, the magnitudes in the interior have to be treated carefully. In order for the simplicial constraint to be respected, certain magnitudes in certain directions are not possible. On the flip side, certain directions might be more likely because it is possible to move further in that direction. Finally, we might extend this work to model all observed random directions instead of just each year's random direction. As suggested by this work, such a model has to change the number of clusters, adjust how the spatial information is assigned, and determine how to associate the mean surfaces from one year to the next.

\section*{Supplementary Material}
The supplementary materials include technical details on how to extract random directions from changes in data that lie on a simplex, a description of the \textit{SvM} model, a section on performing MCMC, simulation results, and more simulation and real data results. They are available from the journals page at \url{https://journals.sagepub.com/home/smj}.

\section*{Acknowledgements}
Any opinions, findings, and conclusions or recommendations expressed in this material are those of the author and do not necessarily reflect the views of the National Science Foundation. We also want to thank Professor Elizabeth Bruch for introducing us to the data set and for her discussions and Lydia Wileden for preparing the data.


\section*{Declaration of conflicting interests}
The authors declared no potential conflicts of interest with respect to the research, authorship and/or publication of this article.

\section*{Funding}
The first author was supported by the NSF Graduate Research Fellowship Program for most of this work (Grant No. DGE 1256260).

The second author was supported in part by NSF grants DMS-1351362, CNS-1409303 and DMS-2015361.




\bibliography{my_refs}

\end{document}


\maketitle

\section{Data overview}
\label{section:overview}
\subsection{Random directions from two dimensional simplices}

\begin{figure}[!tp]
    \centering
    \includegraphics[width = 0.45\textwidth]{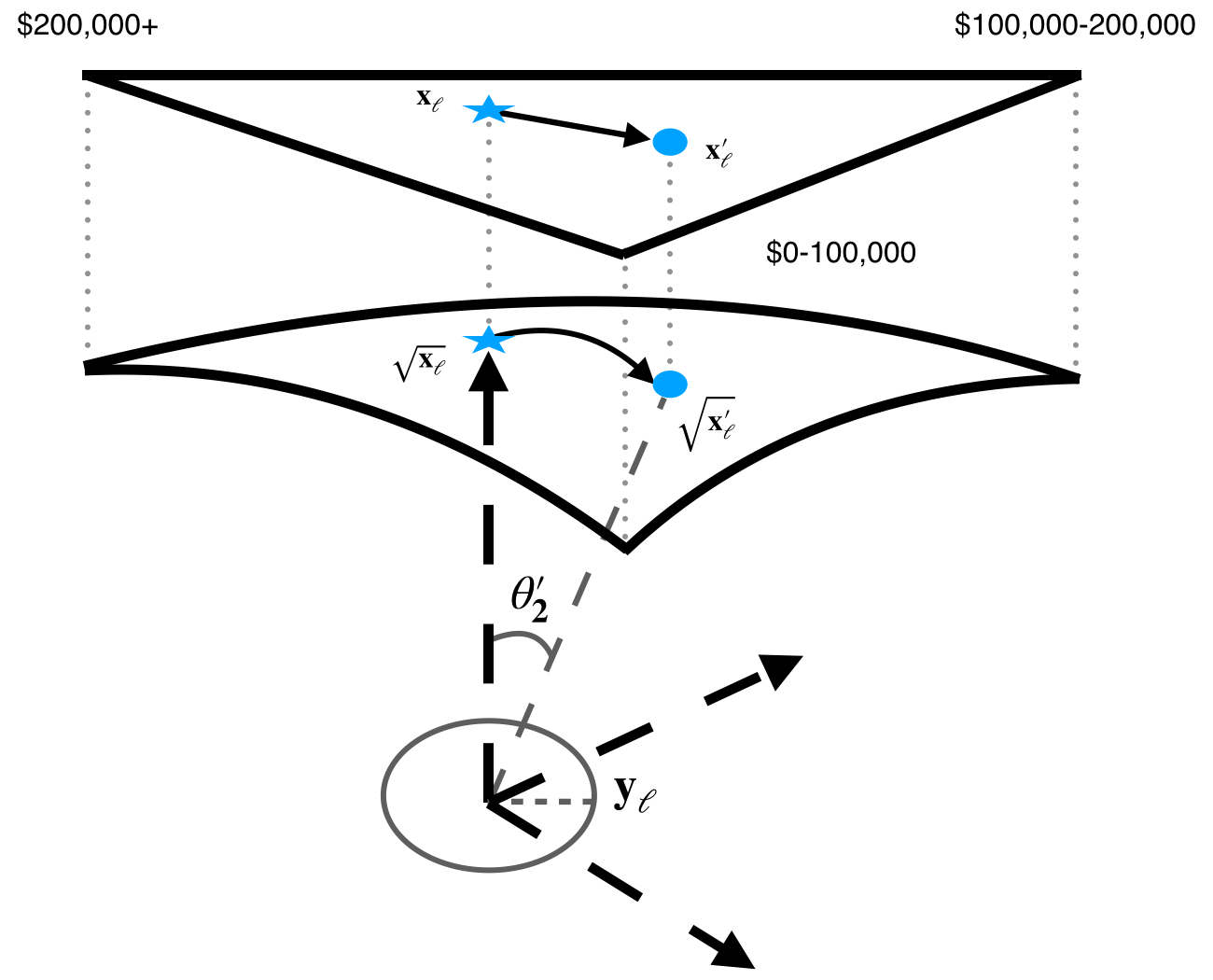}
    \caption{Figure displaying the connection between movement of points on a 2D simplex and random directions. The income proportions observed for a census tract in one year are shown as a star whereas the income proportions observed for a census tract in the next year are shown as a circle. The coordinate system we define in Section \ref{section:overview} is also displayed with dashed lines ending in arrows. Note that $\theta'_2$ is defined in \eqref{eq:rand_dir_info_2D}.}
    \label{fig:rand_dir_overview}
\end{figure}

When there are three income proportions for a census tract for a given year and the next year, we can extract the random
directions using the following procedure. Let $\v{x}_{\ell}$ be the income proportion for one year and $\v{x'}_{\ell}$ be the proportion for the next. First, we need the spherical coordinates for $\sqrt{\v{x}_{\ell}}$. Extract $\theta_1, \phi_1 \in [0,\frac{\pi}{2}]$ such that
\begin{align*}
    \theta_1 = \arccos(\sqrt{x_{\ell, 3}}), & & \phi_1 = \arctan^*\left(\sqrt{x_{\ell, 2}}, \sqrt{x_{\ell, 1}}\right).
\end{align*}
Here, $\arctan^*(\cdot, \cdot)$ is the following modified $\arctan$ function for $z_1, z_2 \in \mathbbm{R}$ such that $z_1^2 + z_2^2 = 1$:
\begin{align}
\textrm{arctan}^*(z_1, z_2) &=
    \begin{cases}
        \textrm{arctan}(\frac{z_2}{z_1}) & z_1 \geq 0, z_2 \geq 0\\
        \textrm{arctan}(\frac{z_2}{z_1}) + 2\pi & z_1 \geq 0, z_2 < 0 \\
        \textrm{arctan}(\frac{z_2}{z_1}) + \pi & z_1 < 0.
    \end{cases}
    \label{fct:arctan_star}
\end{align}
Second, we use the spherical coordinates to construct the following matrix, $\v{\mathcal{O}_p}$:
\begin{align}
\begin{pmatrix}
\cos\theta_1\cos\phi_1 & -\sin\phi_1 & \sin\theta_1\cos\phi_1\\
\cos\theta_1\sin\phi_1 & \cos\phi_1 & \sin\theta_1\sin\phi_1\\
-\sin\theta_1 & 0 & \cos\theta_1\\
\end{pmatrix}
\end{align}
\label{eq:rand_dir_trans_matrix}
Notice that this matrix transforms $(0, 0, 1)$ to $\sqrt{\v{x}_{\ell}}$. In other words, $\sqrt{\v{x}_{\ell}}$ is a "pole". Further, it is orthogonal. The first column's spherical coordinates are $(\theta_1 + \frac{\pi}{2}, \phi_1)$; the second column's are $(\frac{\pi}{2}, \phi_1 + \frac{\pi}{2})$. Consequently, we can use the matrix's inverse to get a new set of "coordinates" for $\sqrt{\v{x'}_{\ell}}$. The final step of our procedure is to extract the spherical coordinates from this new set of "coordinates". Set $\theta'_2 \in [0, \frac{\pi}{2})$ and $y_{\ell} \in [0, 2\pi)$ to be the angles such that
\begin{align}
    \theta'_2 = \arccos(\left(O_p \sqrt{\v{\widetilde{x}'_\ell}}\right)_3), & & y_{\ell} = \arctan^*\left(\left(O_p \sqrt{\v{\widetilde{x}'_\ell}}\right)_1, \left(O_p \sqrt{\v{\widetilde{x}'_\ell}}\right)_2\right).
    \label{eq:rand_dir_info_2D}
\end{align}
Then, $y_{\ell}$ is the random angle and $\theta'_2$ represents how "far" $\v{x}_\ell$ goes in that random direction. An illustration of these variables is shown in Figure \ref{fig:rand_dir_overview}.




This random direction has nice properties. Even if $\v{x}_{\ell}$ is zero for any proportions, the random direction can still be extracted and examined. It is also interpretable. In spherical coordinates, changing $y_{\ell}$ for a fixed $\theta'_2$ only affects the first two coordinates. Due to the transformation defined in the previous paragraph, a change in the first two coordinates changes how the first two columns of the matrix in \eqref{eq:rand_dir_trans_matrix} are weighted. To understand this random direction, we discuss what the first two columns represent and how the random direction interacts with these columns. Because the first column represents the spherical coordinates for ($\theta_1 + \frac{\pi}{2}$, $\phi$), the simplest way to understand this column is to compare $\theta_1 = 0$ and $\theta_1 = \frac{\pi}{2}$. As $\theta_1 = 0$ corresponds to $(0, 0, 1)$ and $\theta_1 = \frac{\pi}{2}$ to $(\cos(\phi), \sin(\phi), 0)$ for some $\phi \in [0, 2\pi)$, this first column represents a push away from the third income category. On the other hand, if we use a similar logic to compare $\theta = \frac{\pi}{2}$ and $\phi_1 = 0$ against $\theta = \frac{\pi}{2}$ and $\phi_1 = \frac{\pi}{2}$, then we see that the second column represents a pull toward the second income category. 

Now that we have explained the meaning of the columns of \eqref{eq:rand_dir_trans_matrix}, we discuss the interpretations of random directions based on how they weight the columns. Because the first and second coordinate include $\cos(y_{\ell})$ and $\sin(y_{\ell})$ respectively, we will examine $0$, $\frac{\pi}{2}$, $\pi$, and $\frac{3\pi}{2}$. Note that at $0$ and $\pi$, the first coordinate will be $1$ and $-1$ and the second will be zero by definition. As the first column represents a push away from the third income category, a random direction of $0$ "confirms" this column and thus is a push away from the third income category. Meanwhile, a random direction of $\pi$ "negates" this column and thus is a pull toward. The reverse is true for $\frac{\pi}{2}$ and $\frac{3\pi}{2}$. While the first coordinate is zero and the second coordinate is $1$ and $-1$ at $\frac{\pi}{2}$ and $\frac{3\pi}{2}$, the second column represents a pull toward the second income category. As a result, a random direction of $\frac{\pi}{2}$ and $\frac{3\pi}{2}$ represent a pull toward and push away from the second income category. 



\subsection{Random directions from simplices of higher dimensions}
At a high level, the procedure to extract random directions when there are more than three income proportions is similar to the procedure to extract random directions when there are three income proportions. Given a vector, $\v{x}_{\ell} \in \Delta^{D}$, and its corresponding vector from the next year, $\v{x}_{\ell} \in \Delta^{D}$, for $D > 2$, we again will use the spherical coordinates in higher dimension of $\sqrt{\v{x}_{\ell}}$ (i.e. $\v{x}_\ell$'s element-wise square root) to define a matrix to transform $\sqrt{\v{x'}_{\ell}}$. The random directions will then be taken from the spherical coordinates of this transform.

However, differences arise because of how spherical coordinates are defined in higher dimension. As a result, we will quickly go over these coordinates for points that lie on a higher dimensional sphere. Let $\v{a} \in [0, \pi]^{D - 1} \times [0, 2\pi)$ denote the spherical coordinates in higher dimensions. Meanwhile, say that $\v{s} \in \mathbbm{R}^{D + 1}$ is a point on the higher dimensional sphere, $\mathbbm{S}^{D}$, if $\sum_{d = 1}^{D + 1} s^2_d = 1$. Then, $\v{s}$ and $\v{a}$ are linked in the following manner:
\begin{equation}
\begin{aligned}
   s_1 &= \cos(a_1),\\
   s_2 &= \sin(a_1)\cos(a_2),\\
   s_3 &= \sin(a_1)\sin(a_2)\cos(a_3),\\
   &\vdots\\
   s_{D} &= \left(\prod_{d = 1}^{D - 1} \sin(a_d)\right)\cos(a_{D}),
   s_{D + 1} &= \left(\prod_{d = 1}^{D - 1} \sin(a_d)\right)\sin(a_{D}).   
\end{aligned}
\label{eq:spherical_transform}    
\end{equation}
For convenience, denote this transformation as $\mathcal{ST}(\cdot)$, i.e. $\v{s} = \mathcal{ST}(\v{a})$. Notice that $\mathcal{ST}(\cdot)$ gives us a point on the sphere for a set of coordinates whereas we are interested in the reverse. Fortunately, this transformation has an unique inverse, $\mathcal{ST}^{-1}(\cdot)$, for points on the sphere with a few exceptions:
\begin{align}
    a_d = \acos\left(\frac{s_d}{\sqrt{\sum_{d' = d}^{D + 1} s^2_{d'}}}\right), d = 1, 2, \ldots, D - 1, & & a_{d} = \begin{cases}
        \acos\left(\frac{s_{D}}{\sqrt{s_D^2 + s_{D + 1}^2}}\right) & s_d \geq 0,\\
        2\pi - \acos\left(\frac{s_{D}}{\sqrt{s_D^2 + s_{D + 1}^2}}\right) & s_d < 0.\\
    \end{cases} 
\end{align}
The exceptions occur when $\sum_{d' = d}^D s^2_{d'} = 0$ for some $d \in 1, 2, \ldots, D$. In those cases, set $a_{d'} = 0$ for $d \leq d' \leq D$. Finally, as an abuse of notation, we may also let $\mathcal{ST}^{-1}(\v{s})$ represent the spherical coordinates themselves.

Now that we have discussed spherical coordinates in higher dimensions, we explain how it affects our procedure. The first step of our procedure is to extract the spherical coordinates of $\sqrt{\v{x}_{\ell}}$. Let $\theta_1, \phi_1, \ldots, \phi_{D - 1} \in [0, \frac{\pi}{2}]$ denote these coordinates. Then, we set these coordinates in the following manner:
\begin{align}
    \theta_1 = \mathcal{ST}^{-1}(\sqrt{\v{x}_{\ell}})_1, & & \phi_d = \mathcal{ST}^{-1}(\sqrt{\v{x}_{\ell}})_{d + 1}, \quad d = 1, 2, \ldots, D - 1.
\end{align}
Notice that in the higher dimensional case, $\theta_1$ refers to the first and not the last angle. While it may be confusing notationally, this choice follows the spirit of the previous section. The coordinate defined by a single $\cos$ function is represented by $\theta_1$. This is the first coordinate for spherical coordinates in higher dimensions. 

Our next step of constructing an orthogonal matrix, $\mathcal{O}_p$, to transform $\sqrt{\v{x'}_{\ell}}$ is also affected. Recall that we want to define a coordinate system so that $\sqrt{\v{x'}_{\ell}}$ serves as a "pole". In particular, the "pole" should be the same coordinate as the coordinate defined by a single $\cos$ function. Hence, the first column of $\mathcal{O}_p$ is $\sqrt{\v{x'}_{\ell}}$. If we follow what we did in the previous subsection, the next two columns should be points on the sphere corresponding to modifications of $\theta_1$ and $\phi_1$. In particular, a natural choice for the second column and third columns are the following: 
\begin{align*}
    \v{\mathcal{O}_{p, \cdot, 2}} &= \mathcal{ST}(\theta_1 + \frac{\pi}{2}, \phi_1, \phi_2, \ldots, \phi_{D - 1}),\\
    \v{\mathcal{O}_{p, \cdot, 3}} &=  \mathcal{ST}(\frac{\pi}{2}, \phi_1 + \frac{\pi}{2}, \phi_2, \ldots, \phi_{D - 1}).
\end{align*}
These column then provide a template for the remaining columns. The difference going from columns two to three is that the angle we previously add $\frac{\pi}{2}$ is set to $\frac{\pi}{2}$. In addition, we add $\frac{\pi}{2}$ to the next angle, $\phi_1$. If we follow this recursively for columns $d = 4, 5, \ldots, D + 1$, set 
\begin{align*}
    \v{\mathcal{O}_{p, \cdot, d}} &= \mathcal{ST}(\frac{\pi}{2}, \frac{\pi}{2}, \ldots, \frac{\pi}{2}, \phi_{d - 2} + \frac{\pi}{2}, \phi_{d - 1}, \ldots, \phi_{D - 1}).
\end{align*}

We now discuss the final step of extracting the random directions from the transformed $\sqrt{\v{x'}_{\ell}}$. Doing so will provide a mathematical explanation for many of the changes induced by the higher dimensional spherical coordinates. Define $\theta_2' \in [0, 2\pi)$ and $\v{y}_\ell \in [0, \pi]^{D - 2} \times [0, 2\pi)$ to be the following quantities:
\begin{align}
    \theta_2' = \mathcal{ST}^{-1}(\mathcal{O}^{-1}_p \sqrt{\v{x}'_{\ell}})_1, & & \v{y}_\ell = \{\mathcal{ST}^{-1}(\mathcal{O}^{-1}_p \sqrt{\v{x}'_{\ell}})_d\}_{d = 2}^{D}.
\end{align}
As hinted by our notation, the random directions in higher dimensions are all the extracted angles except the first angle. While this is inconsistent with the random directions in lower dimensions, the geodesic that we use to represent the movement dictates this choice. Let $v_{1} = 1$ and $v_d = 0$ for $d = 2, 3, \ldots, D$ and $v'_{1} = 0$ and $v'_{d} = \mathcal{ST}(\theta_2', y_{\ell, 1}, y_{\ell, 2}, \ldots, y_{\ell, D - 1})_{d - 1}$ for $d = 2, 3, \ldots, D$. Then, $\v{v}\cos(\theta'_2) + \v{v'}\sin(\theta'_2)$ is the geodesic that connects $\mathcal{O}^{-1}_p\sqrt{\v{x_\ell}}$ and $\mathcal{O}^{-1}_p\sqrt{\v{x'_\ell}}$ \citep{CanaryEtAlFundamentalsHyperbolicManifolds2006}. Notice that we follow this convention when we first extracted the spherical coordinates in higher dimensions from $\sqrt{\v{x_\ell}}$. This ensures the procedure to extract random directions from any dimension is internally consistent. Further, setting $\theta'_2 = 0$ should give us $\sqrt{\v{x_\ell}}$ as a "pole". As $\v{v}\cos(\theta'_2) + \v{v'}\sin(\theta'_2) = \v{v}$ for $\theta'_2 = 0$, it makes sense that the first column of the orthogonal matrix is $\sqrt{\v{x_\ell}}$.

\begin{table}[!t]
\centering
\resizebox{.9\textwidth}{!}{%
 \begin{tabular}{c c c c c c} 
 \hline
 & $\v{O_{p, 2}}$ & $\v{O_{p, 3}}$ & $\v{O_{p, 4}}$ & $\v{O_{p, 5}}$ & $\v{O_{p, 6}}$ \\ [0.5ex] 
 \hline
  & & &\\
 \textbf{Income category} & \$0 - \$25 000 & \$25 - \$50 000 & \$50 000 - \$100 000 & \$100 000 - \$150 000 & \$200 000+ \\ 
 & & &\\
 \textbf{Column Meaning} & Push & Push & Push & Push & Pull \\ 
 & & &\\
 \textbf{Random direction} & $y_{\ell, 1}$ & $y_{\ell, 2}$ & $y_{\ell, 3}$ & $y_{\ell, 4}$ & $y_{\ell, 4}$ \\ 
 & & &\\
 \hline
  & & &\\
 \end{tabular}
 }
 \caption{Table displaying rotation matrix's columns and their associated income category, random direction, and meaning for the six income category data set.}
 \label{table:real_data_random_dir_trans}
\end{table}

We now interpret these random directions. First, we discuss the orthogonal matrix, $\v{\mathcal{O}_p}$, using the same logic from before. We begin with the second column for which we add $\frac{\pi}{2}$ to $\theta_1$ and hold other angles constant. Because $\cos(\theta_1)$ corresponds to the first coordinate and thus the first income category, it represents a push away from the first income category towards higher income categories. The next column sets $\theta_1$ to be $\frac{\pi}{2}$, adds $\frac{\pi}{2}$ to $\phi_1$, and holds other angles constant. Based on the spherical transform defined in \eqref{eq:spherical_transform}, the first coordinate is zero. Meanwhile, the second coordinate can be treated as the "first" coordinate under the logic of the previous subsection because the second coordinate is represented by a single cos function. As a result, the third column of $\mathcal{O}_p$ can be thought of as a push away from the second income category toward higher income categories. Indeed, because we do something similar for $\v{O_{p, \cdot, d}}$ for $d = 3, 4, \ldots, D$, $\v{O_{p, \cdot, d}}$ represents a push away from the $d - 1$ income category towards higher income categories. Finally, the last column of $\v{\mathcal{O}_p}$ can be thought of in similar terms as the second column of $\v{\mathcal{O}_p}$ from the previous subsection. It represents a pull toward whichever coordinate out of the remaining non-zero coordinates corresponds to the sin function. As this is the last coordinate, the last column represents a pull toward the highest income category.

Before we discuss the meaning of the random directions within the context of $\mathcal{O}_p$, we present a more concrete example to clarify our construction. Recall that the higher dimensional data set that we analyze has six income categories. In other words, the data lies on a five dimensional simplex so $D = 5$. The random direction we extract from any movement has dimension $D - 1$ or $4$. Unlike the previous subsection, these angles are the last four angles of $O^{-1}_p\sqrt{\v{x'}_\ell}$'s spherical coordinates in higher dimensions. Speaking of $O^{-1}_p$, there are six columns. The first is the elementwise square root of the initial point, $\sqrt{\v{x}_\ell}$. For the next four columns, each refer to a push away from the income category one less than the column number toward higher income categories. The final column is different because it is a pull toward the highest income category. This is summarized in Table \ref{table:real_data_random_dir_trans}.

We now discuss the interpretation of random direction based on how a dimension's direction interacts $\v{\mathcal{O}_p}$. When we multiply $\v{\mathcal{O}_p}$ by $\mathcal{ST}(\theta_2', y_{\ell, 1}, y_{\ell, 2}, \ldots, y_{\ell, D - 1})$, the first column of the product does not change when we change the random direction. We next examine $\v{y_{\ell, d}}$ for $d = 1, 2, \ldots, D - 2$ because $\v{y_{\ell, d}} \in [0, \pi)$ for the direction in these dimensions. Unlike before, $\v{y_{\ell, d}}$ affects all columns after $d$. If we perturb $\v{y_{\ell, d}}$, we have that for some constants $c, c_{d'} \in \mathbbm{R}$ for $d' \in {d + 1, d + 2, \ldots, D - 2}$, 
\begin{align*}
  \v{\mathcal{O}_{p, d + 1}} \cos(y_{\ell, d}) c & & \v{\mathcal{O}_{p, d'}} \sin(y_{\ell, d}) c_{d'}, \quad d' \in {d + 1, d + 2, \ldots, D - 2}.
\end{align*}
Since $\cos(0) = 1$ and $\cos(\pi) = -1$ and $\sin(0) = \sin(\pi) = 0$, $\v{y_{\ell, d}}$ switches between a push away from income category $d$ to a pull towards income category $d$ as $\v{y_{\ell, d}}$ increases from $0$ to $\pi$. In addition, because $\cos(\frac{\pi}{2}) = 0$ and $\sin(\frac{\pi}{2}) = 1$, the random direction can indicate "support" for the changes occuring to higher income categories. These directions also reflect how much change there is for its income category relative to higher income categories. Finally, note that $\v{y}_{\ell, D - 1} \in [0, 2\pi)$ and it is associated with the last two columns. In particular, we multiply the penultimate column by $\cos(\v{y}_{\ell, D - 1})$ and the last column by $\sin(\v{y}_{\ell, D - 1})$. As a result, $\v{y}_{\ell, D - 1} = 0$ again represents a push away from the third highest income category and $\v{y}_{\ell, D - 1} = \pi$ represents a pull toward the third highest income category. However, because the last column of $O_{p}$ is a pull toward the highest income categories, the meaning of $\frac{\pi}{2}$ and $\frac{3\pi}{2}$ is different. Now, $\v{y}_{\ell, D - 1} = \frac{\pi}{2}$ is a pull toward the highest income category and $\v{y}_{\ell, D - 1} = \frac{3\pi}{2}$ is a push away from that income category. Table \ref{table:real_data_random_dir_trans} also shows which random directions are directly related to which columns for the six income category data set.

\section{Homogeneous Spatial Random Direction Model}
\label{section:svm}

For notational clarity, we discuss \textit{SvM} or \textit{SvM-c} with one component. First, in the two dimensional case, each observation $y_\ell$ has a mean parameter, $m_\ell \in [0, 2\pi)$, and a concentration parameter, $\rho_\ell \in \mathbbm{R}^+$. We use the modified arctan function defined in \eqref{fct:arctan_star} to element-wise transform two draws from the Gaussian process, $\v{z_1}, \v{z_2} \in \mathbbm{R}^N$, into the mean parameters, $\v{m} \in [0, 2\pi)^N$. The positive-valued concentration parameter, $\rho_\ell$, is transformed via exponentiation from a real-valued $\varphi_\ell$, which is then endowed with a normal prior.  
In summary, \textit{SvM} is the following.
\begin{align*}
    \v{z_1} &\sim \textrm{GP}(\cdot \mid \mu_1, \v{\Sigma}),\\
    \v{z_2} &\sim \textrm{GP}(\cdot \mid \mu_2, \v{\Sigma}),\\
    \v{m} &= \textrm{arctan}^*(\v{z_1}, \v{z_2}), \stepcounter{equation}\tag{\theequation}\label{model:SvM}\\
    \v{\varphi} &\stackrel{iid}{\sim} \textrm{N}(\cdot \mid \nu, \varsigma^2),\\
    \v{\rho} &= \exp{\v{\varphi}},\\
    y_\ell \mid m_\ell, \rho_\ell &\sim \vM{\cdot}{m_\ell}{\rho_\ell}, & \ell = 1, 2, \ldots, N.
\end{align*}

In higher dimensions, we work with points on a sphere. This means that while we again transform two draws from the Gaussian process, $\v{z_1}, \v{z_2} \in \mathbbm{R}^N$, element-wise into the mean parameters, we use the $L_2$ norm to do so. Set $m_{\ell, d} = \frac{z_{d, \ell}}{\sqrt{z^2_{1, \ell} + z^2_{2, \ell}}}$ for $d = 1, 2$ and $\ell = 1, 2, \ldots, N$. Otherwise, the setup for the concentration parameter is the same as before. In summary, \textit{SvM} is the following in higher dimensions.
\begin{align*}
    \v{z_1} &\sim \textrm{GP}(\cdot \mid \mu_1, \v{\Sigma}),\\
    \v{z_2} &\sim \textrm{GP}(\cdot \mid \mu_2, \v{\Sigma}),\\
    m_{\ell, d} &= \frac{z_{d, \ell}}{\sqrt{\sum_{d' = 1}^2 z^2_{d', \ell}}} & \ell = 1, 2, \ldots, N,
    & & \d = 1, 2,\stepcounter{equation}\tag{\theequation}\label{model:SvM-higher}\\
    \v{\varphi} &\stackrel{iid}{\sim} \textrm{N}(\cdot \mid \nu, \varsigma^2),\\
    \v{\rho} &= \exp{\v{\varphi}},\\
    y_\ell \mid m_\ell, \rho_\ell &\sim \vMF{\cdot}{\v{m_\ell}}{\rho_\ell}, & \ell = 1, 2, \ldots, N.
\end{align*}

Note that in the model description, we drop $\v{\zeta}$, $\v{\lambda}$, and $k$ from the indices. This is done because there is only one von Mises distribution and set of Gaussian processes.

\section{MCMC Sampling}
\subsection{SvM and SvM-c Models (Two dimensions)}
\label{ssection:model_fit_svm_svm_c}
We use the following Gibbs approach: 


\begin{itemize}
    \item For $\ell = 1, 2, \ldots, N$, sample $\zeta_\ell$ given $y_\ell$; $z_{1, 1, \ell}, z_{1, 2, \ell}, z_{2, 1, \ell}, z_{2, 2, \ell} \ldots, z_{K, 1, \ell}, z_{K, 2, \ell}$; $\v{\varphi}_{1, \ell}, \v{\varphi}_{2, \ell}, \ldots, \v{\varphi}_{K, \ell}$; and $\nu_1, \nu_2, \ldots, \nu_K$ according to the following probability:
    \[
    p(\zeta_\ell = k \mid y_\ell; z_{k, 1, \ell}, z_{k, 2, \ell}; \v{\varphi}_\ell; \nu_\ell) \propto \lambda_k\vM{y_\ell}{m_{k, \ell}}{\rho_{k, \ell}}.
    \]
    Note that this step is skipped for \textit{SvM}.
    \item For $k = 1, 2, \ldots K$, sample $\v{z_{k, 1}}, \v{z_{k, 2}}$ given $\v{y}$, $\zeta_1, \zeta_2, \ldots, \zeta_N$; $\v{\varphi}_1, \v{\varphi}_2,$ $\ldots \v{\varphi}_K$; and $\nu_1, \nu_2, \ldots, \nu_K$ using the elliptical slice sampling outlined in Algorithm 2 \citep{MurrayEtAlEllipticalSliceSampling2010}. We restate how the algorithm is used to sample $\v{z_{k, 1}} - \v{\mu_{k, 1}}, \v{z_{k, 2}} - \v{\mu_{k, 2}}$:
    \begin{itemize}
        \item Draw
        \begin{align*}
            \v{u} &\sim \textrm{N}(\cdot \mid \v{0}, \v{\mathbbm{I}_{2 \times 2}} \otimes \v{\Sigma_k}),\\ 
            c &\sim \textrm{Unif}(0, 1),\\
            a &\sim \textrm{Unif}(0, 2\pi).
        \end{align*}
        Set $a_{\min} = a - 2\pi$ and $a_{\max} = a$.
        \item Propose a new $\v{z'_{k, 1}}$ and $\v{z'_{k, 2}}$ such that
        \[
        \begin{pmatrix} 
            \v{z'_{k, 1}} - \v{\mu_{k,1}}\\ 
            \v{z'_{k, 2}} - \v{\mu_{k,2}}\\ 
        \end{pmatrix} =
        \begin{pmatrix} 
            \v{z_{k, 1}} - \v{\mu_{k,1}}\\ 
            \v{z_{k, 2}} - \v{\mu_{k,2}}\\ 
        \end{pmatrix}\cos(a) + \v{u}\sin(a).
        \]
        The notation is the same as that in Section 5.
        \item Accept this proposal if $\frac{\prod_{\ell = 1}^N L(y_\ell \mid \v{z'_{k, 1}}, \v{z'_{k, 2}}, \rho_\ell, \zeta_\ell)}{\prod_{\ell = 1}^N L(y_\ell \mid \v{z_{k, 1}}, \v{z_{k, 2}}, \rho_\ell, \zeta_\ell)} \geq c$. Here,
        \[
        L(y_\ell \mid \v{z_{k, 1}}, \v{z_{k, 2}}, \v{\rho}, \v{\zeta}) = \prod_\ell (\vM{y_\ell}{\arctan^*(z_{k, 1, \ell}, z_{k, 2, \ell})}{\rho_{k, \ell}})^{\indFct{\zeta_\ell = k}}.
        \]
        \item If the proposal is rejected, set $a_{\min} = a$ if $a < 0$ and $a_{\max} = a$ otherwise. Draw $a$ from $\textrm{Unif}(a_{\min}, a_{\max})$ and repeat the previous two steps until the proposal is accepted.
    \end{itemize}
    Note that for \textit{SvM}, $k = 1$ and all observations are included in the likelihood.
    
    \item Sample $\v{\varphi}_1, \v{\varphi}_2, \ldots, \v{\varphi}_K$ and $\nu_1, \nu_2, \ldots, \nu_K$ given $\v{y}$; $\zeta_1, \zeta_2, \ldots, \zeta_N$; and $\v{z_{1, 1}}, \v{z_{1, 2}}, \ldots,$ $\v{z_{K, 1}}, \v{z_{K, 2}}$ using HMC and the following probability:
    \begin{align*}
        p(\v{\varphi}_1, & \v{\varphi}_2, \ldots, \v{\varphi}_K, \nu_1, \nu_2, \ldots, \nu_K \mid \v{y}; \zeta_1, \zeta_2, \ldots, \zeta_N; \v{z_{1, 1}}, \v{z_{1, 2}}, \ldots, \v{z_{K, 1}}, \v{z_{K, 2}}) \propto\\
        & \prod_{\ell} (\vM{y_\ell}{m_{k, \ell}}{\rho_{k, \ell}})^{\indFct{\zeta_\ell = k}} \prod_k \left(\prod_\ell \textrm{N}(\varphi_{k, \ell} \mid \nu_k, \varsigma^2)\right) \textrm{N}(\nu_k \mid 0, \tau^2).
    \end{align*}
    Here, we let 
    \[
    U(\cdot \mid \cdot) =  -\log(p(\v{\varphi}_1, \v{\varphi}_2, \ldots, \v{\varphi}_K, \nu_1, \nu_2, \ldots, \nu_K \mid \v{y}; \zeta_1, \zeta_2, \ldots, \zeta_N; \v{z_{1, 1}}, \v{z_{1, 2}}, \ldots, \v{z_{K, 1}}, \v{z_{K, 2}})).
    \]  
    We also sample $\v{\varphi_k}$ and $\nu_k$ using the following parameterization for $k = 1, 2, \ldots, K$: 
    \begin{align*}
        \nu_k &\sim \textrm{N}(\cdot \mid 0, \tau^2),\\
        \widetilde{\varphi}_{k, \ell} &\sim \textrm{N}(\cdot \mid 0, 1) & \ell = 1, 2, \ldots, N,\\
        \varphi_{k, \ell} &= \varsigma \widetilde{\varphi}_{k, \ell} + \nu_k & \ell = 1, 2, \ldots, N.\\
    \end{align*}
    Then, the gradients are the following for $\nu_k$ and $\v{\widetilde{\rho}_{k}}$:
    \begin{itemize}
       \item For $k = 1, 2, \dots, K$ and $\ell = 1, 2, \dots N$, the derivative of $-U(\cdot \mid \cdot)$ with respect to $\varphi_{k, \ell}$ is:
        \begin{equation*}
            \begin{aligned}
                -\frac{\partial U(\cdot \mid \cdot)}{\partial \varphi_{k, \ell}} &= \indFct{\zeta_\ell = k}\rho_{k, \ell} \left(\cos(y_\ell - m_{k, \ell}) - \frac{I_{-1}(\rho_{k, \ell})}{I_{0}(\rho_{k, \ell})}\right)\varsigma - \widetilde{\varphi}_{k, \ell}.
            \end{aligned}
        \end{equation*}
        \item For $k = 1, 2, \dots K$, the derivative of $-U(\cdot \mid \cdot)$ with respect to $\nu_k$ is:
        \begin{align*}
            -\frac{\partial U(\cdot \mid \cdot)}{\partial \nu_k} &= \sum_{\ell =1}^N \indFct{\zeta_\ell = k}\rho_{k, \ell} \left(\cos(y_\ell - m_{k, \ell}) - \frac{I_{-1}(\rho_{k, \ell})}{I_{0}(\rho_{k, \ell})}\right) - \frac{\nu_k}{\tau^2}.
        \end{align*}
    \end{itemize}
\end{itemize}

\subsection{SvM and SvM-c Models (Higher dimensions)}
\label{ssection:model_fit_svm_svm_c_higher_dim}
We use the following Gibbs approach: 


\begin{itemize}
    \item For $\ell = 1, 2, \ldots, N$, sample $\zeta_\ell$ given $y_\ell$; $z_{1, 1, \ell}, z_{1, 2, \ell}, z_{2, 1, \ell}, z_{2, 2, \ell} \ldots, z_{K, 1, \ell}, z_{K, 2, \ell}$; $\v{\varphi}_{1, \ell}, \v{\varphi}_{2, \ell}, \ldots, \v{\varphi}_{K, \ell}$; and $\nu_1, \nu_2, \ldots, \nu_K$ according to the following probability:
    \[
    p(\zeta_\ell = k \mid y_\ell; z_{k, 1, \ell}, z_{k, 2, \ell}; \v{\varphi}_\ell; \nu_\ell) \propto \lambda_k\vMF{y_\ell}{\widetilde{m_{k, \ell}}}{\rho_{k, \ell}}.
    \]
    Note that this step is skipped for \textit{SvM}.
    \item For $k = 1, 2, \ldots K$ and $d = 1, 2, \ldots, D$, sample $\v{z_{k, d}}$ given $\v{y}$, $\zeta_1, \zeta_2, \ldots, \zeta_N$; $\v{z_{k, 1}}, \v{z_{k, 2}}, \ldots, \v{z_{k, d - 1}}, \v{z_{k, d + 1}}, \ldots, \v{z_{k, D}}$; $\v{\varphi}_1, \v{\varphi}_2,$ $\ldots \v{\varphi}_K$; and $\nu_1, \nu_2, \ldots, \nu_K$ using the elliptical slice sampling outlined in Algorithm 2 \citep{MurrayEtAlEllipticalSliceSampling2010}. We restate how the algorithm is used to sample $\v{z_{k, d}} - \v{\mu_{k, d}}$:
    \begin{itemize}
        \item Draw
        \begin{align*}
            \v{u} &\sim \textrm{N}(\cdot \mid \v{0}, \v{\Sigma_k}),\\ 
            c &\sim \textrm{Unif}(0, 1),\\
            a &\sim \textrm{Unif}(0, 2\pi).
        \end{align*}
        Set $a_{\min} = a - 2\pi$ and $a_{\max} = a$.
        \item Propose a new $\v{z'_{k, d}}$ such that $\v{z'_{k, d}} = (\v{z_{k, d}} - \v{\mu_{k, d}})\cos(a) + \v{u}\sin(a)$.
        \item Accept this proposal if $\frac{L(y_\ell \mid \v{z_{k, 1}}, \v{z_{k, 2}}, \ldots, \v{z'_{k, d}}, \ldots, \v{z_{k, D}}, \v{\rho}, \v{\zeta})}{L(y_\ell \mid \v{z_{k, 1}}, \v{z_{k, 2}}, \ldots, \v{z_{k, D}}, \v{\rho}, \v{\zeta})} \geq c$. Here,
        \[
        L(y_\ell \mid \v{z_{k, 1}}, \v{z_{k, 2}}, \ldots, \v{z_{k, D}}, \v{\rho}, \v{\zeta}) = \prod_\ell (\vMF{y_\ell}{\v{\widetilde{m}_{k, \ell}}}{\rho_{k, \ell}})^{\indFct{\zeta_\ell = k}}.
        \]
        \item If the proposal is rejected, set $a_{\min} = a$ if $a < 0$ and $a_{\max} = a$ otherwise. Draw $a$ from $\textrm{Unif}(a_{\min}, a_{\max})$ and repeat the previous two steps until the proposal is accepted.
    \end{itemize}
    Note that for \textit{SvM}, $k = 1$ and all observations are included in the likelihood.
    
    \item Sample $\v{\varphi}_1, \v{\varphi}_2, \ldots, \v{\varphi}_K$ and $\nu_1, \nu_2, \ldots, \nu_K$ given $\v{y}$; $\zeta_1, \zeta_2, \ldots, \zeta_N$; and $\v{z_{1, 1}}, \v{z_{1, 2}}, \ldots, \v{z_{1, D}}, \v{z_{2, 1}}, \ldots, \v{z_{K, D}}$ using HMC and the following probability:
    \begin{align*}
        p(\v{\varphi}_1, & \v{\varphi}_2, \ldots, \v{\varphi}_K, \nu_1, \nu_2, \ldots, \nu_K \mid \v{y}; \zeta_1, \zeta_2, \ldots, \zeta_N; \v{z_{1, 1}}, \v{z_{1, 2}}, \ldots, \v{z_{K, 1}}, \v{z_{K, 2}}) \propto\\
        & \prod_{\ell} (\vMF{y_\ell}{\widetilde{m}_{k, \ell}}{\rho_{k, \ell}})^{\indFct{\zeta_\ell = k}} \prod_k \left(\prod_\ell \textrm{N}(\varphi_{k, \ell} \mid \nu_k, \varsigma^2)\right) \textrm{N}(\nu_k \mid 0, \tau^2).
    \end{align*}
    Here, we let 
    \[
    U(\cdot \mid \cdot) =  -\log(p(\v{\varphi}_1, \v{\varphi}_2, \ldots, \v{\varphi}_K, \nu_1, \nu_2, \ldots, \nu_K \mid \v{y}; \zeta_1, \zeta_2, \ldots, \zeta_N; \v{z_{1, 1}}, \v{z_{1, 2}}, \ldots, \v{z_{K, 1}}, \v{z_{K, 2}})).
    \]  
    We also sample $\v{\varphi_k}$ and $\nu_k$ using the following parameterization for $k = 1, 2, \ldots, K$: 
    \begin{align*}
        \nu_k &\sim \textrm{N}(\cdot \mid 0, \tau^2),\\
        \widetilde{\varphi}_{k, \ell} &\sim \textrm{N}(\cdot \mid 0, 1) & \ell = 1, 2, \ldots, N,\\
        \varphi_{k, \ell} &= \varsigma \widetilde{\varphi}_{k, \ell} + \nu_k & \ell = 1, 2, \ldots, N.\\
    \end{align*}
    Then, the gradients are the following for $\nu_k$ and $\v{\widetilde{\rho}_{k}}$:
    \begin{itemize}
       \item For $k = 1, 2, \dots, K$ and $\ell = 1, 2, \dots N$, the derivative of $-U(\cdot \mid \cdot)$ with respect to $\varphi_{k, \ell}$ is:
        \begin{equation*}
            \begin{aligned}
                -\frac{\partial U(\cdot \mid \cdot)}{\partial \varphi_{k, \ell}} &= \indFct{\zeta_\ell = k}\rho_{k, \ell} \left(\v{\widetilde{m}_{k, \ell}}^T \v{y}_\ell - \frac{I_{D / 2}(\rho_{k, \ell})}{I_{D / 2 - 1}(\rho_{k, \ell})}\right)\varsigma - \widetilde{\varphi}_{k, \ell}.
            \end{aligned}
        \end{equation*}
        \item For $k = 1, 2, \dots K$, the derivative of $-U(\cdot \mid \cdot)$ with respect to $\nu_k$ is:
        \begin{align*}
            -\frac{\partial U(\cdot \mid \cdot)}{\partial \nu_k} &= \sum_{\ell =1}^N \indFct{\zeta_\ell = k}\rho_{k, \ell} \left(\v{\widetilde{m}_{k, \ell}}^T \v{y}_\ell - \frac{I_{D / 2}(\rho_{k, \ell})}{I_{D / 2 - 1}(\rho_{k, \ell})}\right) - \frac{\nu_k}{\tau^2}.
        \end{align*}
    \end{itemize}
\end{itemize}

\subsection{Regularized Expectation Maximization (Two dimension)}
To help with the sampling, we initialized it with results from applying Expectation Maximization (EM) to the unnormalized posterior or a regularized EM. Because $\v{\Sigma}$ might be close to singular, we ran regularized EM on the non-centered parametrization for our updates.

Within the regularized EM framework, we let $\zeta_\ell$ for $\ell = 1, 2, \ldots, N$ be the latent variable and $\lambda_k$, $\v{\varphi_k}$, $\nu_k$, and $\v{\widetilde{z}_k}$ for $k = 1, 2, \ldots, K$ be the parameters. If we summarize the parameters as $\Theta$, the posterior can be written as following.
\begin{align*}
    \widetilde{p}(\v{\zeta}, \Theta \mid \v{y}, \v{x}) &:= \prod_{\ell = 1}^N \prod_{k = 1}^K \left(\lambda_k \vM{y_\ell}{m_{k, \ell}}{\rho_{k, \ell}}\right)^{\indFct{\zeta_\ell = k}} \prod_{k = 1}^K \textrm{N}(\widetilde{z} \mid 0, \mathbbm{I}_n)\\ &\qquad\prod_{\ell = 1}^N \prod_{k = 1}^K \textrm{N}(\varphi_{k, \ell} \mid \nu_k, \varsigma^2) \prod_{k = 1}^K \textrm{N}(\nu_k \mid 0, \tau^2).
\end{align*}

For the expected conditional log unnormalized posterior, we need a term to represent $P(\zeta_{\ell} = k \mid y_{\ell}, \lambda_k, \v{\widetilde{z}_k}, \v{\varphi_{., \ell}})$. We let $r_{k, \ell}$ be this term. To be explicit, we are attaching the draws from the Gaussian Process, $\v{\widetilde{z}_k}$, to $\v{m_{., \ell}}$. Then, it is defined in the following manner.
\begin{align}
    r_{k, \ell} &:= \frac{\lambda_k \vM{y_\ell}{m_{k, \ell}}{\rho_{k, \ell}}}{\sum_{k'} \lambda_{k'} \vM{y_\ell}{m_{k', \ell}}{\rho_{k', \ell}}}.
\label{eq:SvM-c_r_k_m}
\end{align}
With this term, the expected conditional log unnormalized posterior is the following.
\begin{equation}
    \begin{aligned}
    \E{\log(\widetilde{p}(\v{\zeta}, \Theta \mid \v{y}, \v{x}))} &= \sum_{\ell = 1}^N \sum_{k = 1}^K r_{k, \ell}\left(\log(\lambda_k) + \log(\vM{y_\ell}{m_{k, \ell}}{\rho_{k, \ell}})\right) +\\
    & \qquad\sum_{\ell = 1}^N \sum_{k = 1}^K\log(\textrm{N}(\widetilde{z}_{k, \ell} \mid 0, 1)) +\\ 
    & \qquad \sum_{\ell = 1}^N\sum_{k = 1}^K \log(\textrm{N}(\varphi_{k, \ell} \mid \nu_k, \varsigma^2)) + \sum_{k = 1}^K \log(\textrm{N}(\nu_k \mid 0, \tau^2)).
\end{aligned}    
\label{eq:SvM-c_exp_cond_log_post}
\end{equation}

With this, we can now state the regularized EM algorithm.
\begin{itemize}
    \item \textbf{E step:} Calculate the expected conditional log unnormalized posterior from \eqref{eq:SvM-c_exp_cond_log_post} and $r_{k, \ell}$ from \eqref{eq:SvM-c_r_k_m}.
    \item \textbf{M step:}
    \begin{itemize}
        \item For $k = 1, 2, \dots, K$, set $\lambda_k = \frac{1}{N} \sum\limits_{\ell = 1}^N r_{k, \ell}$.
        \item For $k = 1, 2, \dots, K$, alternate updating $\v{\widetilde{z}_{k, 1, .}}$ and $\v{\widetilde{z}_{k, 2, .}}$ using coordinate gradient ascent and the following gradients.
        \begin{equation*}
            \begin{aligned}
            \frac{\partial}{\partial \widetilde{z}_{k, 1, \ell}} &  \E{\log(\widetilde{p}(\v{\zeta}, \Theta \mid \v{y}, \v{x}))} =\\
            & \quad\sum_{\ell' = 1}^N L_{\ell', \ell} r_{k, \ell'} \frac{-z_{k, 2, \ell'}}{z^2_{k, 1, \ell'} + z^2_{k, 2, \ell'}}\rho_{k, \ell'}\sin(y_{\ell'} - m_{k, \ell'}) - \widetilde{z}_{k, \ell}.\\
            \frac{\partial}{\partial \widetilde{z}_{k, 2, \ell}} & \E{\log(\widetilde{p}(\v{\zeta}, \Theta \mid \v{y}, \v{x}))} =\\
            & \quad \sum_{\ell' = 1}^N L_{\ell', \ell} r_{k, \ell'} \frac{z_{k, 1, \ell'}}{z^2_{k, 1, \ell'} + z^2_{k, 2, \ell'}}\rho_{k, \ell'}\sin(y_{\ell'} - m_{k, \ell'}) - \widetilde{z}_{k, \ell}.
            \end{aligned}
        \end{equation*}
        \item For $k = 1, 2, \dots, K$ and $\ell = 1, 2, \dots N$, update $\varphi_{k, \ell}$ using gradient ascent and the gradient below.
        \begin{equation*}
            \begin{aligned}
                \frac{\partial}{\partial \varphi_{k, \ell}} & \E{\log(\widetilde{p}(\v{\zeta}, \Theta \mid \v{y}, \v{x}))} =\\
                & r_{k, \ell}\rho_{k, \ell} \left(\cos(y_\ell - m_{k, \ell}) - \frac{I_{-1}(\rho_{k, \ell})}{I_{0}(\rho_{k, \ell})}\right) - \frac{1}{\varsigma^2}(\varphi_{k, \ell} - \nu_k).
            \end{aligned}
        \end{equation*}
        \item For $k = 1, 2, \dots K$, set $\nu_k = \frac{\sum_\ell \frac{\varphi_{k, \ell}}{\varsigma^2}}{\frac{N}{\varsigma^2} + \frac{1}{\tau^2}}$.
    \end{itemize}
\end{itemize}

\subsection{Regularized Expectation Maximization (Higher dimensions)}

We now discuss the multidimensional setting. Within the regularized EM framework, we let $\zeta_\ell$ for $\ell = 1, 2, \ldots, N$ be the latent variable and $\lambda_k$, $\v{\varphi_k}$, $\nu_{k}$, and $\v{\widetilde{z}_{k, 1}}, \v{\widetilde{z}_{k, 2}}, \ldots, \v{\widetilde{z}_{k, D}}$ for $k = 1, 2, \ldots, K$ be the parameters. If we summarize the parameters as $\Theta$, the posterior can be written as following.
\begin{align*}
    \widetilde{p}(\v{\zeta}, \Theta \mid \v{y}, \v{x}) &:= \prod_{\ell = 1}^N \prod_{k = 1}^K \left(\lambda_k \vMF{y_\ell}{\widetilde{m}_{k, \ell}}{\rho_{k, \ell}}\right)^{\indFct{\zeta_\ell = k}} \prod_{k = 1}^K  \prod_{d = 1}^D \textrm{N}(\widetilde{z}_{k, d} \mid 0, \mathbbm{I}_n)\\ &\qquad\prod_{\ell = 1}^N \prod_{k = 1}^K  \textrm{N}(\varphi_{k, \ell} \mid \nu_k, \varsigma^2) \prod_{k = 1}^K  \textrm{N}(\nu_k \mid 0, \tau^2).
\end{align*}

For the expected conditional log unnormalized posterior, we need a term to represent $P(\zeta_{\ell} = k \mid y_{\ell}, \lambda_k, \v{\widetilde{z}_k}, \v{\varphi_{., \ell}})$. We let $r_{k, \ell}$ be this term. To be explicit, we are attaching the draws from the Gaussian Process, $\v{\widetilde{z}_k}$, to $\v{m_{., \ell}}$. Then, it is defined in the following manner.
\begin{align}
    r_{k, \ell} &:= \frac{\lambda_k \vMF{y_\ell}{\widetilde{m}_{k, \ell}}{\rho_{k, \ell}}}{\sum_{k'} \lambda_{k'} \vMF{y_\ell}{\widetilde{m}_{k', \ell}}{\rho_{k', \ell}}}.
\label{eq:SvM-c_r_k_m_high_dim}
\end{align}
With this term, the expected conditional log unnormalized posterior is the following.
\begin{equation}
    \begin{aligned}
    \E{\log(\widetilde{p}(\v{\zeta}, \Theta \mid \v{y}, \v{x}))} &= \sum_{\ell = 1}^N \sum_{k = 1}^K r_{k, \ell}\left(\log(\lambda_k) + \log(\vMF{y_\ell}{\widetilde{m}_{k, \ell}}{\rho_{k, \ell}})\right) +\\
    & \qquad\sum_{\ell = 1}^N \sum_{k = 1}^K \sum_{d = 1}^D \log(\textrm{N}(\widetilde{z}_{k, d, \ell} \mid 0, 1)) +\\ 
    & \qquad \sum_{\ell = 1}^N\sum_{k = 1}^K \log(\textrm{N}(\varphi_{k, \ell} \mid \nu_k, \varsigma^2)) + \sum_{k = 1}^K \log(\textrm{N}(\nu_k \mid 0, \tau^2)).
\end{aligned}    
\label{eq:SvM-c_exp_cond_log_post_high_dim}
\end{equation}

With this, we can now state the regularized EM algorithm.
\begin{itemize}
    \item \textbf{E step:} Calculate the expected conditional log unnormalized posterior from \eqref{eq:SvM-c_exp_cond_log_post_high_dim} and $r_{k, \ell}$ from \eqref{eq:SvM-c_r_k_m_high_dim}.
    \item \textbf{M step:}
    \begin{itemize}
        \item For $k = 1, 2, \dots, K$, set $\lambda_k = \frac{1}{N} \sum\limits_{\ell = 1}^N r_{k, \ell}$.
        \item For $k = 1, 2, \dots, K$ and $d = 1, 2, \ldots, D$, alternate updating $\v{\widetilde{z}_{k, d, \cdot}}$ using coordinate gradient ascent. Let $r \in \mathbbm{R}^+$ be the value such that $r\widetilde{m}_{\ell, d} = z_{k, d, \ell}$. The gradient is the following:
        \begin{equation*}
            \begin{aligned}
            \frac{\partial}{\partial \widetilde{z}_{k, d, \ell}} &  \E{\log(\widetilde{p}(\v{\zeta}, \Theta \mid \v{y}, \v{x}))} =\\
            & \quad\sum_{\ell'} L_{\ell', \ell} r_{k, \ell'} \rho_{k, \ell'}\v{y_{\ell'}} \left(\frac{1}{r}\v{\mathbbm{I}_{D \times D}} - (\{z_{k, d, \ell'}\}_{d = 1}^D)^T (\{z_{k, d, \ell'}\}_{d = 1}^D)_{\cdot, d}\right)  - \widetilde{z}_{k, \ell}.\\
            \end{aligned}
        \end{equation*}
        \item For $k = 1, 2, \dots, K$ and $\ell = 1, 2, \dots N$, update $\varphi_{k, \ell}$ using gradient ascent and the gradient below.
        \begin{equation*}
            \begin{aligned}
                \frac{\partial}{\partial \varphi_{k, \ell}} & \E{\log(\widetilde{p}(\v{\zeta}, \Theta \mid \v{y}, \v{x}))} =\\
                & r_{k, \ell}\rho_{k, \ell} \left(\cos(\v{m_{k, \ell}}^T \v{y}_\ell) - \frac{I_{D / 2}(\rho_{k, \ell})}{I_{D / 2 - 1}(\rho_{k, \ell})}\right) - \frac{1}{\varsigma^2}(\varphi_{k, \ell} - \nu_k).
            \end{aligned}
        \end{equation*}
        \item For $k = 1, 2, \dots K$, set $\nu_k = \frac{\sum_\ell \frac{\varphi_{k, \ell}}{\varsigma^2}}{\frac{N}{\varsigma^2} + \frac{1}{\tau^2}}$.
    \end{itemize}
\end{itemize}

\subsection{Calculating posterior predictive probability}
Let $x^*$ represent the withheld locations, $y^*$ the withheld data, $\vartheta$ a posterior draw for the parameters based on $x$ and $y$, and $\vartheta^*$ a draw for the parameters for $x^*$ and $y^*$. Set $N$ to be the number of locations and $N^*$ to be the number of withheld locations. The posterior predictive probability is given as follows.
\begin{align}
p(\v{y^*} \mid x, x^*, \v{y}) := \int \int \textrm{p}(y^* \mid \vartheta^*) \textrm{p}(\vartheta^* \mid \vartheta, x, x^*)  \textrm{p}(\vartheta \mid x, y) d\vartheta^* d\vartheta.
\label{eq:post_pred_prob_a}    
\end{align}
We now discuss how to calculate it for our models.

\noindent\textbf{SvM-c:} Here, $\vartheta = \{\{\v{z}_{k, 1}, \v{z}_{k, 2}\}_{k = 1}^K, \{\v{\varphi_{k}}\}_{k = 1}^K, \{\nu_k\}_{k = 1}^K, {\lambda}_{k = 1}^K\}$. Further, for the Gaussian processes of component k, let $\v{\Sigma^*_{k, 1}}$ and $\v{\Sigma^*_{k, 2}}$ be the covariance matrices based on $x^*$ and $\v{\widetilde{\Sigma}_{k, 1}}$ and $\v{\widetilde{\Sigma}_{k, 2}}$ be the covariance matrices based on $x$ and $x^*$. In other words, $\widetilde{\Sigma}_{k, 1, i, j} = K(x_i, x^*_j)$ for some kernel function $K(\cdot, \cdot)$. Then, the posterior predictive probability for posterior draw $i$ can be found in the following procedure:
\begin{itemize}
    \item Draw $\v{z^*_{k, 1}} \sim \textrm{N}(\v{\mu_{k, 1}} + \v{\widetilde{\Sigma}_{k, 1}}^T \v{\Sigma_{k, 1}}^{-1} (\v{z_{k, 1}} - \v{\mu_{k, 1}}), \v{\v{\Sigma}^*_{k, 1}} - \v{\widetilde{\Sigma}}_{k, 1}^T \v{\Sigma_{k, 1}}^{-1} \v{\widetilde{\Sigma}}_{k, 1})$ and \\ $\v{z^*_{k, 2}} \sim \textrm{N}(\v{\mu_{k, 2}} + \v{\widetilde{\Sigma}}_{k, 2}^T \v{\Sigma_{k, 2}}^{-1} (\v{z_{k, 2}} - \v{\mu_{k, 2}}), \v{\Sigma^*_{k, 2}} - \v{\widetilde{\Sigma}_{k, 2}}^T \v{\Sigma_{k, 2}}^{-1} \v{\widetilde{\Sigma}_{k, 2}})$ for $k = 1, 2, \ldots, K$. 
    \item Set $\v{m^*_k} = \arctan^*(\v{z^*_{k, 1}}, \v{z^*_{k, 2}})$ if $D = 2$. Otherwise, set $\v{m^*_k} = \{\frac{z^*_{k, 2,d}}{\norm{\v{z^*_{k, 2}}}}_2\}_{d = 1}^D$.
    \item Draw $\v{\varphi^*_k} \sim \textrm{N}(\nu_k, \varsigma^2)$. Set $\v{\rho^*} = \exp(\v{\varphi^*_k})$.
    \item Let $p^*_{i, 1} = \prod_{n^* = 1}^{N^*} \left(\sum_{k = 1}^K \lambda_k\vM{y^*_{n^*}}{m^*_{k, n^*}}{\rho^*_{k, n^*}}\right)$ if $D = 2$. Otherwise, let $p^*_{i, 1} = \prod_{n^* = 1}^{N^*} \left(\sum_{k = 1}^K \lambda_k\vMF{y^*_{n^*}}{m^*_{k, n^*}}{\rho^*_{k, n^*}}\right)$
    \item Repeat the previous steps $M - 1$ more times for some $M \in \mathbbm{N}$ and denote the $m^{th}$ result as $p^*_{i, m}$.
    \item If there are $I$ iterations, return $\frac{1}{M} \sum_m \frac{1}{I} \sum_i p^*_{i, m}$ for the posterior predictive probability.
\end{itemize}

.

\section{Simulations}
\label{section:sim}
In this section we shall present a simulation study for the introduced models. The simulation study is conducted with the following goals in mind. First, we wanted evidence that when used for inference, our models could correctly recover the model parameters for data generated according to their respective models. Next, with this menagerie of simulated data, we could examine how these models behave in mis-specified settings and how to address model selection. Then, while fitting these models, we might develop better approaches to address inefficiencies in sampling. Finally, inspecting the results might yield additional insights about these models' behaviors.

\subsection{One dimension simulation study}
\label{ssection:sim_results}

We first used a uniform distribution to generate 500 points residing in the simplex $\Delta^2$. We created observations in five different ways. First, we generated observations using Von Mises($\pi$, 5). Next, we created them using a mixture of Von Mises($\frac{\pi}{2}$, 5) and Von Mises($\frac{3\pi}{2}$, 10) with mixing probability of 0.3 and 0.7 respectively. We denote these models as \textit{vM} and \textit{iVM}. While these two methods did not use the spatial information on the simplex, the remaining methods do. Third, we simulated observations according to \textit{SvM} specified in the supplementary material with a prior mean of $\pi$ for all locations. Fourth, we created them according to \textit{SvM-c}. One component's prior mean was set at $\frac{\pi}{2}$ and the other component's mean at $\frac{3\pi}{2}$. There was equal probability of using either component. Finally, we simulated according to \textit{SvM} specified in the supplementary material with a prior mean of $0$ for all locations. When the random directions are plotted, it will appear as if there are two mean surfaces. However, if we add $2\pi$ to all directions less than $\pi$ or subtract $2\pi$ to all directions greater than $\pi$, there will be one contiguous mean surface. This will test our model's ability to handle data that are near the end points, but are actually close together due to the geometry of the circle.


\begin{figure}[!tb]
\captionsetup[subfigure]{justification=centering}
\centering
\begin{subfigure}[t]{.3\textwidth}
\centering
\includegraphics[width = .8\textwidth]{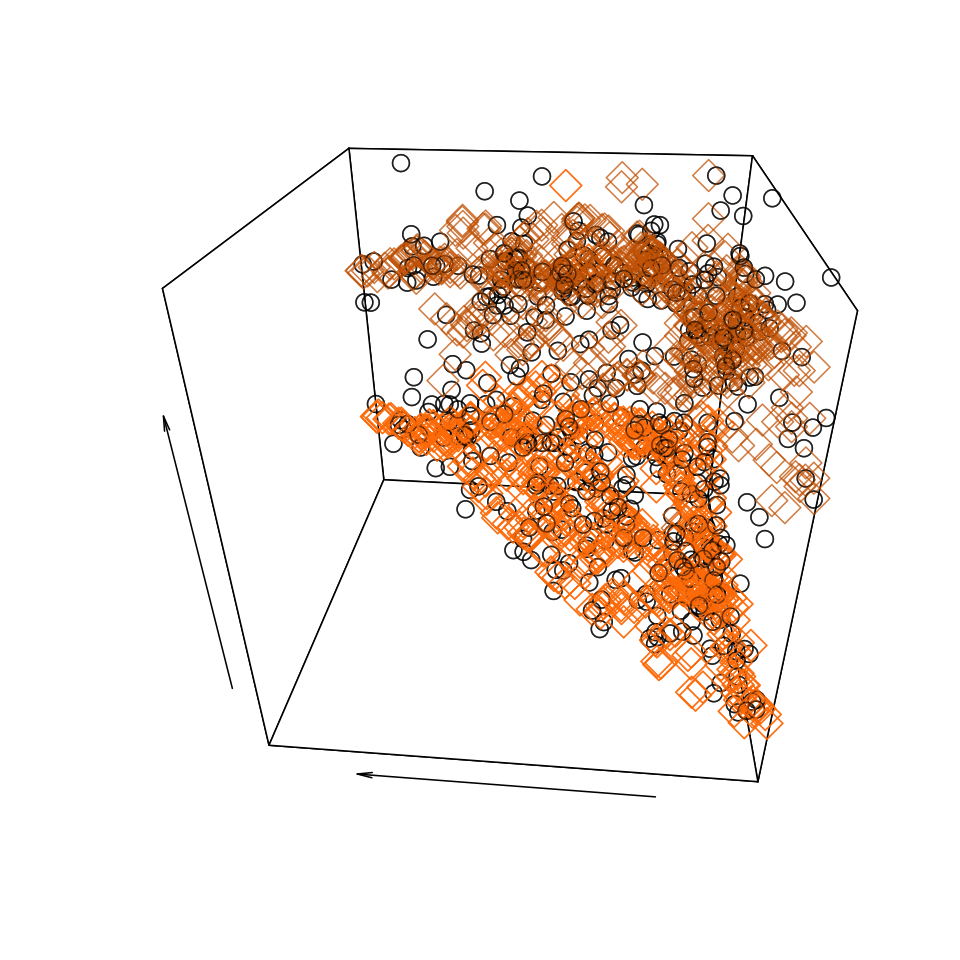}
\caption{Simulated data with\\means in orange}
\end{subfigure}
\begin{subfigure}[t]{.3\textwidth}
\centering
\includegraphics[width = .8\textwidth]{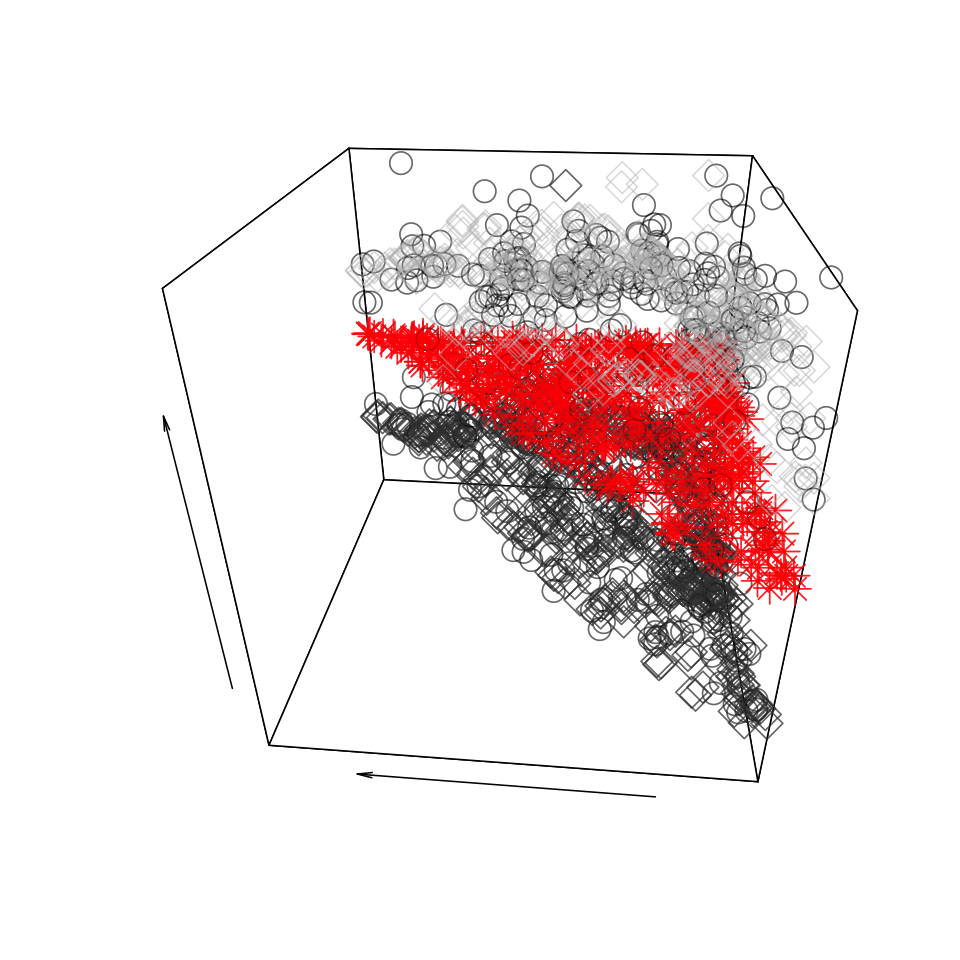}
\caption{SvM model \\fitted means in red}
\end{subfigure}
\begin{subfigure}[t]{.3\textwidth}
\centering
\includegraphics[width = .8\textwidth]{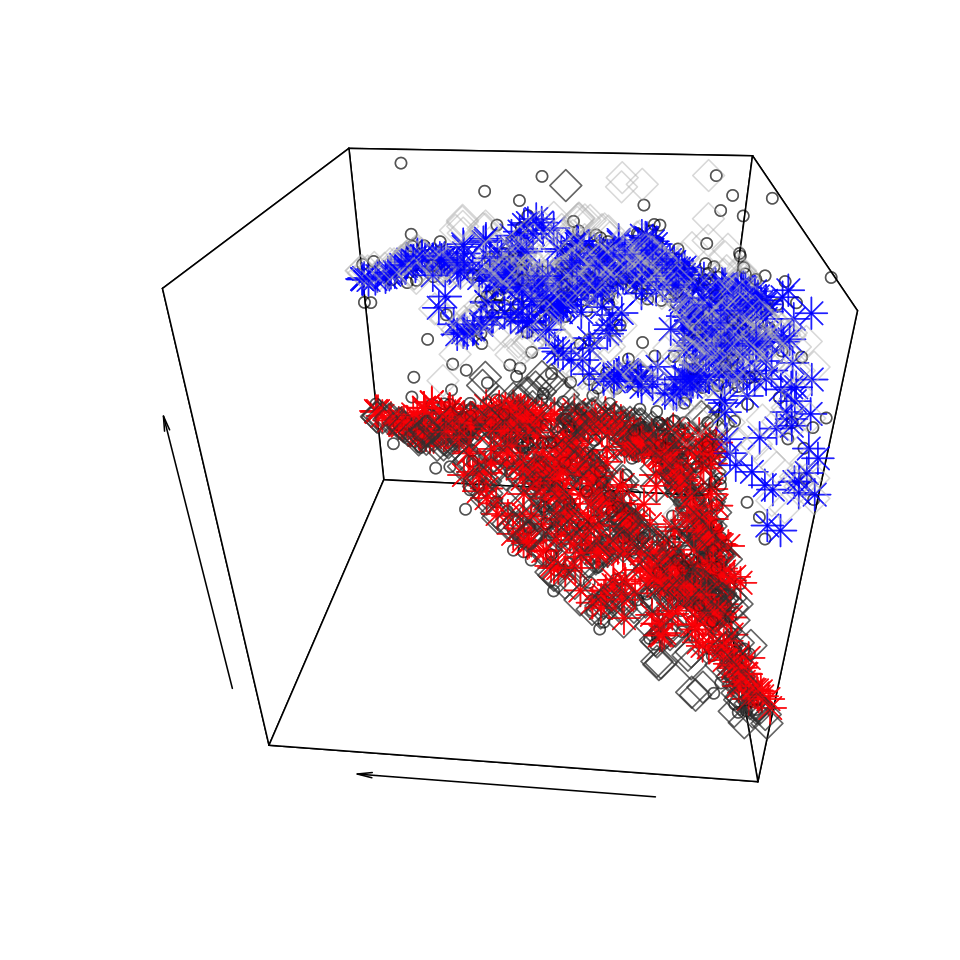}
\caption{SvM-c model fitted\\ means in red and blue}
\end{subfigure}
\caption{\small The right two plots show the mean component fitted by different models for random directions simulated according to \textit{SvM-c}. The fitted mean components for the different models are shown in colored asterisks with the simulated random directions as circles and the simulated means as grey rhombuses. The leftmost plot is just the simulated data and the mean surface used to generate that data. The x-y axis show the locations in the first two dimensions of a three dimensional simplex; the z-axis represent the direction.}
\label{fig:model_fitted_SvM-c_plots}

\resizebox{0.8\textwidth}{!}{
\centering
\begin{tabular}{@{\extracolsep{5pt}} cccc} 
\\[-1.8ex]\hline 
\hline \\[-1.8ex] 
 & $\bm{m}$ & $\bm{\rho}$ & $\bm{\lambda}$ \\ 
\hline \\[-1.8ex] 
\multirow{2}{*}{\textbf{Simulation}} & $\overline{m}_1$ = 1.36 (0.41, 2.16) & $\overline{\rho}_1 = 3.01 (2.56, 3.50)$ & $\lambda_1 = 0.50$\\
& $\overline{m}_2$ = 4.85 (3.53, 5.64) & $\overline{\rho}_2 = 8.00 (6.98, 9.00)$ & $\lambda_2 = 0.50$\\[1ex]
\textit{SvM}  & $\overline{m}$ = 3.15 (1.93, 4.35) & $\overline{\rho}$ = 0.01 (0.00, 0.04) & ---\\[1ex] 
\multirow{2}{*}{\shortstack[c]{\textit{SvM-c}}} & $\overline{m}_1$ = 1.42 (0.90, 1.97) & $\overline{\rho}_1$ = 2.79 (2.17, 3.52) & $\lambda_1$ =  0.54 (0.49, 0.58)\\
& $\overline{m}_2$ = 4.79 (4.38, 5.20) & $\overline{\rho}_2$ = 8.01 (6.11, 10.52) & $\lambda_2$ = 0.46 (0.42, 0.51)\\
& & &\\
\hline \\[-1.8ex] 
\end{tabular} 
}
\captionof{table}{\small Circular mean posterior values and 95\% credible intervals for the parameters in von Mises distribution and mixing probability are shown for the simulation data in Figure \ref{fig:model_fitted_SvM-c_plots}. Parameters with bars over them were averaged across all locations.}
\end{figure}

\noindent\textbf{Simulation model fitting} We fit the \textit{SvM} and \textit{SvM-c} models to these simulated observations using the approaches outlined in the previous section. Because there was at most two components, we only fit the two component versions of \textit{SvM-c} to our simulated observations. Then, for the Gaussian processes, we used the squared exponential kernel, i.e.
\begin{align}
k(\v{x}_{\ell}, \v{x_{\ell'}}) &= \sigma^2 \exp{-\frac{(\v{x}_{\ell} - \v{x_{\ell'}})^2}{2\omega^2}}.
\label{eq:sq_exp_kernel}
\end{align}
for some $\omega, \sigma^2 \in \mathbbm{R}^+$. For \textit{SvM} and \textit{SvM-c} model, we set $\sigma = 0.5$. We set $\mu = (-1, 0)$ for \textit{SvM} and $\mu_1 = (0, 1)$ and $\mu_2 = (0, -1)$ for \textit{SvM-c}. These values correspond to $\pi$, $\frac{\pi}{2}$, and $\frac{3\pi}{2}$ respectively. We set $\varsigma = 0.05$ and $\tau = 5$ for \textit{SvM} and \textit{SvM-c} concentration parameter's hierarchical prior. We wanted these components to be more distinct. 

For posterior inference, any step involving HMC was implemented in Stan \citep{StanDevelopmentTeam.StanCoreLibrary2018}. Otherwise, we implemented our approach in R. Using this code that can be found on a Github repo: \url{https://github.com/rayleigh/housing\_price}, we ran four chains and 10000 iterations, using every fifth iteration after the first 5000 iterations. On a cluster using 2x 3.0 GHz Intel Xeon Gold 6154 as its processor and with 192 GB as its RAM, \textit{SvM-c} took between four and a half to eight and a half hours. We checked for convergence by primarily looking at trace plots of parameters and examining Rhat values for non-circular values as computed by Stan \citep{GelmanRubinInferenceIterativeSimulation1992}.

\noindent\textbf{Simulation model findings} These are a number of interesting results to report from these experiments. First, as exemplified by Figure \ref{fig:model_fitted_SvM-c_plots} and other such figures in the Appendix, the models appeared to correctly capture the mean surfaces for data simulated from their respective models. By "mean surface", we mean the average direction for each location within the simplex. For example, the fitted mean surface for \textit{SvM} matches the simulated mean surface, even when the simulated surface has a mean of $0$. While the numerical summaries of the random variables were slightly off, the posterior credible intervals are being averaged across all locations, possibly affecting the summaries. Next, homogeneous models have difficulty capturing heterogeneous patterns. For instance, the fitted mean surface from \textit{SvM} essentially gives up and becomes a uniform distribution in the case of \textit{SvM-c}. It does not recognize the simulated mean pattern. Interestingly enough, in the \textit{SvM} case, \textit{SvM-c} matches one fitted mean surface to the simulated mean surface and "zeroes" out the other surface.

\noindent\textbf{Simulation model selection} We decided to compute the posterior predictive probability for another 50 random locations and their observations simulated according to the different scenarios in order to model select. Let $x^*$ represent the withheld locations, $y^*$ the withheld data, $\vartheta$ a posterior draw for the parameters based on $x$ and $y$, and $\vartheta^*$ a draw for the parameters for $x^*$ and $y^*$. The posterior predictive probability is:
\begin{align}
p(\v{y^*} \mid x, x^*, \v{y}) := \int \int \textrm{p}(y^* \mid \vartheta^*) \textrm{p}(\vartheta^* \mid \vartheta, x, x^*)  \textrm{p}(\vartheta \mid x, y) d\vartheta^* d\vartheta.
\label{eq:post_pred_prob}    
\end{align}
We discuss how to compute this probability in the supplementary material because it is straightforward to compute $\textrm{p}(y^* \mid \vartheta^*)$ according to our models defined in \textit{SvM-c} and \textit{SvM}. It is also simple to sample from $\textrm{p}(\vartheta^* \mid \theta, x, x^*)$ due to the conditional formula for the Multivariate normal distribution to generate these draws from the Gaussian processes. We take advantage of this fact by sampling 100 times from $\textrm{p}(\vartheta^* \mid \vartheta, x, x^*)$ to create our posterior predictive draws. To ensure that both the posterior draws and the posterior predictive draws speak equally, we first averaged across posterior draws. We then averaged these means across the posterior predictive draws. This gives us a sense how likely our posterior is on the withheld data.

%

\begin{table}[!t]
    \centering
    \resizebox{.7\textwidth}{!}{%
    \begin{tabular}{c c c c c c }
    \hline \\[-1.8ex]
    & iV & iVM & \textit{SvM} Mean $\pi$ & \textit{SvM-c} & \textit{SvM} Mean $0$\\
    \hline \\[-1.8ex] 
    \textit{iV} & \textbf{-39.85} & -78.20 & -68.08 & -91.99 &  -$89.20$\\
    & \multicolumn{4}{c}{}\\
    \textit{iVM} & \textbf{-39.92} & \textbf{-47.60} & -68.24 & -86.47 & -85.69\\
    & \multicolumn{4}{c}{}\\
    \textit{SvM} & -42.42 & -84.28 & \textbf{-59.34} & -92.43 & -63.49 \\
    & \multicolumn{4}{c}{}\\
    \textit{SvM-c} & -45.78 & -56.79 & -62.19 & \textbf{-73.40} & \textbf{-60.67}\\
    & \multicolumn{4}{c}{}\\
\end{tabular}
}
\caption{\small The log of the posterior predictive probability given in Equation \eqref{eq:post_pred_prob} for 50 draws averaged across posterior draws and then posterior predictive draws. Rows indicate which models were fitted to the simulated data and columns denote which models were used to generate the simulated data.}
\label{table:sim_post_pred_results}
\end{table}

Table \ref{table:sim_post_pred_results} shows the results from doing so. There are a several interesting observations to make. First, the posterior predictive probability confirms that the homogeneous models performs poorly in any heterogeneous scenarios. Next, the posterior predictive probability calculated after fitting \textit{iV} and \textit{iVM} is higher when the means of the generating von Mises are constant. On the other hand, the posterior predictive probability calculated after fitting the \textit{SvM} and \textit{SvM-c} models is higher when the von Mises components' means are spatially correlated. We might expect the former because spatially correlated observations might be overfitted to the data. For the latter, \textit{iV} and \textit{iVM} have difficulty in capturing the variability and thus cannot accurately predict. Finally, while we do not report these cases, \textit{SvM-c} can occasionally perform poorly according to the predictive posterior probability. When we plot them, we see that the worst performing observations lie between the fitted mean surface. Even though the fitted mean surface and concentration parameter may be mostly correct for the observed data, the model is being penalized for being "over-certain" about the withheld data. This means that if we desire well-separated components, the observations must also be well-separated. If not, this issue might be addressed by adding more components. Even with this caveat, we use the posterior predictive probability to model select because it allows us to compare models that are different in their approaches and generally picks the model used to generate the data in simulation.

\begin{table}[!t]
    \centering
    \resizebox{.5\textwidth}{!}{%
    \begin{tabular}{c c c c c }
    \hline \\[-1.8ex]
    & iV & iVM & \textit{SvM} & \textit{SvM-c}\\
    \hline \\[-1.8ex] 
    \textit{iV} & \textbf{-96.28} & -137.68 & -157.62 & -163.71\\
    & \multicolumn{4}{c}{}\\
    \textit{iVM} & \textbf{-96.04} & \textbf{-98.74} & -157.61 & -163.73\\
    & \multicolumn{4}{c}{}\\
    \textit{SvM} & -112.91 & -163.54 & \textbf{-127.95} & -163.54 \\
    & \multicolumn{4}{c}{}\\
    \textit{SvM-c} & -151.17 & -123.91 & -160.05 & \textbf{-138.40}\\
    & \multicolumn{4}{c}{}\\
\end{tabular}
}
\caption{\small The log of the posterior predictive probability given in Equation \eqref{eq:post_pred_prob} for 50 draws averaged across posterior draws and then posterior predictive draws. Rows indicate which models were fitted to the simulated data and columns denote which models were used to generate the simulated data.}
\label{table:sim_post_pred_results_higher_dim}
\end{table}

\subsection{Multidimension simulation study}
\label{ssection:sim_results_higher_dim}
For our multidimensional simulation study, we assumed a five dimensional simplex and generated four data sets in a similar manner as before. We drew 500 "locations" for the training set and 50 "locations" for the test set uniformly from the five dimensional simplex. Then, we generated the random directions in the following ways. First, we used the von Mises-Fisher distribution with a mean vector of $(\frac{1}{\sqrt{5}}, \frac{1}{\sqrt{5}}, \frac{1}{\sqrt{5}}, \frac{1}{\sqrt{5}}, \frac{1}{\sqrt{5}})$ and concentration parameter of $5$. Second, we generated random direction using a mixture of two von Mises-Fisher distributions. The first distribution has mean vector of $(\frac{1}{\sqrt{5}}, \frac{1}{\sqrt{5}}, -\frac{1}{\sqrt{5}}, -\frac{1}{\sqrt{5}}, -\frac{1}{\sqrt{5}})$ and concentration parameter of $8$ whereas the second distribution has mean vector of $(-\frac{1}{\sqrt{5}}, --\frac{1}{\sqrt{5}}, \frac{1}{\sqrt{5}}, \frac{1}{\sqrt{5}}, \frac{1}{\sqrt{5}})$ and concentration parameter of $3$. The probability of selecting the first distribution is 0.7. Again, we denote the first and second model used to generate the data sets as \textit{iV} and \textit{iVM} respectively. Next, we considered "spatial information" for our next two data sets. To generate data according to \textit{SvM}, we assume that each location's dimension has a prior mean of $1$ and the covariance matrix is a squared exponential kernel with $\omega = 0.1$ and $\sigma = 0.5$. Finally, for \textit{SvM-c}, each location's prior mean is either $(1, 1, -1, -1, -1)$ with probability 0.7 or $(-1, -1, 1, 1, 1)$ with probability 0.3. The same squared exponential kernel was used for \textit{SvM-c}.

Posterior inference was done differently depending on the model. We used HMC as implemented in Stan for the $\textit{iV}$ and $\textit{iVM}$ models. The sampler for the multidimensional versions of \textit{SvM} and \textit{SvM-c} was written in R. This included the Hamiltonian Monte Carlo step for the concentration parameters, which was tuned to an acceptance rate of roughly 75\% for data generated according to the correct model. We ran the sampler for 25 000 iterations with 20 000 as burn-in and keeping only every fifth iteration. It took roughly between 69 and 169 minutes on average to run that many iterations for \textit{SvM} whereas it roughly took between 177 and 323 minutes for \textit{SvM-c}.

To check how our samplers performed, we again computed the posterior predictive probabilities and examine the posterior credible intervals. As seen in Table \ref{table:sim_post_pred_results_higher_dim}, the posterior predictive probabilities are smallest when \textit{SvM} and \textit{SvM-c} are fitted to data generated from their models. This suggests that the samplers are able to recover the parameters used to generate the data. Indeed, when examining \textit{SvM}'s credible intervals or regions for data generated by the model, the generating hierarchical prior concentration mean and at least 95\% of the generating mean angles are in the intervals or regions. Note that to come up with the credible regions for an angle, we used a multivariate normal distribution parameterized by the MCMC samples' mean and covariance matrix to approximate the posterior of an angle. This seemed sensible because of the von Mises-Fisher and projected Gaussian processes' connections to the normal distribution. Further, we adjust the last dimension of the angle to be on the interval of $2\pi$ centered at the mean of that dimension's angle. This is the only angle in which the endpoints, $0$ and $2\pi$, are actually the same point. Meanwhile, we achieve similar coverage with \textit{SvM-c}'s credible intervals and regions even with the extra component. On the other hand, when \textit{SvM} and \textit{SvM-c} performed poorly according to the posterior predictive, the methods are still able to recover the mean surface when the number of components match. For instance, \textit{SvM}'s overall average of the posterior mean points on the five sphere for the \textit{iV} data set is close to uniform mean point on the five sphere used to generate the data set. A similar result holds for \textit{SvM-c}'s overall average of the posterior mean points on the five sphere for the \textit{iVM} data set. Even though it appears that it is more important to correctly identify whether there is a spatial pattern and the number of components in higher dimensions, our samplers for \textit{SvM} and \textit{SvM-c} are able to correctly estimate the parameters.

\section{Additional Simulation and Results}
\subsection{Additional simulation plots and tables (one dimension)}
\begin{table}
\resizebox{0.95\textwidth}{!}{
\begin{tabular}{@{\extracolsep{5pt}} ccccccc} 
\\[-1.8ex]\hline 
\hline \\[-1.8ex] 
 & \textit{iV} & \textit{iVM} & \textit{SvM} mean $\pi$ & \textit{SvM-c} & \textit{SvM} mean $0$\\ 
\hline \\[-1.8ex] 
\textit{SvM} & 4528 & 3039 & 5209 & 2663 & 5488\\
& & & & &\\
\textit{SvM-c} & 23738
& 16442 & 29486 & 22624 & 30571\\
& & & & &\\
\hline \\[-1.8ex] 
\end{tabular} 
}
\caption{Time it takes for our models to run four chains for different models in seconds according to the sampling scheme outlined in Section \ref{ssection:sim_results} on four cores of a cluster using 2x 3.0 GHz Intel Xeon Gold 6154 as its processor.}
\end{table}

\begin{figure}[!h]
\captionsetup[subfigure]{justification=centering}
\centering
\begin{subfigure}{.23\textwidth}
\centering
\includegraphics[width = 1\textwidth]{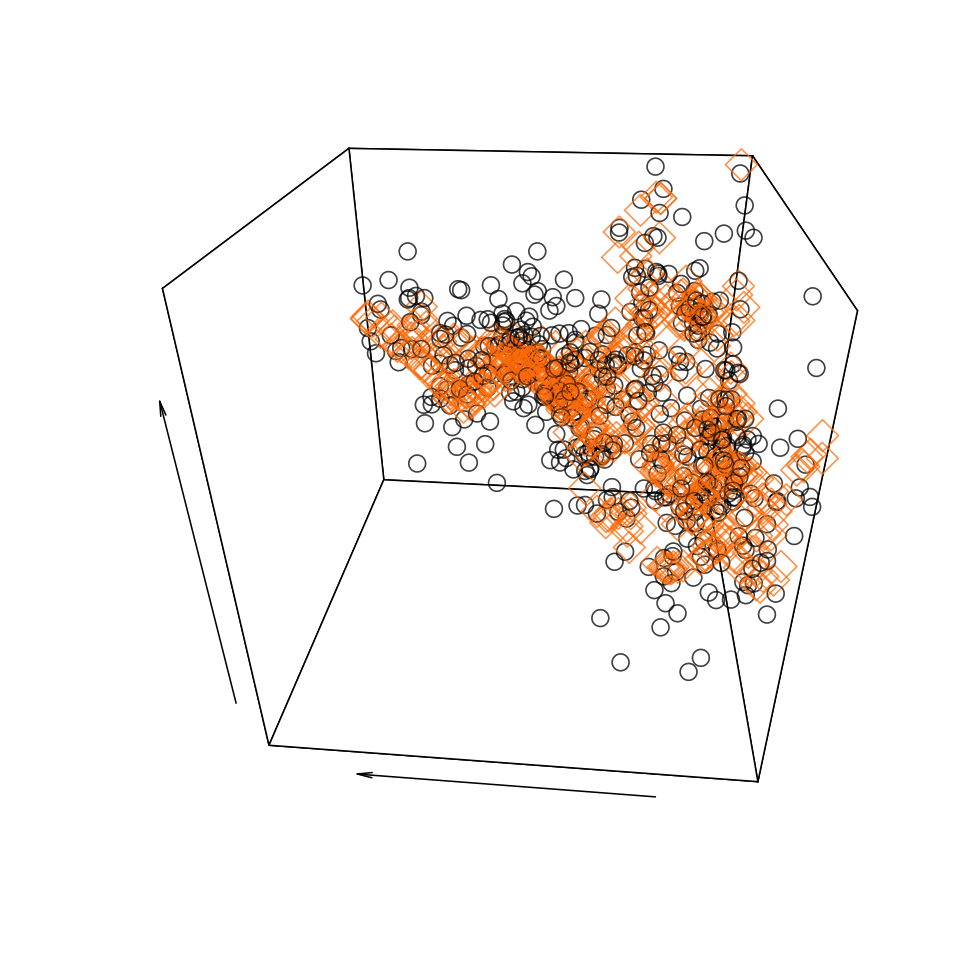}
\caption{Simulated data with\\means in orange}
\end{subfigure}
\begin{subfigure}{.23\textwidth}
\centering
\includegraphics[width = 1\textwidth]{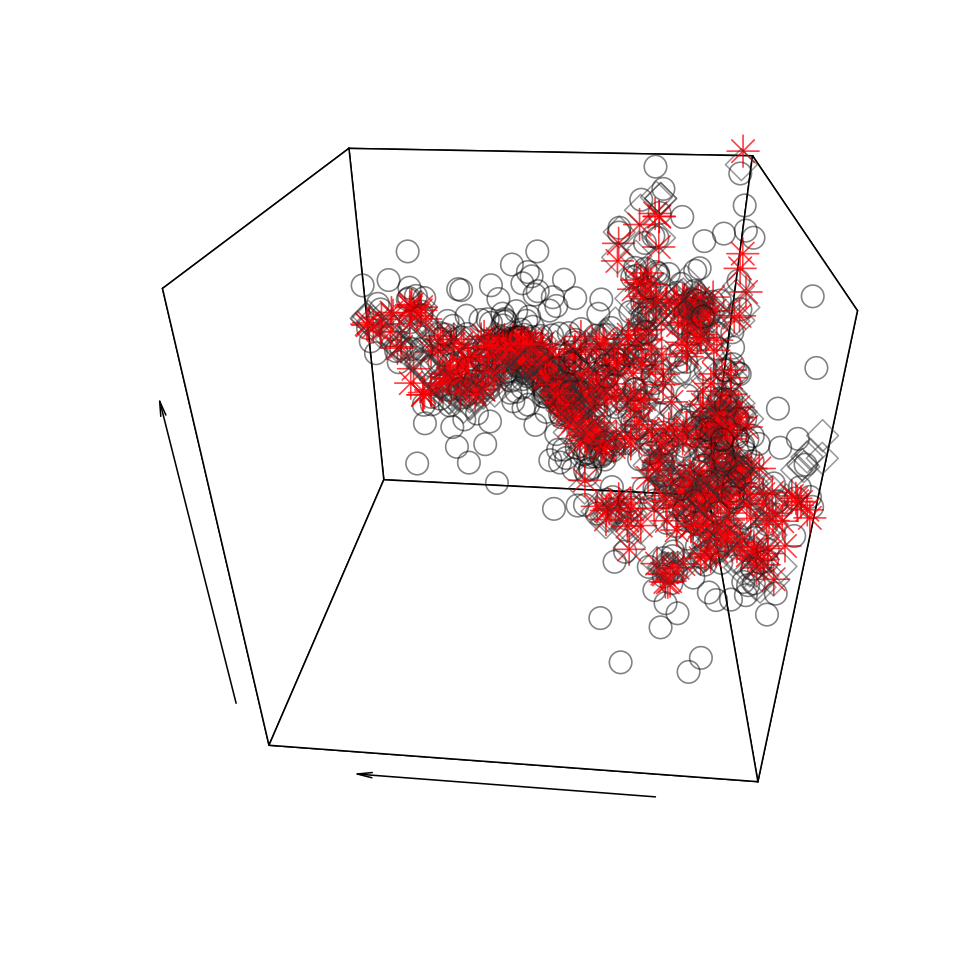}
\caption{SvM model \\fitted means in red}
\end{subfigure}
\begin{subfigure}{.23\textwidth}
\centering
\includegraphics[width = 1\textwidth]{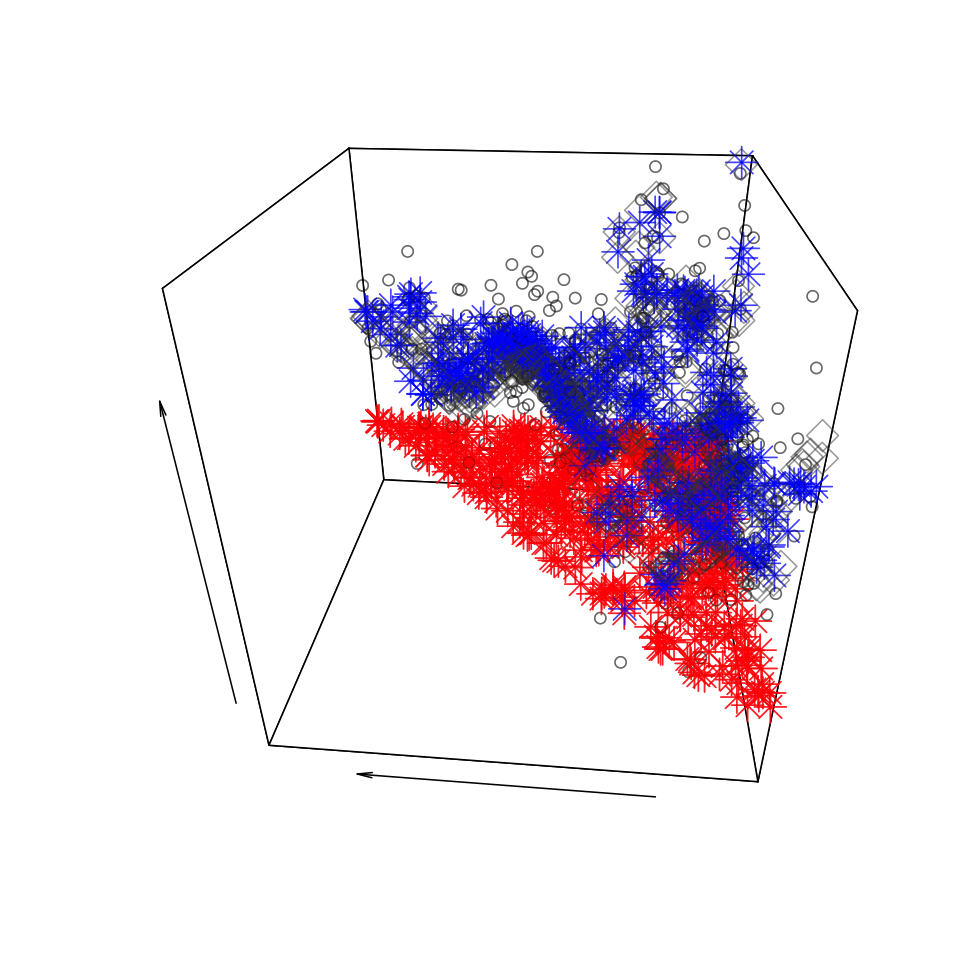}
\caption{SvM-c model\\fitted means in red and blue}
\end{subfigure}
\caption{Plots of the mean component fitted by different models for observations simulated according to the $\textit{SvM}$ model in \eqref{model:SvM}. The fitted mean components are shown in colored asterisks with the simulated observations as circles and the simulated means as grey rhombuses for different scenarios. The x-y axis are the observations' location in the first two dimensions of a three dimensional simplex; the z-axis represent the observations.}
\label{fig:model_fitted_SvM_plots}

\resizebox{0.95\textwidth}{!}{
\begin{tabular}{@{\extracolsep{5pt}} cccc} 
\\[-1.8ex]\hline 
\hline \\[-1.8ex] 
 & $\bm{m}$ & $\bm{\rho}$ & $\bm{\lambda}$ \\ 
\hline \\[-1.8ex] 
\textbf{Simulation} & $\overline{m}$ = 3.23 (1.78, 4.86)  & $\overline{\rho} = 3.00 (2.54, 3.50)$ & ---\\
& & &\\
\textit{SvM}  & $\overline{m}$ = 3.28 (2.89, 3.68) & $\overline{\rho}$ = 3.14 (2.70, 3.63) & ---\\
& & &\\
\multirow{2}{*}{\shortstack[c]{\textit{SvM-c}}} & $\overline{m}_1$ = 1.58 (0.39, 2.78) & $\overline{\rho}_1$ = 0.37 (0.00, 89.98) & $\lambda_1$ = 0.02 (9.07e-4, 0.04)\\
& $\overline{m}_2$ = 3.32 (2.90, 3.76) & $\overline{\rho}_2$ = 3.45 (2.93, 4.09) & $\lambda_2$ = 0.98 (0.96, 0.9991)\\
& & &\\
\hline \\[-1.8ex] 
\end{tabular} 
}
\captionof{table}{Circular mean posterior values and 95\% credible intervals for the parameters in the Von Mises distribution and mixing probability are shown for the models fitted to the simulation data in Figure \ref{fig:model_fitted_SvM_plots}. Parameters with bars over them were averaged across all locations.}
\end{figure}

\begin{figure}[!h]
\captionsetup[subfigure]{justification=centering}
\centering
\begin{subfigure}{.23\textwidth}
\centering
\includegraphics[width = 1\textwidth]{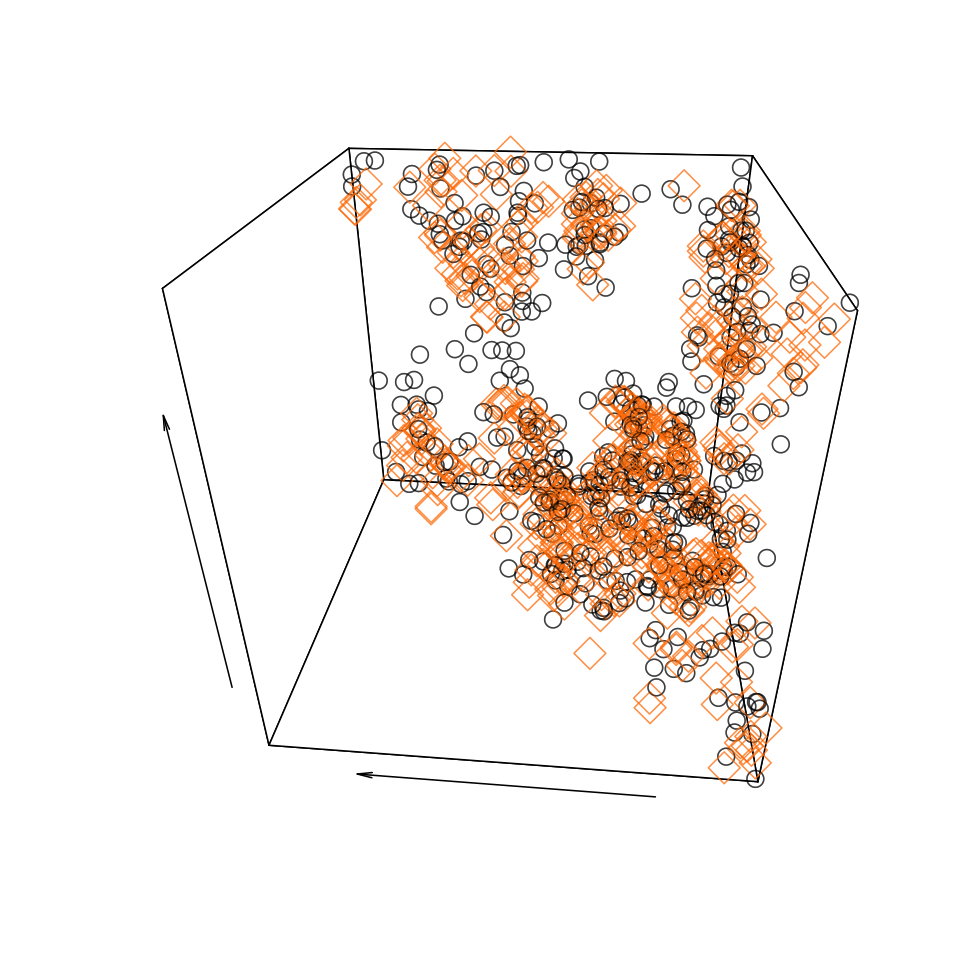}
\caption{Simulated data with\\means in orange}
\end{subfigure}
\begin{subfigure}{.23\textwidth}
\centering
\includegraphics[width = 1\textwidth]{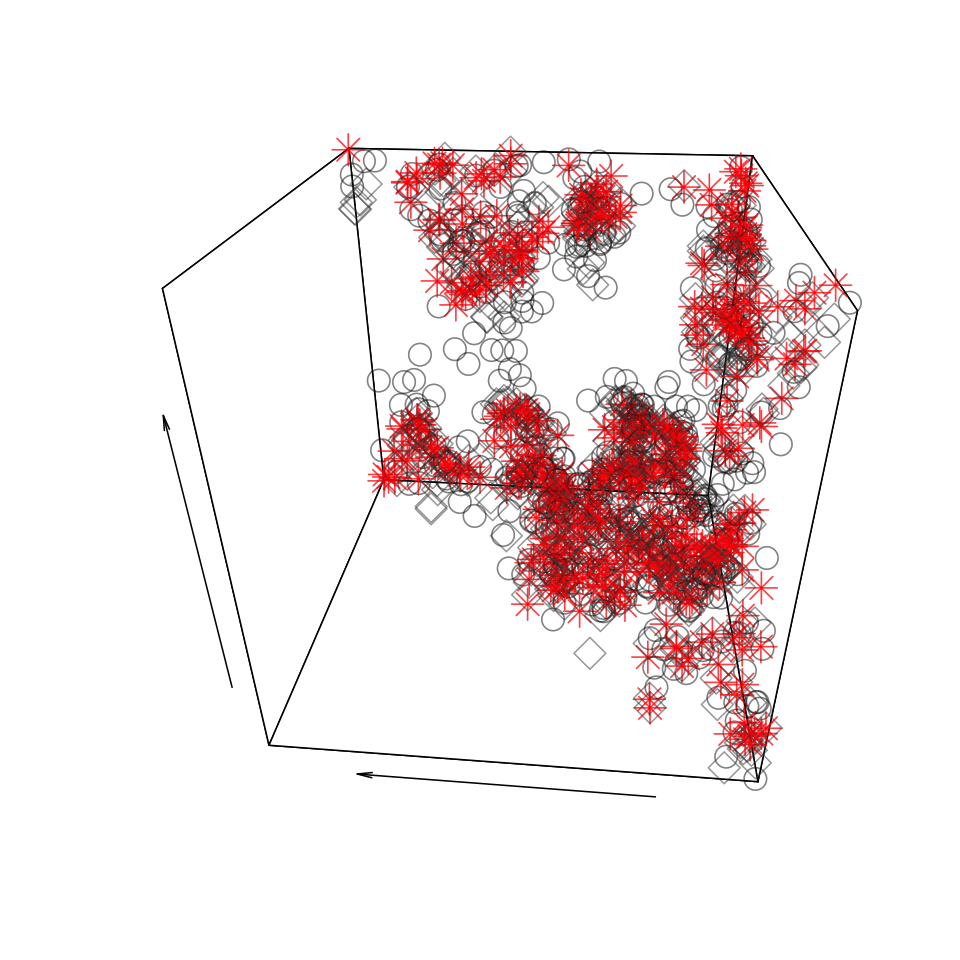}
\caption{SvM model \\fitted means in red}
\end{subfigure}
\begin{subfigure}{.23\textwidth}
\centering
\includegraphics[width = 1\textwidth]{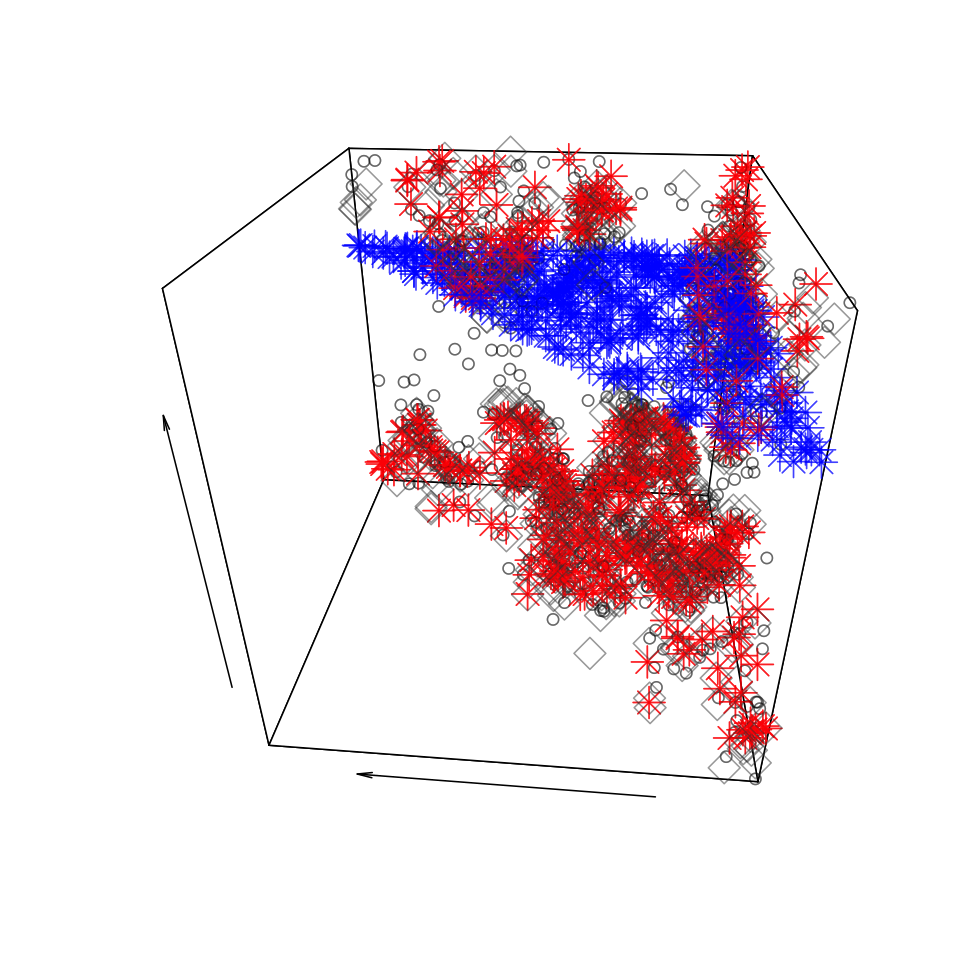}
\caption{SvM-c model\\fitted means in red and blue}
\end{subfigure}
\caption{Plots of the mean component fitted by different models for observations simulated according to the $\textit{SvM}$ model in \eqref{model:SvM}. The fitted mean components are shown in colored asterisks with the simulated observations as circles and the simulated means as grey rhombuses for different scenarios. The x-y axis are the observations' location in the first two dimensions of a three dimensional simplex; the z-axis represent the observations.}
\label{fig:model_fitted_alt_SvM_plots}

\resizebox{0.9\textwidth}{!}{
\begin{tabular}{@{\extracolsep{5pt}} cccc} 
\\[-1.8ex]\hline 
\hline \\[-1.8ex] 
 & $\bm{m}$ & $\bm{\rho}$ & $\bm{\lambda}$ \\ 
\hline \\[-1.8ex] 
\textbf{Simulation} & $\overline{m}$ = 0.68 (4.17, 2.38)  & $\overline{\rho} = 2.99 (2.51, 3.45)$ & ---\\
& & &\\
\textit{SvM} & $\overline{m}$ = 0.64 (6.27, 1.27) & $\overline{\rho}$ = 2.65 (2.24, 3.12) & ---\\
& & &\\
\multirow{2}{*}{\shortstack[c]{\textit{SvM-c}}} & $\overline{m}_1$ = 0.67 (0.10, 1.24) & $\overline{\rho}_1$ = 2.66 (2.24, 3.19)  & $\lambda_1$ = 0.99 (0.95, 0.9996)\\
& $\overline{m}_2$ = 4.72 (3.52, 5.91) & $\overline{\rho}_2$ = 0.12 (3.40e-5, 225.56) & $\lambda_2$ = 0.01 (4.51e-4, 0.05\\
& & &\\
\hline \\[-1.8ex] 
\end{tabular} 
}
\captionof{table}{Circular mean posterior values and 95\% credible intervals for the parameters in the Von Mises distribution and mixing probability are shown for the models fitted to the simulation data in Figure \ref{fig:model_fitted_SvM_plots}. Parameters with bars over them were averaged across all locations.}
\end{figure}

\newpage

\subsection{Additional simulation tables (multidimensions)}
\begin{table}
\resizebox{0.95\textwidth}{!}{
\begin{tabular}{@{\extracolsep{5pt}} cccccc} 
\\[-1.8ex]\hline 
\hline \\[-1.8ex] 
 & \textit{iV} & \textit{iVM} & \textit{SvM} mean & \textit{SvM-c}\\ 
\hline \\[-1.8ex] 
\textit{SvM} & 9989 & 4287 & 10181 & 4127\\
& & & & &\\
\textit{SvM-c} & 10615
& 19368 & 10786 & 18682\\
& & & & &\\
\hline \\[-1.8ex] 
\end{tabular} 
}
\caption{Time it takes for our models to run four chains for different models in seconds according to the sampling scheme outlined in Section \ref{ssection:model_fit_svm_svm_c_higher_dim} on four cores of a cluster using 2x 3.0 GHz Intel Xeon Gold 6154 as its processor.}
\end{table}

\subsection{Real data hyperparameter and model selection}
\begin{figure}[!tbp]
\centering
\begin{subfigure}{.3\textwidth}
\centering
\includegraphics[width = 1\textwidth]{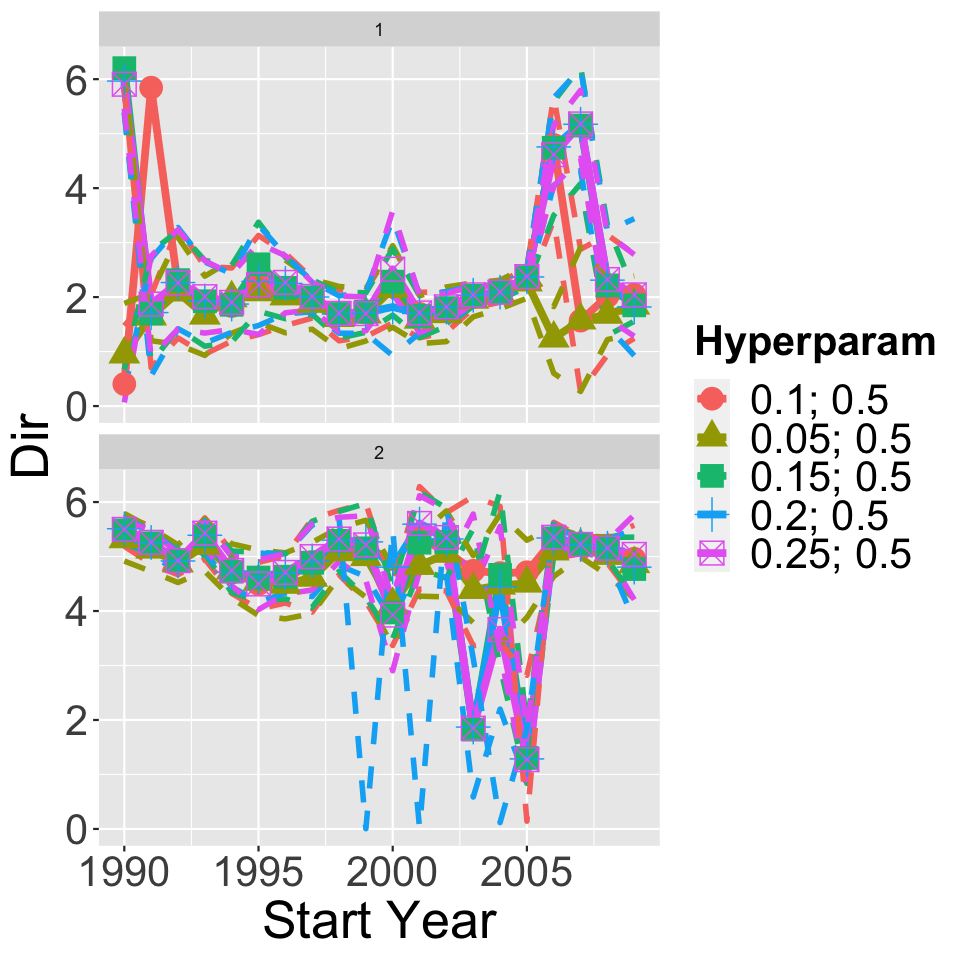}
\end{subfigure}
\qquad
\begin{subfigure}{.3\textwidth}
\centering
\includegraphics[width = 1\textwidth]{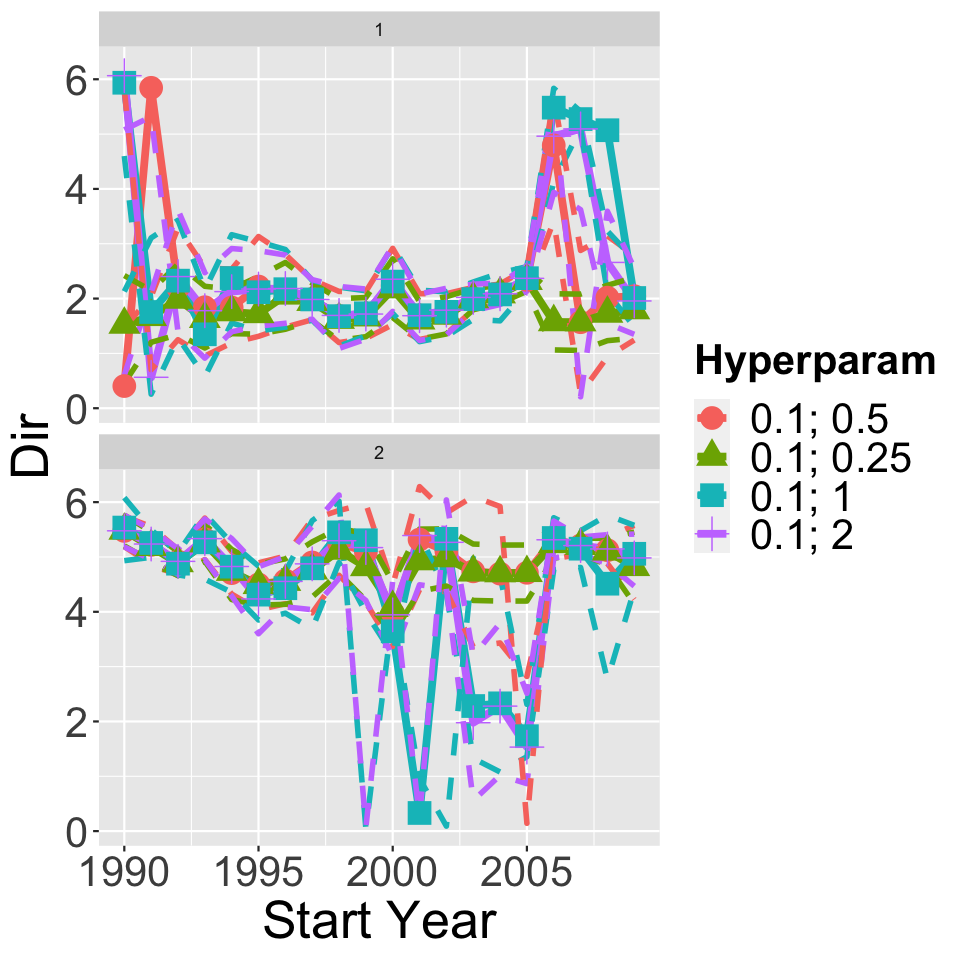}
\end{subfigure}
\caption{Plots showing the mean direction and credible intervals averaged over all locations for various hyperparameters of the squared exponential kernel for \textit{SvM-c} compared to $\omega = 0.1$ and $\sigma = 0.5$ on the left. The value for $\omega$ is given first.}
\label{fig:real_data_exp_results}
\end{figure}

While we used the same priors that we utilized for simulation, it was still unclear what to set the hyperparameters for the squared exponential kernels. To pick, we used $\omega = 0.2$ and $\omega = 0.5$ as a baseline for \textit{SvM-c-2}. This choice was based on running our analysis on a bootstrapped sample of the data. We then compared it against $\sigma = 0.25, 1$, and $2$ and $\omega = 0.05, 0.1$, and $0.5$ for \textit{SvM-c}. Some of our findings from these experiments can be seen in Figure \ref{fig:real_data_exp_results}. When we examine \textit{SvM-c}, changing the hyperparameters does not appear to affect the first or second mean averaged by location too much except around 2005 to 2007 for the first mean and 2003 to 2005 for the second mean. Still, there are differences in the mean surfaces hidden by this statistic. If we focus on the results from 1996 to 1997, we see that while the lower mean surface is mostly similar when we change $\v{\Sigma}$, the upper mean surface varies more when $\v{\Sigma}$ increases. For instance, the upper mean surface for richer neighborhoods expands to become what appears to be a series of thinly connected pillars that span from zero to $2\pi$ as $\v{\Sigma}$ increases. This makes sense that neighboring values on the mean surface can vary drastically when $\v{\Sigma}$ grows. On the other hand, the lower mean surface in 1996 to 1997 becomes more concentrated as we increase $\omega$ from 0.05 to 0.5. On the other hand, the upper mean surface is similar in shape. However, the mean surface for neighborhoods with income proportions largely in the second one become more concentrated as there is no longer a chunk peaking out from zero.

\begin{figure}[!tbp]
\centering
\begin{subfigure}{.3\textwidth}
\centering
\includegraphics[width = 1\textwidth]{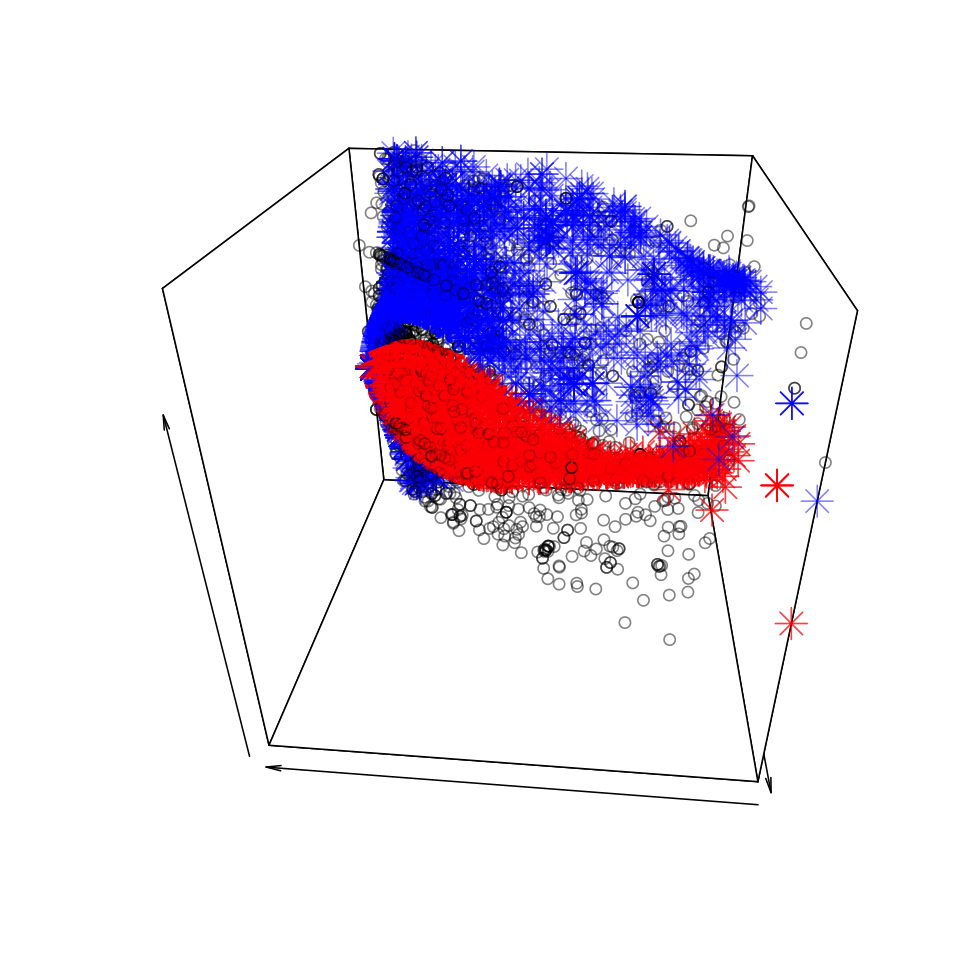}
\caption{Squared exponential}
\end{subfigure}
\qquad
\begin{subfigure}{.3\textwidth}
\centering
\includegraphics[width = 1\textwidth]{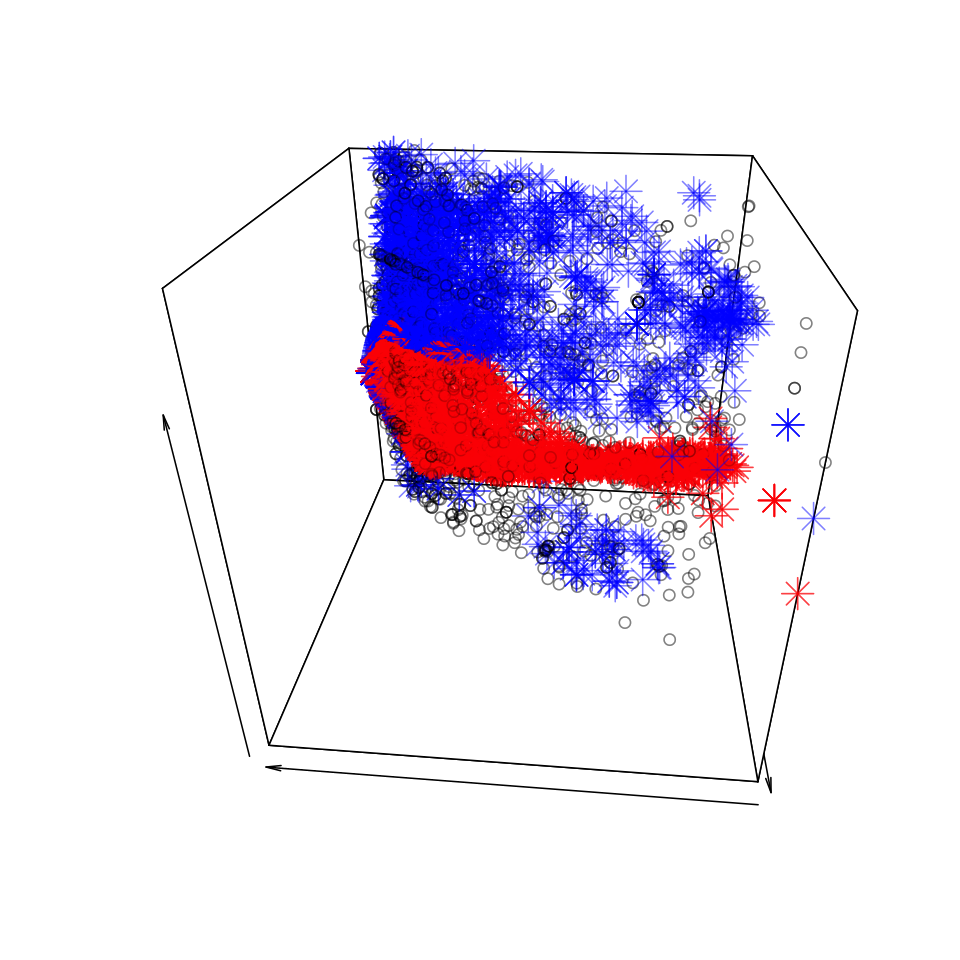}
\caption{Matern $v = \frac{3}{2}$}
\end{subfigure}
\begin{subfigure}{.3\textwidth}
\centering
\includegraphics[width = 1\textwidth]{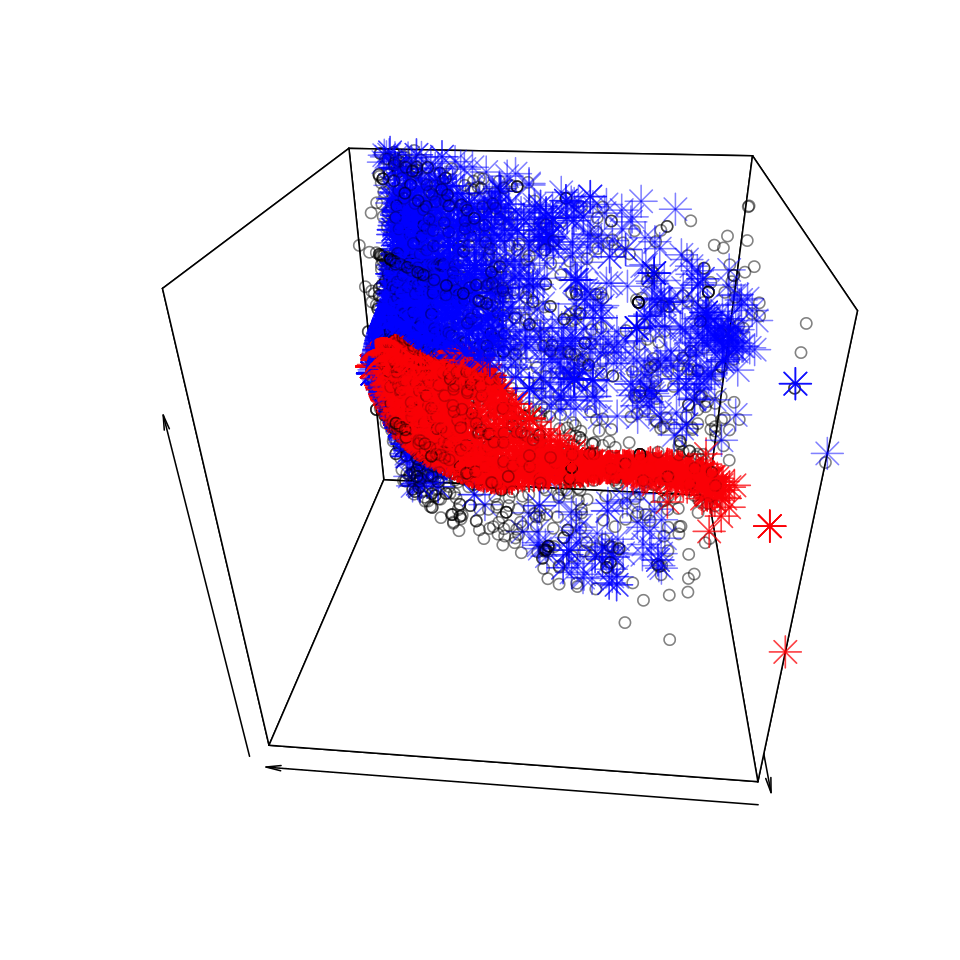}
\caption{Matern $v = \frac{5}{2}$}
\end{subfigure}
\caption{Plots showing the fitted mean surface of the random directions for \textit{SvM-c} under different kernel choices for the income proportions observed in 1996-1997. For the squared exponential kernel, $\sigma = 0.5$ and $\omega = 0.5$. Meanwhile, for the Matern kernel, $v$ is given in the caption and $\omega = 0.2$.}
\label{fig:real_data_kernel_comparison}
\end{figure}

After running these experiments, we selected the hyperparameters that had the average highest log posterior predictive probability across all years. Here, we take the average with respect to the number of observations. For \textit{SvM-c}, this was $\sigma = 0.5$ and $\omega = 0.5$. We then used the respective $\omega$'s to explore the choice of kernel. In particular, we compared against the Matern kernel with $v = \frac{3}{2}$ and $v = \frac{5}{2}$. The Matern kernel is defined as following:
\begin{align}
    k_{\mathcal{M}}(\v{x}_{\ell}, \v{x_{\ell'}}) &= \frac{2^{1 - v}}{\Gamma(v)}\left(\sqrt{2v}\frac{\sqrt{(\v{x}_{\ell} - \v{x_{\ell'}})^2}}{\omega}\right)^v K_v\left(\sqrt{2v}\frac{\sqrt{(\v{x}_{\ell} - \v{x_{\ell'}})^2}}{\omega}\right).
    \label{eq:matern_kernel_a}
\end{align}
The Matern with $v = \frac{1}{2}$ was not compared because previous experiments demonstrated that kernels leading to results that are less smooth did not perform as well and the Matern kernel with one half is such a kernel \citep{RasmussenWilliamsGaussianProcessesMachine2006}. Meanwhile, choosing the Matern kernels or the optimal squared exponential kernel for \textit{SvM-c} result in similar mean directions averaged by location. However, if we again look at the mean surface from 1996 to 1997, we notice differences. While the lower mean surface is similar from the Matern kernel with five halves degrees of freedom and squared exponential kernel, the surface is more dispersed for the Matern kernel with three half degrees of freedom. However, for non-lower income neighborhoods, the top mean surface is more scattered for the Matern kernel than for the squared exponential kernel. Interestingly enough, the surface from the Matern kernel with five halves degrees of freedom is more diffuse than the surface from the Matern kernel with three halves degrees of freedom. One other key difference is that the \textit{SvM-c} models with the Matern kernels performed worse than the same model with the squared exponential kernel. This suggests that while the mean surface for the random direction might be sensitive to the choice of kernel, regions that are likely to move in the same direction are not sensitive.

Finally, we also wanted to compare the results from the square exponential hyperparameter experiments against $\textit{SvM}$ and \textit{SvM-c-3} using the hyperparameters selected for \textit{SvM-c} and \textit{SvM-p-3} with the hyperparameters selected according to \textit{SvM-p-2}. According to the log posterior predictive probability given in \eqref{eq:post_pred_prob_a} to select the number of components and model, \textit{SvM-c-3} perform the best for all years except 1990-1991, 1997-1999, and 2004-2005. For all other years, \textit{SvM-c-2} performs the best. Further, the posterior predictive probability for \textit{SvM-c-2} is close to \textit{SvM-c-3} in 1990-1991 and 2006-2007.


\subsection{Real data results}
\begin{figure}[!tb]
\centering
\begin{subfigure}{.18\textwidth}
\centering
\includegraphics[width = 1\textwidth]{logistic_gp_dir_mixture_comp_hier_p/l_05_no_duplicates_real_results/gp_dir_1_results.png}
\caption{1990-1991\\ (\textit{SvM-c-2})}
\end{subfigure}
\begin{subfigure}{.18\textwidth}
\centering
\includegraphics[width = 1\textwidth]{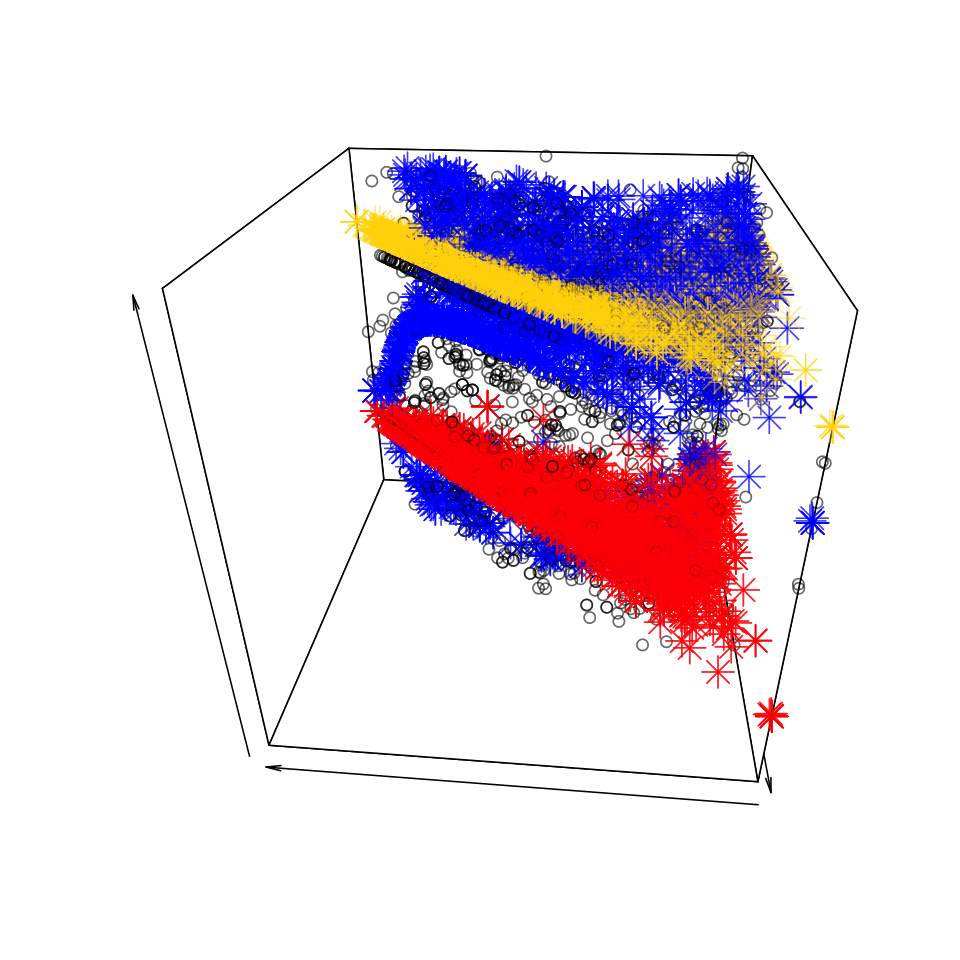}
\caption{1991-1992\\ (\textit{SvM-c-3})}
\end{subfigure}
\begin{subfigure}{.18\textwidth}
\centering
\includegraphics[width = 1\textwidth]{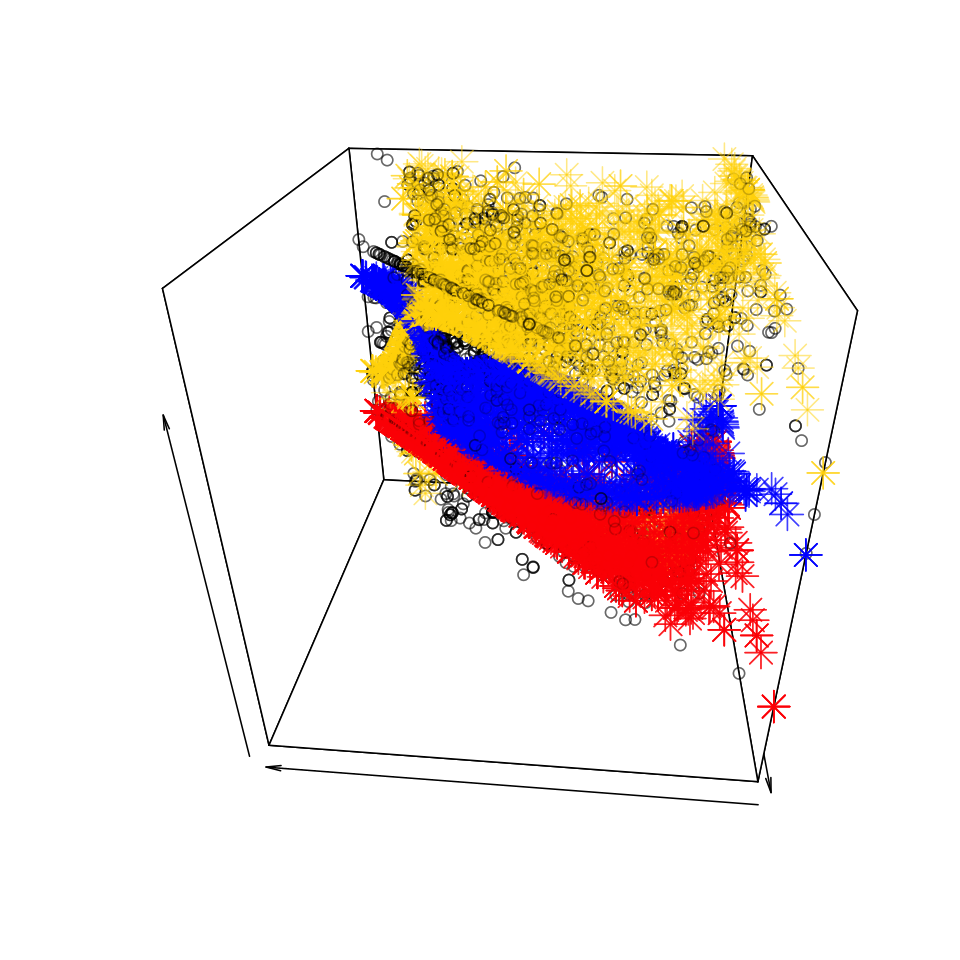}
\caption{1992-1993\\ (\textit{SvM-c-3})}
\end{subfigure}
\begin{subfigure}{.18\textwidth}
\centering
\includegraphics[width = 1\textwidth]{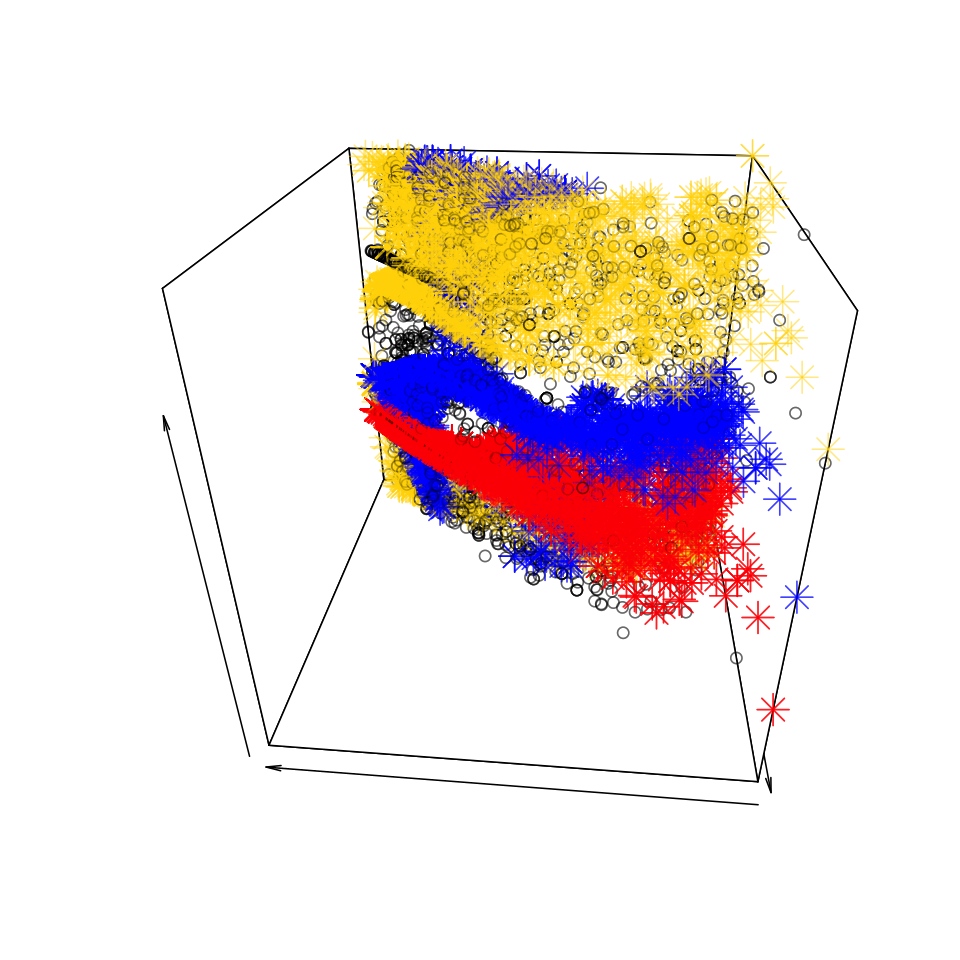}
\caption{1993-1994\\ (\textit{SvM-c-3})}
\end{subfigure}
\begin{subfigure}{.18\textwidth}
\centering
\includegraphics[width = 1\textwidth]{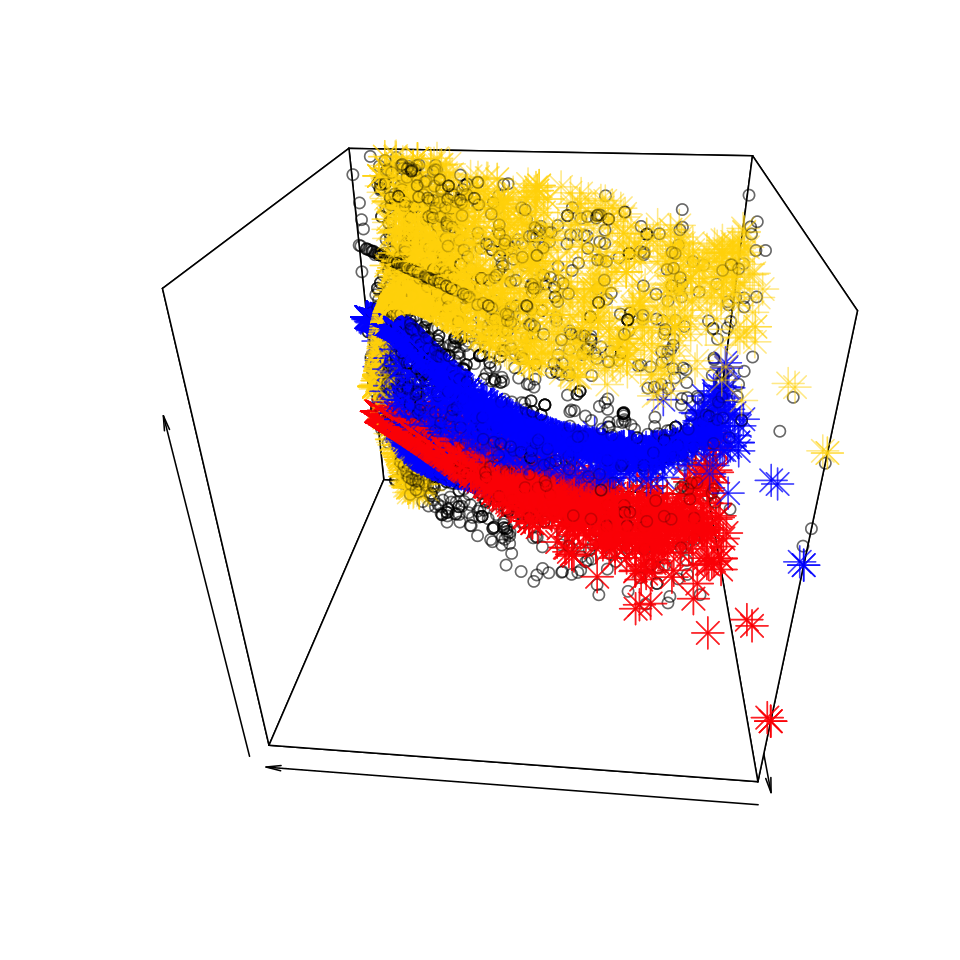}
\caption{1994-1995\\ (\textit{SvM-c-3})}
\end{subfigure}\\
\begin{subfigure}{.18\textwidth}
\centering
\includegraphics[width = 1\textwidth]{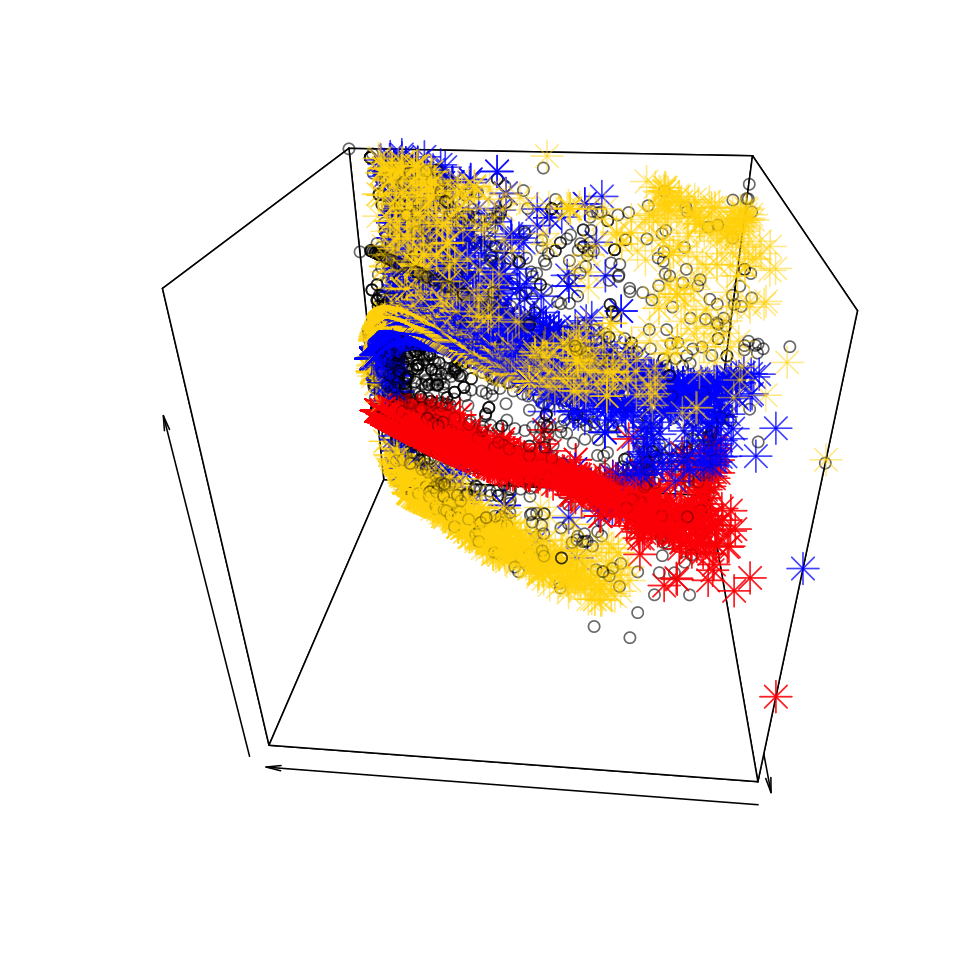}
\caption{1995-1996\\ (\textit{SvM-c-3})}
\end{subfigure}
\begin{subfigure}{.18\textwidth}
\centering
\includegraphics[width = 1\textwidth]{logistic_gp_dir_mixture_comp_hier_p/l_05_no_duplicates_3_real_results/gp_dir_7_results.png}
\caption{1996-1997\\ (\textit{SvM-c-3})}
\end{subfigure}
\begin{subfigure}{.18\textwidth}
\centering
\includegraphics[width = 1\textwidth]{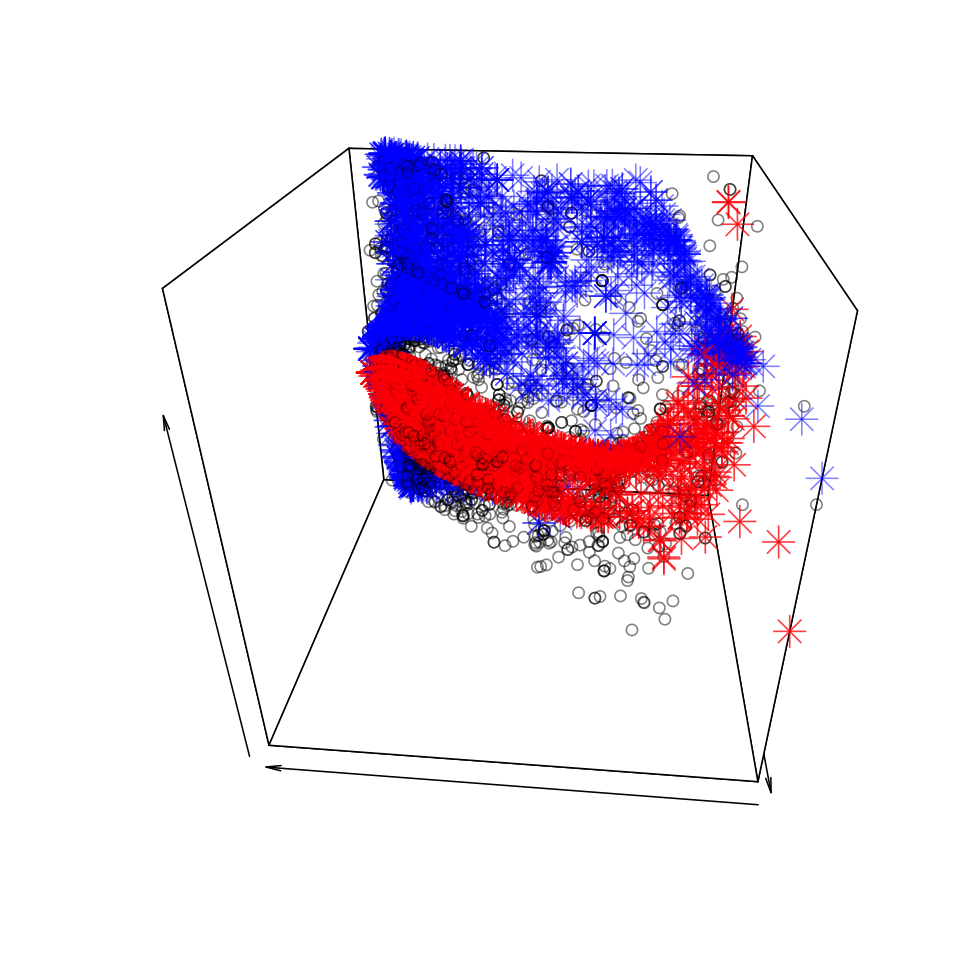}
\caption{1997-1998\\ (\textit{SvM-c-2})}
\end{subfigure}
\begin{subfigure}{.18\textwidth}
\centering
\includegraphics[width = 1\textwidth]{logistic_gp_dir_mixture_comp_hier_p/l_05_no_duplicates_real_results/gp_dir_9_results.png}
\caption{1998-1999\\ (\textit{SvM-c-2})}
\end{subfigure}
\begin{subfigure}{.18\textwidth}
\centering
\includegraphics[width = 1\textwidth]{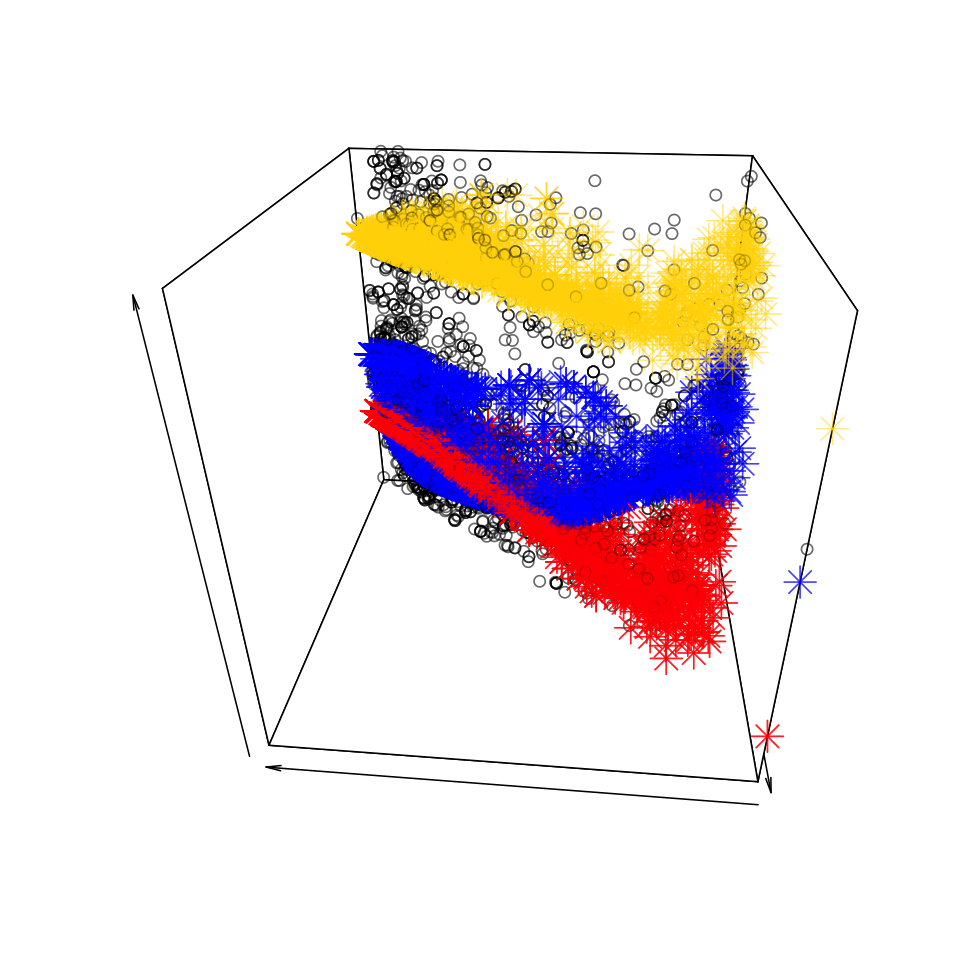}
\caption{1999-2000\\ (\textit{SvM-c-3})}
\end{subfigure}\\
\begin{subfigure}{.18\textwidth}
\centering
\includegraphics[width = 1\textwidth]{logistic_gp_dir_mixture_comp_hier_p/l_05_no_duplicates_3_real_results/gp_dir_11_results.png}
\caption{2000-2001\\ (\textit{SvM-c-3})}
\end{subfigure}
\begin{subfigure}{.18\textwidth}
\centering
\includegraphics[width = 1\textwidth]{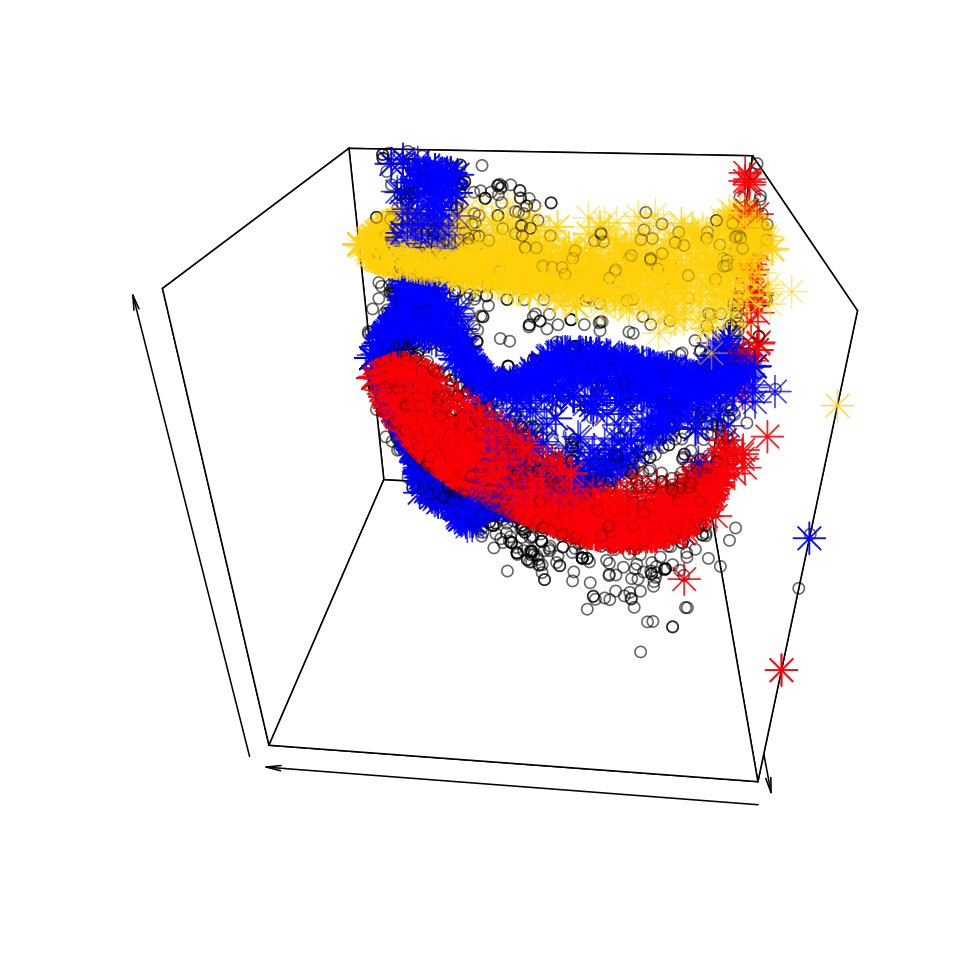}
\caption{2001-2002\\ (\textit{SvM-c-3})}
\end{subfigure}
\begin{subfigure}{.18\textwidth}
\centering
\includegraphics[width = 1\textwidth]{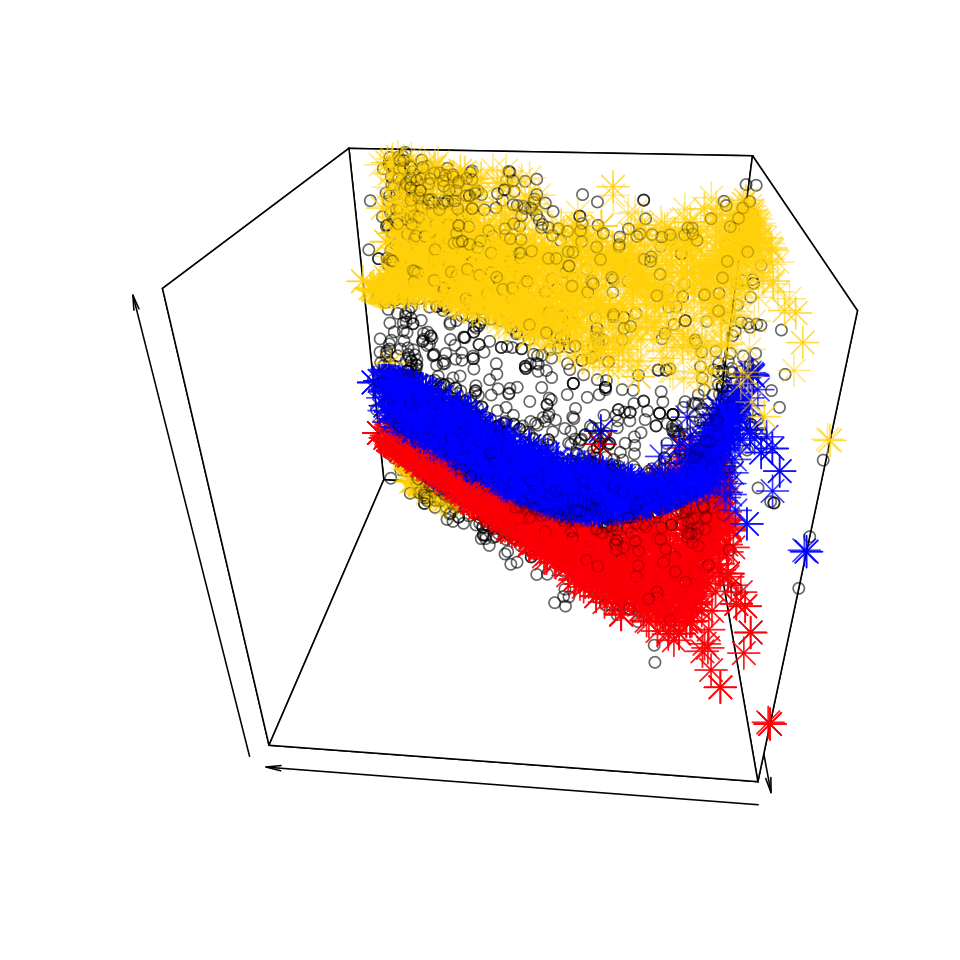}
\caption{2002-2003\\ (\textit{SvM-c-3})}
\end{subfigure}
\begin{subfigure}{.18\textwidth}
\centering
\includegraphics[width = 1\textwidth]{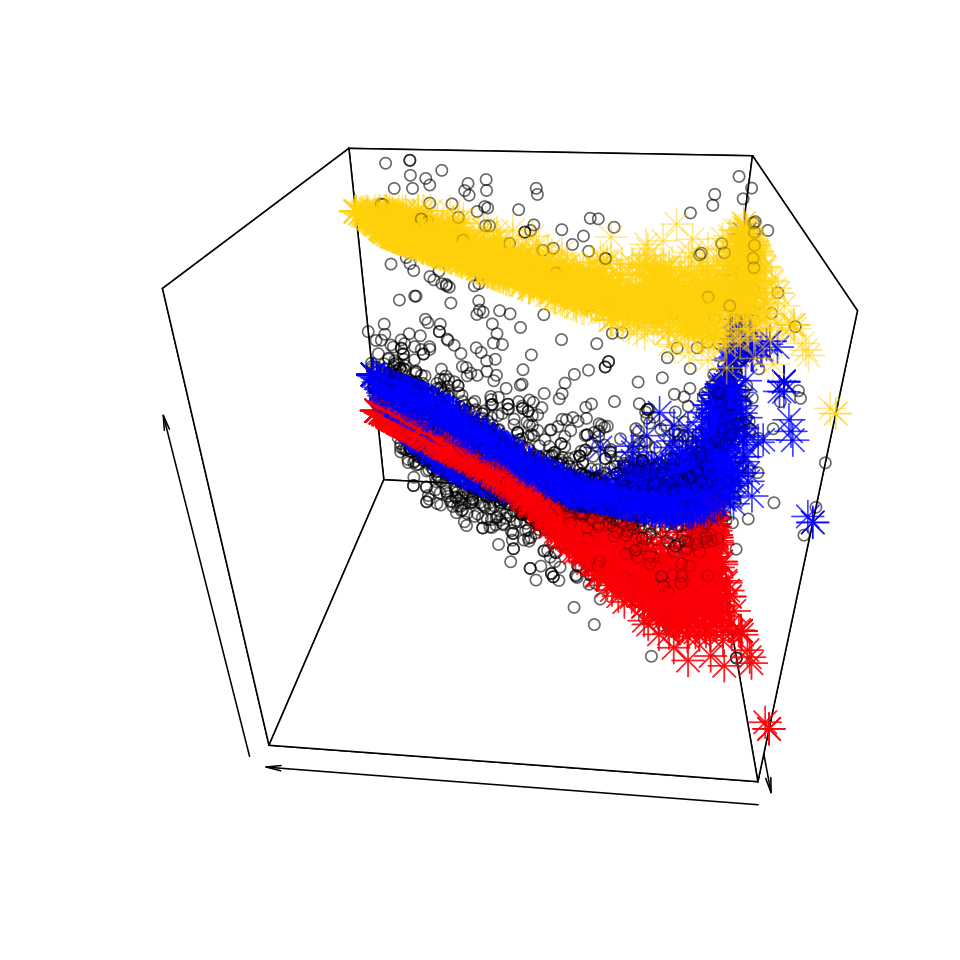}
\caption{2003-2004\\ (\textit{SvM-c-3})}
\end{subfigure}
\begin{subfigure}{.18\textwidth}
\centering
\includegraphics[width = 1\textwidth]{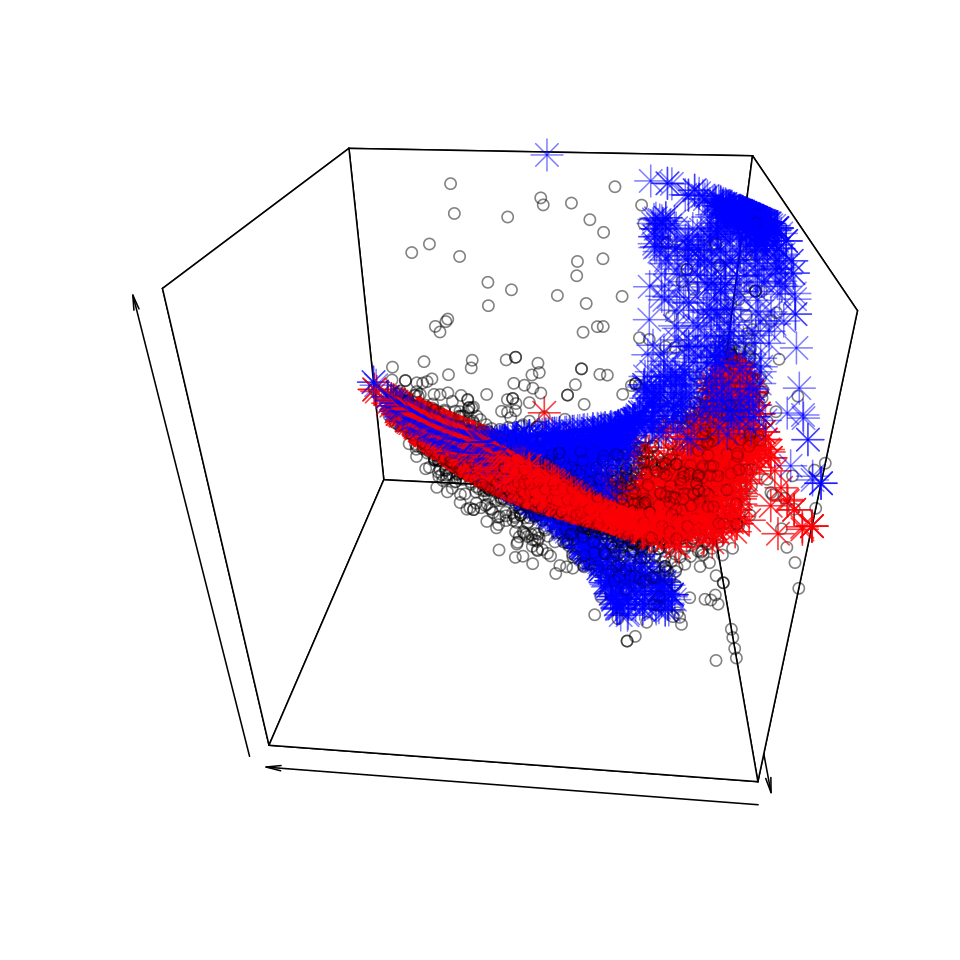}
\caption{2004-2005\\ (\textit{SvM-c-2})}
\end{subfigure}\\
\begin{subfigure}{.18\textwidth}
\centering
\includegraphics[width = 1\textwidth]{logistic_gp_dir_mixture_comp_hier_p/l_05_no_duplicates_real_results/gp_dir_16_results.png}
\caption{2005-2006\\ (\textit{SvM-c-2})}
\end{subfigure}
\begin{subfigure}{.18\textwidth}
\centering
\includegraphics[width = 1\textwidth]{logistic_gp_dir_mixture_comp_hier_p/l_05_no_duplicates_real_results/gp_dir_17_results.png}
\caption{2006-2007\\ (\textit{SvM-c-2})}
\end{subfigure}
\begin{subfigure}{.2\textwidth}
\centering
\includegraphics[width = 1\textwidth]{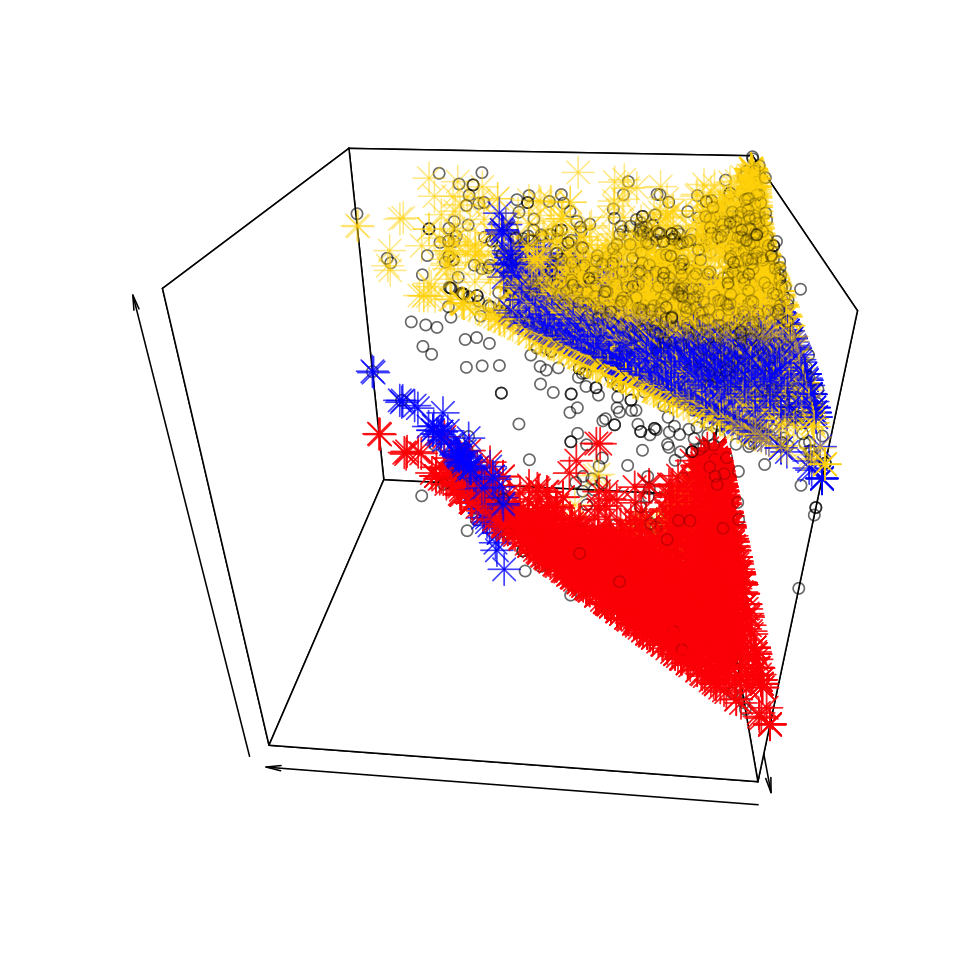}
\caption{2007-2008\\(\textit{SvM-c-3})}
\end{subfigure}
\begin{subfigure}{.18\textwidth}
\centering
\includegraphics[width = 1\textwidth]{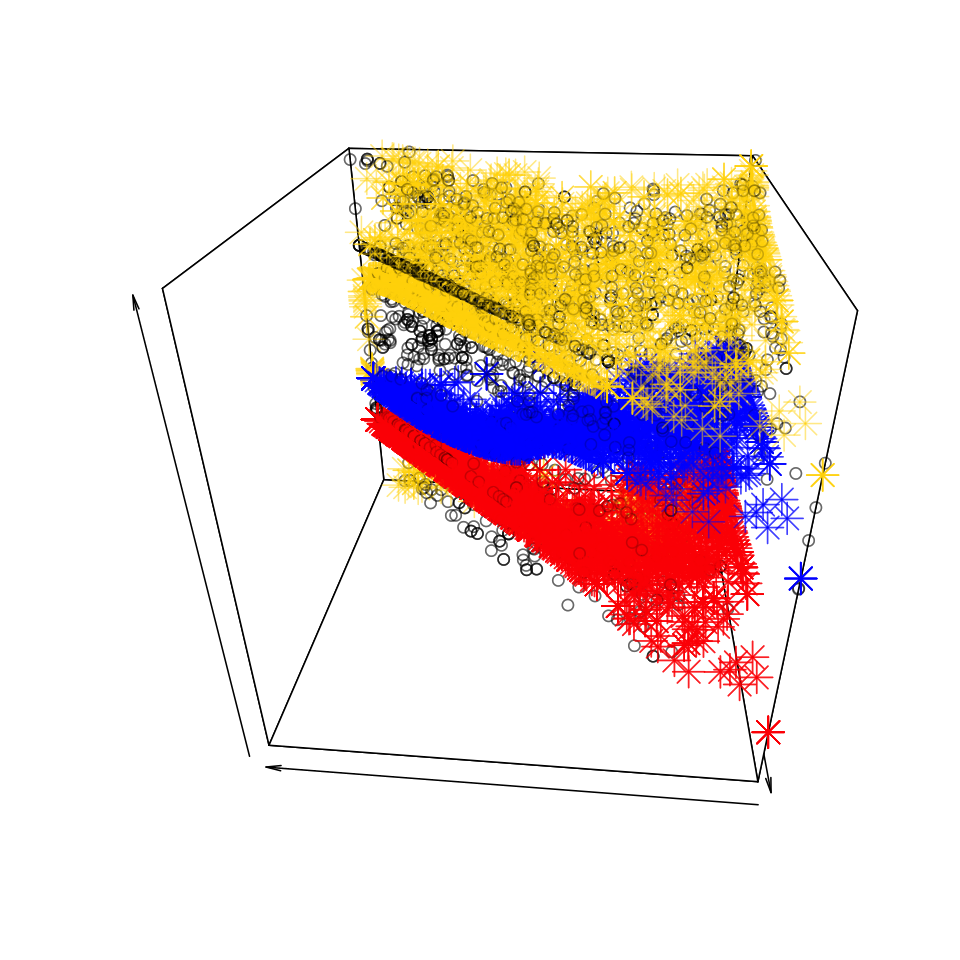}
\caption{2008-2009\\ (\textit{SvM-c 3})}
\end{subfigure}
\begin{subfigure}{.18\textwidth}
\centering
\includegraphics[width = 1\textwidth]{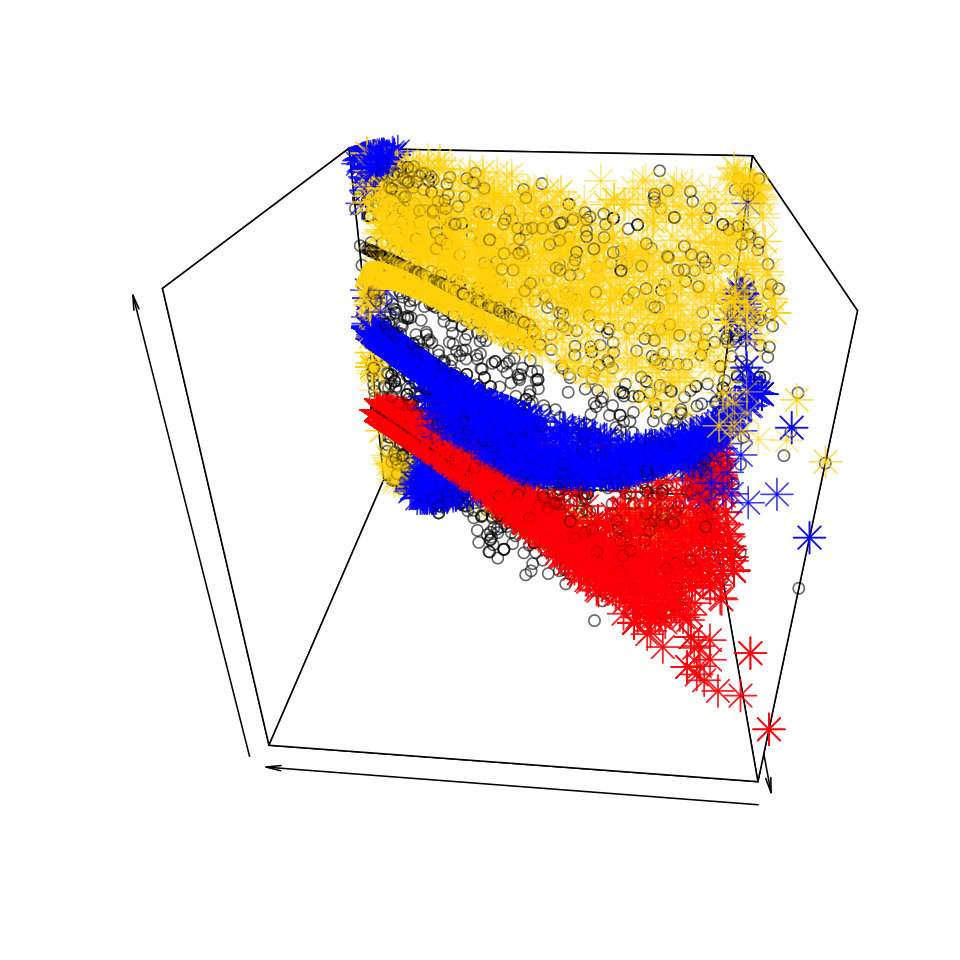}
\caption{2009-2010\\ (\textit{SvM-c-3})}
\end{subfigure}
\caption{Plots showing the observed random direction not withheld and the fitted mean surface of the model selected by the posterior predictive log probability in \eqref{eq:post_pred_prob_a}. The front axis represents the proportion in the first income category and the side axis represents the proportion in the second income category. The up-down axis represents the direction. The start of the arrow indicates a value of zero whereas the end indicates a value of $2\pi$ for angles and 1 for proportions.}
\label{fig:real_data_fitted_results_all}
\end{figure}

\begin{table}[!htbp]
\centering
\resizebox{0.45\textwidth}{!}{
\begin{tabular}{@{\extracolsep{5pt}} cccc} 
\\[-1.8ex]\hline 
\hline \\[-1.8ex] 
\multirow{2}{*}{\shortstack[c]{Year\\(Model)}} & \multirow{2}{*}{\shortstack[c]{ $\bm{m}$}} & \multirow{2}{*}{\shortstack[c]{ $\bm{\rho}$}} & \multirow{2}{*}{\shortstack[c]{ $\bm{\lambda}$}} \\ 
& & &\\
\hline \\[-1.8ex] 
& & &\\
\multirow{2}{*}{\shortstack[c]{1990-1991\\ \textit{SvM-c-2}}} & $\overline{m}_1$ = 5.72 (5.42, 6.05) & $\overline{\rho}_1$ = 1.23 (1.03, 1.45) & $\lambda_1$ = 0.56 (0.45, 0.66)\\
& $\overline{m}_2$ = 5.48 (5.34, 5.61) & $\overline{\rho}_2$ = 6.44 (4.59, 9.64) & $\lambda_2$ = 0.44 (0.34, 0.55)\\
\multirow{3}{*}{\shortstack[c]{1991-1992\\ \textit{SvM-c-3}}} & $\overline{m}_1$ = 1.54 (1.37, 1.71) & $\overline{\rho}_1$ = 84.63 (68.60, 97.13) & $\lambda_1$ = 0.04 (0.03, 0.06)\\
& $\overline{m}_2$ = 5.24 (5.10, 5.38) & $\overline{\rho}_2$ = 2.28 (1.42, 3.04) & $\lambda_2$ = 0.78 (0.73, 0.84)\\
& $\overline{m}_3$ = 5.17 (4.71, 5.73) & $\overline{\rho}_3$ = 0.03 (6.18e-6, 92.21) & $\lambda_3$ = 0.17 (0.12, 0.23)\\
\multirow{3}{*}{\shortstack[c]{1992-1993\\ \textit{SvM-c-3}}} & $\overline{m}_1$ = 1.43 (1.27, 1.56) & $\overline{\rho}_1$ = 81.35 (57.32, 96.79) & $\lambda_1$ = 0.03 (0.02, 0.05)\\
& $\overline{m}_2$ = 2.98 (2.24, 3.61) & $\overline{\rho}_2$ = 1.63 (1.06, 2.52) & $\lambda_2$ = 0.18 (0.12, 0.25)\\
& $\overline{m}_3$ = 4.95 (4.84, 5.05) & $\overline{\rho}_3$ = 2.75 (2.34, 3.26) & $\lambda_3$ = 0.79 (0.71, 0.84)\\
\multirow{3}{*}{\shortstack[c]{1993-1994\\ \textit{SvM-c-3}}} & $\overline{m}_1$ = 1.61 (1.37, 1.77) & $\overline{\rho}_1$ = 84.72 (67.83, 97.02) & $\lambda_1$ = 0.06 (0.04, 0.07)\\
& $\overline{m}_2$ = 2.89 (2.53, 3.27) & $\overline{\rho}_2$ = 3.14 (1.77, 6.62) & $\lambda_2$ = 0.19 (0.11, 0.27)\\
& $\overline{m}_3$ = 5.47 (5.25, 5.67) & $\overline{\rho}_3$ = 1.98 (1.67, 2.35) & $\lambda_3$ = 0.75 (0.68, 0.83)\\
\multirow{3}{*}{\shortstack[c]{1994-1995\\ \textit{SvM-c-3}}} & $\overline{m}_1$ = 1.63 (1.37, 1.98) & $\overline{\rho}_1$ = 85.39 (72.12, 96.97) & $\lambda_1$ = 0.06 (0.05, 0.08)\\
& $\overline{m}_2$ = 2.51 (1.96, 3.03) & $\overline{\rho}_2$ = 1.32 (1.02, 1.79) & $\lambda_2$ = 0.38 (0.26, 0.50)\\
& $\overline{m}_3$ = 4.97 (4.75, 5.20) & $\overline{\rho}_3$ = 2.16 (1.72, 2.78)& $\lambda_3$ = 0.56 (0.44, 0.67)\\
\multirow{3}{*}{\shortstack[c]{1995-1996\\ \textit{SvM-c-3}}} & $\overline{m}_1$ = 1.79 (1.59, 2.00) & $\overline{\rho}_1$ = 20.91 (7.86, 62.77) & $\lambda_1$ = 0.08 (0.05, 0.11)\\
& $\overline{m}_2$ = 3.91 (3.66, 4.13) & $\overline{\rho}_2$ = 2.01 (1.59, 2.68) & $\lambda_2$ = 0.49 (0.38, 0.58)\\
& $\overline{m}_3$ = 5.62 (5.15, 5.97) & $\overline{\rho}_3$ = 1.57 (1.20, 2.11) & $\lambda_3$ = 0.43 (0.34, 0.55)\\
\multirow{3}{*}{\shortstack[c]{1996-1997\\ \textit{SvM-c-3}}} & $\overline{m}_1$ = 1.24 (0.98, 1.49) & $\overline{\rho}_1$ = 48.26 (19.21, 90.10) & $\lambda_1$ = 0.04 (0.02, 0.06)\\
& $\overline{m}_2$ = 2.46 (2.00, 2.93) & $\overline{\rho}_2$ = 2.02 (1.47, 2.94) & $\lambda_2$ = 0.35 (0.27, 0.42)\\
& $\overline{m}_3$ = 4.72 (4.45, 4.91) & $\overline{\rho}_3$ = 1.64 (1.37, 1.97) & $\lambda_3$ = 0.61 (0.52, 0.71)\\
\multirow{2}{*}{\shortstack[c]{1997-1998\\ \textit{SvM-c-2}}} & $\overline{m}_1$ = 2.00 (1.80, 2.19) & $\overline{\rho}_1$ = 2.14 (1.69, 2.74) & $\lambda_1$ = 0.62 (0.53, 0.72)\\
& $\overline{m}_2$ = 4.92 (4.48, 5.40) & $\overline{\rho}_2$ = 1.37 (1.02, 1.96) & $\lambda_2$ = 0.38 (0.47, 0.28)\\
\multirow{2}{*}{\shortstack[c]{1998-1999\\ \textit{SvM-c-2}}} & $\overline{m}_1$ = 1.67 (1.42, 1.91) & $\overline{\rho}_1$ = 1.56 (1.30, 1.90) & $\lambda_1$ = 0.66 (0.59, 0.72)\\
& $\overline{m}_2$ = 5.29 (5.02, 5.56) & $\overline{\rho}_2$ = 1.75 (1.31, 2.35) & $\lambda_2$ = 0.34 (0.28, 0.41)\\
\multirow{3}{*}{\shortstack[c]{1999-2000\\ \textit{SvM-c-3}}} & $\overline{m}_1$ = 1.36 (1.20, 1.53) & $\overline{\rho}_1$ = 62.07 (13.20, 93.85) & $\lambda_1$ = 0.05 (0.03, 0.07)\\
& $\overline{m}_2$ = 1.74 (1.55, 1.94) & $\overline{\rho}_2$ = 1.59 (1.30, 1.90) & $\lambda_2$ = 0.82 (0.76, 0.89)\\
& $\overline{m}_3$ = 5.07 (4.66, 5.52) & $\overline{\rho}_3$ = 2.52 (1.35, 17.50) & $\lambda_3$ = 0.13 (0.07, 0.19)\\
\multirow{3}{*}{\shortstack[c]{2000-2001\\ \textit{SvM-c-3}}} & $\overline{m}_1$ = 1.00 (0.50, 1.52) & $\overline{\rho}_1$ = 0.20 (2.61 e-5, 79.84) & $\lambda_1$ = 0.02 (6.66e-4, 0.08)\\
& $\overline{m}_2$ = 2.55 (2.24, 2.86) & $\overline{\rho}_2$ = 1.97 (1.54, 2.51) & $\lambda_2$ = 0.50 (0.43, 0.58)\\
& $\overline{m}_3$ = 4.26 (3.91, 4.62) & $\overline{\rho}_3$ = 1.46 (1.16, 1.88) & $\lambda_3$ = 0.48 (0.40, 0.55)\\
\multirow{3}{*}{\shortstack[c]{2001-2002\\ \textit{SvM-c-3}}} & $\overline{m}_1$ = 1.49 (1.25, 1.81) & $\overline{\rho}_1$ = 2.18 (1.75, 2.71) & $\lambda_1$ = 0.58 (0.49, 0.68)\\
& $\overline{m}_2$ = 3.63 (4.40, 5.74) & $\overline{\rho}_2$ = 1.50 (1.05, 2.56) & $\lambda_2$ = 0.35 (0.22, 0.48)\\
& $\overline{m}_3$ = 5.15 (4.53, 5.84) & $\overline{\rho}_3$ = 1.79 (8.63e-4, 18.20) & $\lambda_3$ = 0.07 (2.03e-3, 0.16)\\
\multirow{3}{*}{\shortstack[c]{2002-2003\\ \textit{SvM-c-3}}} & $\overline{m}_1$ = 1.04 (0.32, 1.56) & $\overline{\rho}_1$ = 0.12 (2.03e-5, 76.01) & $\lambda_1$ = 0.03 (7.63e-4, 0.09)\\
& $\overline{m}_2$ = 1.74 (1.51, 1.97) & $\overline{\rho}_2$ = 1.77 (1.44, 2.19) & $\lambda_2$ = 0.69 (0.61, 0.77)\\
& $\overline{m}_3$ = 5.25 (4.83, 5.67) & $\overline{\rho}_3$ = 1.41 (1.00, 2.03) & $\lambda_3$ = 0.29 (0.21, 0.37)\\
\multirow{3}{*}{\shortstack[c]{2003-2004\\ \textit{SvM-c-3}}} & $\overline{m}_1$ = 1.38 (0.98, 1.74) & $\overline{\rho}_1$ = 24.51 (5.91, 82.95) & $\lambda_1$ = 0.07 (0.04, 0.11)\\
& $\overline{m}_2$ = 2.03 (1.91, 2.14) & $\overline{\rho}_2$ = 3.21 (2.71, 3.82) & $\lambda_2$ = 0.81 (0.75, 0.86)\\
& $\overline{m}_3$ = 5.24 (4.68, 5.88) & $\overline{\rho}_3$ = 0.20 (2.62e-4, 1.40) & $\lambda_3$ = 0.12 (0.08, 0.18)\\
\multirow{2}{*}{\shortstack[c]{2004-2005\\ \textit{SvM-c-2}}} & $\overline{m}_1$ = 2.10 (2.01, 2.19) & $\overline{\rho}_1$ = 4.76 (4.12, 5.52) & $\lambda_1$ = 0.86 (0.82, 0.90)\\
& $\overline{m}_2$ = 1.52 (0.85, 2.23) & $\overline{\rho}_2$ = 0.78 (0.48, 1.13) & $\lambda_2$ = 0.14 (0.10, 0.18)\\
\multirow{2}{*}{\shortstack[c]{2005-2006\\ \textit{SvM-c-2}}} & $\overline{m}_1$ = 2.39 (2.28, 2.49) & $\overline{\rho}_1$ = 4.10 (3.50, 4.80) & $\lambda_1$ = 0.76 (0.72, 0.80)\\
& $\overline{m}_2$ = 1.29 (0.84, 1.84) & $\overline{\rho}_2$ = 1.05 (0.78, 1.37) & $\lambda_2$ = 0.24 (0.18, 0.20)\\
\multirow{2}{*}{\shortstack[c]{2006-2007\\ \textit{SvM-c-2}}} & $\overline{m}_1$ = 4.00 (2.81, 5.22) & $\overline{\rho}_1$ = 0.99 (0.73, 1.41) & $\lambda_1$ = 0.33 (0.21, 0.48)\\
& $\overline{m}_2$ = 5.40 (5.17, 5.61) & $\overline{\rho}_2$ = 2.42 (1.90, 3.31) & $\lambda_2$ = 0.67 (0.52, 0.79)\\
\multirow{3}{*}{\shortstack[c]{2007-2008\\ \textit{SvM-c-3}}} & $\overline{m}_1$ = 1.04 (0.52, 1.56) & $\overline{\rho}_1$ =  2.77e-03 (4.92e-6, 0.12) & $\lambda_1$ = 0.09 (0.06, 0.11)\\
& $\overline{m}_2$ = 4.93 (4.76, 5.12) & $\overline{\rho}_2$ = 49.64 (22.77, 88.37) & $\lambda_2$ = 0.14 (0.11, 0.18)\\
& $\overline{m}_3$ = 5.26 (5.18, 5.35) & $\overline{\rho}_3$ = 6.79 (5.64, 8.30) & $\lambda_3$ = 0.77 (0.73, 0.81)\\
\multirow{3}{*}{\shortstack[c]{2008-2009\\ \textit{SvM-c-3}}} & $\overline{m}_1$ = 1.26 (0.58, 1.74) & $\overline{\rho}_1$ = 0.87 (3.78e-5, 93.44) & $\lambda_1$ = 0.08 (0.03, 0.15)\\
& $\overline{m}_2$ = 2.75 (1.92, 3.45) & $\overline{\rho}_2$ = 12.64 (1.10, 94.47) & $\lambda_2$ = 0.07 (0.04, 0.12)\\
& $\overline{m}_3$ = 5.16 (5.07, 5.24) & $\overline{\rho}_2$ = 3.30 (2.86, 3.82) & $\lambda_2$ = 0.85 (0.80, 0.89)\\
\multirow{3}{*}{\shortstack[c]{2009-2010\\ \textit{SvM-c-3}}} & $\overline{m}_1$ = 1.32 (1.22, 1.42) & $\overline{\rho}_1$ = 87.23 (77.20, 97.49) & $\lambda_1$ = 0.10 (0.08, 0.11)\\
& $\overline{m}_2$ = 2.60 (2.24, 3.01) & $\overline{\rho}_2$ = 1.53 (1.17, 2.42) & $\lambda_2$ = 0.48 (0.33, 0.58)\\
& $\overline{m}_3$ = 5.18 (4.88, 5.49) & $\overline{\rho}_3$ = 2.29 (1.57, 3.34) & $\lambda_3$ = 0.43 (0.32, 0.57)\\
\hline \\[-1.8ex] 
\end{tabular} 
}
\caption{Circular mean posterior values and 95\% credible intervals for the parameters in the Von Mises distribution and mixing probability are shown for the models fitted to the income proportions in Los Angeles County from 1990 to 2010. Parameters with bars over them were averaged across all locations.}
\label{table:real_data_results_param_summary}
\end{table}

\newpage
\bibliography{my_refs}